\documentclass[12pt]{article}
\usepackage[utf8]{inputenc}
\usepackage{amsmath}
\usepackage{amssymb}
\usepackage{graphicx}

\makeatletter

\DeclareTextSymbolDefault{\textquotedbl}{T1}

\numberwithin{equation}{section}

\@ifundefined{date}{}{\date{}}

\usepackage{braket}
\usepackage{bm}
\usepackage{bbm}

\usepackage{epsfig}
\usepackage{pstricks}
\usepackage{color}
\usepackage{tikz}

\usepackage{hyperref}

\newcommand{\punc}{K}
\newcommand {\nn} {\nonumber}

\@ifundefined{showcaptionsetup}{}{%
 \PassOptionsToPackage{caption=false}{subfig}}
\usepackage{subfig}
\makeatother

\begin{document}
\begin{titlepage}
\renewcommand{\thefootnote}{\fnsymbol{footnote}}

\begin{flushright} 
KEK-TH-2209
\end{flushright} 


\begin{center}
{\bf \large Complex Langevin analysis of
2D U(1) gauge theory\\
on a torus with a $\theta$ term}
\end{center}



\begin{center}
         Mitsuaki H{\sc irasawa}$^{1)}$\footnote
          { E-mail address : mitsuaki@post.kek.jp},
         Akira M{\sc atsumoto}$^{1)}$\footnote
          { E-mail address : akiram@post.kek.jp},\\
         Jun N{\sc ishimura}$^{1,2)}$\footnote
          { E-mail address : jnishi@post.kek.jp} and
         Atis Y{\sc osprakob}$^{1)}$\footnote
          { E-mail address : ayosp@post.kek.jp}


\vspace{1cm}

$^{1)}$\textit{Department of Particle and Nuclear Physics,}\\
\textit{School of High Energy Accelerator Science,}\\
{\it Graduate University for Advanced Studies (SOKENDAI),\\
1-1 Oho, Tsukuba, Ibaraki 305-0801, Japan} 

~

$^{2)}$\textit{KEK Theory Center,
Institute of Particle and Nuclear Studies,}\\
{\it High Energy Accelerator Research Organization,\\
1-1 Oho, Tsukuba, Ibaraki 305-0801, Japan} 
\end{center}

\vspace{0.5cm}

\begin{abstract}
\noindent Monte Carlo simulation of gauge theories with a $\theta$ term is 
known to be extremely difficult due to the sign problem.
Recently there has been major progress in solving this problem
based on the idea of complexifying dynamical variables.
Here we consider the complex Langevin method (CLM), which is 
a promising approach for its low computational cost.
The drawback of this method, however, is 
the existence of a condition that has to be met in order for the 
results to be correct.
As a first step, we apply the method to
2D U(1) gauge theory on a torus with a $\theta$ term, 
which can be solved analytically.
We find that a naive implementation of the method fails because of the
topological nature of the $\theta$ term.
In order to circumvent this problem, 
we simulate the same theory on 
a punctured torus, which 
is equivalent to the original model
in the infinite volume limit for $ |\theta| < \pi$.
Rather surprisingly, we find that the CLM works 
and reproduces the exact results for a punctured torus
even at large $\theta$, where the link variables near the puncture
become very far from being unitary.
%
\end{abstract}
\vfill
\end{titlepage}
\vfil\eject


\setcounter{footnote}{0}

\section{Introduction}

The $\theta$ term
provides an interesting avenue of research in quantum field theories.
Due to its topological nature,
its effects on physics should be genuinely nonperturbative, if present at all.
In particular, it does not affect the equation of motion,
which implies that $\theta$ is a parameter that does not exist
in the corresponding classical theory.
For instance, the $\theta$ term in QCD is given by 
$S_{\rm \theta}= - i \theta Q$,
where $Q$ is the topological charge defined by
\begin{equation}
Q = \frac{1}{32\pi^2}\epsilon_{\mu\nu\rho\sigma} \int d^4 x 
\, {\rm tr} F_{\mu\nu }F_{\rho\sigma} \ ,
\label{def-theta-term}
\end{equation}
which takes integer values on a compact space.
This term is renormalizable by power-counting and hence 
it is a term which is perfectly sensible to add in the action.
However, it breaks parity and time-reversal symmetries,
and hence the CP symmetry.
This leads to a non-vanishing electric dipole moment of a neutron, 
which is severely restricted by experiments.
The upper bound on $\theta$ thus obtained
is $|\theta| \lesssim 10^{-10}$ \cite{Baker:2006ts},
which is extremely small although there is no reason 
for it theoretically.
This is a naturalness problem known as the strong CP problem.

A popular solution to this problem is 
the Peccei-Quinn 
mechanism \cite{Peccei:1977hh, Peccei:1977ur, Weinberg:1977ma, Wilczek:1977pj},
which introduces axions as a pseudo Nambu-Goldstone boson
of a hypothetical global ${\rm U}(1)_{\rm PQ}$ symmetry.
In this mechanism, the potential for the axions induced by QCD
chooses a CP invariant vacuum automatically.
Recently, gauge theories with a $\theta$ term
have attracted attention also from the viewpoint of
the 't Hooft anomaly matching 
condition \cite{Gaiotto:2017yup, Kitano:2017jng, Kan:2019rsz}
and the gauge-gravity 
correspondence \cite{Parnachev:2008fy, Dubovsky:2011tu, Bigazzi:2015bna, Arean:2016hcs}.
In particular, there is an interesting prediction
for a phase transition at $\theta = \pi$,
which claims that either spontaneous CP breaking or deconfinement
should occur there \cite{Gaiotto:2017yup,Kitano:2017jng}.
In order to investigate gauge theories from first principles
in the presence of a $\theta$ term
motivated either by the physics related to axions or by the
recent predictions,
one needs to perform nonperturbative calculations based
on Monte Carlo methods.
However, this is known to be extremely difficult
because the $\theta$ term
appears as a purely imaginary term in the 
Euclidean action $S$.
The Boltzmann weight ${\rm e}^{-S}$ becomes complex, and 
one cannot interpret it as the probability distribution
as one does in Monte Carlo methods.

One can still use the reweighting method by treating
the phase of the complex weight as a part of the observable.
In the case at hand, this amounts to obtaining
the histogram of the topological charge
at $\theta=0$ and taking an average over the topological sectors
characterized by the integer $Q$
with the weight ${\rm e}^{i \theta Q}$.
Various results obtained in this way
are nicely reviewed in Ref.~\cite{Vicari:2008jw}.
Clearly, the calculation becomes extremely difficult
due to huge cancellations between topological sectors
when topological sectors with $|Q| \gg \pi/|\theta|$ 
make significant contributions to the partition function,
which occurs either for $|\theta| \sim \pi$ or for 
smaller $|\theta|$ with sufficiently large volume.
This is the so-called sign problem
which occurs in general for systems with a complex action.

There are actually a few promising methods that 
can handle systems which suffer from this problem
such as
the complex Langevin method (CLM) \cite{Klauder:1983sp, Parisi:1984cs, Aarts:2009uq, Aarts:2011ax, Nagata:2015uga, Nagata:2016vkn}, 
the generalized Lefschetz thimble 
method \cite{Witten:2010cx, Cristoforetti:2012su, Fujii:2013sra, Alexandru:2015sua, Fukuma:2017fjq, Fukuma:2019uot}, 
the path optimization method \cite{Mori:2017pne, Mori:2017nwj, Kashiwa:2018vxr, Kashiwa:2019lkv} 
and 
the tensor network 
method \cite{Levin:2006jai, PhysRevB.86.045139, PhysRevLett.115.180405, Adachi:2019paf, Kadoh:2019kqk}.
Each method has its pros and cons, however.
In this work, we focus on the CLM, which can be applied to 
various physically interesting models with large system size 
in a straightforward manner.
(See, for instance, Refs.~\cite{Tsutsui:2019suq,Nishimura:2019qal,Anagnostopoulos:2020xai}.)
The only drawback of the method is that it can give wrong results
depending on the system, the parameter region, and even on the choice
of the dynamical variables.
Recently, the reason for this behavior has been understood 
theoretically \cite{Aarts:2009uq,Aarts:2011ax,Nagata:2015uga,Nagata:2016vkn,Scherzer:2018hid,Scherzer:2019lrh}.
In particular, Ref.~\cite{Nagata:2016vkn} proposed
a practical criterion for correct convergence, which 
made the CLM a method of choice for complex action systems
as long as the criterion is met.
There are indeed many successful applications to
lattice quantum
field theory 
\cite{Berges:2005yt,Berges:2006xc,Aarts:2008wh,Aarts:2009hn,Sexty:2013ica,Fodor:2015doa,Aarts:2016qrv,Nagata:2018net,Nagata:2018mkb,Ito:2018jpo,Kogut:2019qmi,Sexty:2019vqx,Tsutsui:2019suq}
and matrix models 
\cite{Mollgaard:2014mga,Ito:2016efb,Bloch:2017sex,Anagnostopoulos:2017gos,Nishimura:2019qal,Nagata:2018net,Basu:2018dtm,Joseph:2019sof}
with complex actions.

As a first step towards its application to 4D
non-Abelian gauge theories,
here we apply it to 2D U(1) lattice gauge theory 
with a $\theta$ term, which suffers from a severe sign problem 
despite its simplicity.\footnote{Abelian gauge theories have also
been discussed in the context of a non-standard lattice discretization,
which is not only theoretically beneficial \cite{Sulejmanpasic:2019ytl}
but also enables a dual simulation \cite{Gattringer:2018dlw}, 
which is free from the sign problem.}
In fact, the model can be solved analytically 
with finite lattice spacing and finite volume on an arbitrary 
manifold \cite{Wiese:1988qz, Rusakov:1990rs, Bonati:2019ylr},
which makes it a useful testing ground
for new
methods \cite{Wiese:1988qz, Hassan:1995dn, Plefka:1996tz, Kuramashi:2019cgs}
aiming at solving the sign problem.
By using the reweighting method \cite{Plefka:1996tz}, for instance,
one can only reach $\theta\sim2.2$ with a $16\times16$ lattice,
and in particular, it seems almost impossible to approach $\theta=\pi$ by 
this method.
Note also that
the region of $\theta$ that can be explored by this method 
shrinks to zero
as one increases the lattice size.

We find that a naive implementation of the CLM fails.
The reason for this is that the configurations that
appear when the topology change occurs during the Langevin process
necessarily result in a large drift term.
Due to this fact, the criterion \cite{Nagata:2016vkn}
for the validity of the method based on the histogram of the drift term
cannot be satisfied. If one tries to suppress the appearance of
the problematic configurations by approaching the continuum limit,
the criterion can be satisfied, but the topology change does not occur
during the Langevin process, hence the ergodicity is lost.

In order to cure this problem, we introduce a puncture
on the torus, which makes the base manifold noncompact.
We have obtained exact results for this punctured model as well.
Even in the continuum limit,
the topological charge is no longer restricted
to integer values 
and the $2\pi$ periodicity in $\theta$ does not hold.
However, if we take the infinite volume limit with $|\theta| < \pi$,
one cannot distinguish the model from the original non-punctured model
as far as the observables that make sense in that limit are concerned.
Note that in that limit, the topological charge can take 
arbitrarily large values and therefore it does not
really matter whether it is an integer or not.

On the other hand, the situation of the complex Langevin simulation changes 
drastically for the punctured model.
The topology change occurs freely and the 
appearance of the problematic configurations can be suppressed
by simply approaching the continuum limit.
Thus the criterion for the validity of the CLM is met
without losing the ergodicity,
and we are able to reproduce the exact results
for the punctured model.

The most striking aspect of our results is that
the CLM works
even if
the link variables close to the puncture become very far from being unitary.
This can happen because
the direct effect of the $\theta$ term on the complex Langevin dynamics
is actually concentrated on these link variables.
While the link variables are allowed to be non-unitary in the CLM 
in general in order to include the effects of the complex action,
all the previous work suggested that
the condition for the validity cannot be satisfied
unless the non-unitarity is sufficiently suppressed.
Precisely for this reason, the gauge 
cooling \cite{Seiler:2012wz,Nagata:2015uga, Nagata:2016vkn}
was invented as a crucial technique in applying the CLM to
gauge theories.
In fact, we also use the gauge cooling in our simulation, 
but 
the link variables close to the puncture
nevertheless 
become far from being unitary when $\theta$ or the physical volume gets large.
Yet the criterion for the drift term is not violated
and the exact results are perfectly reproduced.

The rest of this paper is organized as follows.
In Section \ref{sec:2d_u1}, we make a brief review on
2D U(1) gauge theory on a torus with a $\theta$ term
and discuss how to put it on a lattice.
In Section \ref{sec:CLM}, we
apply the CLM to this theory and 
show that a naive implementation fails.
In Section \ref{sec:introduce-puncture},
we introduce a puncture on the torus in order to circumvent the 
problem encountered in Section \ref{sec:CLM}, and show
the equivalence of the punctured model and 
the original non-punctured model
in the infinite volume limit for $|\theta|<\pi$.
In Section \ref{sec:results}, we apply the CLM to the 
punctured model and show that it works perfectly even at large $\theta$,
where the link variables near the puncture
become very far from being unitary.
Section \ref{sec:summary} is devoted to a summary and discussions.
In Appendix \ref{sec:analytic_calc}, we review how one can solve
the theory analytically for various boundary conditions,
and obtain exact results for the punctured and non-punctured models,
in particular.
In Appendix \ref{sec:results-punctured}, we show our results
of the CLM for the punctured model using a different lattice definition 
of the topological charge.

\paragraph{Note added.}
When this paper was about to be completed,
we encountered a preprint \cite{Gattringer:2020mbf}
on the arXiv, which investigates the same theory 
numerically by the density of states method.
The exact results are reproduced
for $|\theta|<\pi$ up to $L=24$,
which goes far beyond 
the reweighting method \cite{Plefka:1996tz}.
It is crucial to use an open boundary condition,
which is similar in spirit to introducing a puncture 
in our work 
although the purpose is quite different.

\section{2D U(1) lattice gauge theory with a $\theta$ term}
\label{sec:2d_u1}

In this section, we review
2D U(1) gauge theory with a $\theta$ term
and discuss how to define it on a lattice.

In the continuum 2D U(1) gauge theory on a Euclidean space,
the action for the gauge field $A_\mu (x)$ ($\mu=1,2$) is given by 
\begin{equation}
S_{g}=\frac{1}{4g^{2}}\int d^{2}x \, (F_{\mu\nu})^2 \ ,
\label{gauge-action-cont}
\end{equation}
where $g$ is the gauge coupling constant and $F_{\mu\nu}$ is the field
strength defined as
\begin{equation}
F_{\mu\nu}=\partial_{\mu}A_{\nu}(x) - \partial_{\nu}A_{\mu}(x) \ .
\label{def-F}
\end{equation}
We add a $\theta$ term 
\begin{equation}
S_{\theta}=-i\theta Q
\label{def-theta-term-2d}
\end{equation}
in the action, where $Q$ is the topological charge defined by 
\begin{equation}
Q=\frac{1}{4\pi}\int d^{2}x \, \epsilon_{\mu\nu}F_{\mu\nu} \ ,
\label{def-Q}
\end{equation}
which takes integer values if the space is compact.


We put this theory on a 2D torus, which is discretized into
an $L\times L$ periodic lattice
with the lattice spacing $a$.
On the lattice, we
define the link variables $U_{n,\mu} \in\mathrm{U(1)}$,
where $n$ labels the lattice site as $x_\mu =a n_\mu $.
We also define the plaquette
\begin{equation}
P_{n}=U_{n,\hat{1}}U_{n+\hat{1},2}U_{n+\hat{2},1}^{-1}U_{n,2}^{-1} \ ,
\label{def-P}
\end{equation}
which is a gauge invariant object.
Here we write $U_{n,\mu}^{-1}$ instead of $U_{n,\mu}^{\dagger}$,
which will be important later 
in applying 
the CLM,
where we complexify the dynamical variables respecting holomorphicity.

The lattice counterpart of the field strength (\ref{def-F}) 
can be defined as
\begin{equation}
F_{n , 12} = \frac{1}{i a^2} \log P_{n} \ ,
\end{equation}
where we take the principal value for the complex log;
namely $\log z=\log|z|+i\arg z$ with $\qquad-\pi<\arg z\leq\pi$.
Since the plaquette can then be written in terms of $F_{n , \mu\nu}$ as
\begin{align}
P_{n} & =e^{ia^{2}F_{n,12}} \ ,
\label{P-F-rel}
\end{align}
the lattice counterpart of the gauge action
(\ref{gauge-action-cont}) can be defined as
\begin{equation}
S_{g}
= -\frac{\beta}{2}\sum_{n}
\left(P_{n}+P_{n}^{-1}\right)
=-\beta\sum_{n}\cos\left(a^{2}F_{n,12}\right) \ ,
\label{eq:S_beta}
\end{equation}
which approaches
\begin{equation}
S_{g} \simeq \frac{1}{4g^{2}}\sum_{n}a^{2} (F_{n,\mu\nu})^2 \ ,
\end{equation}
in the continuum limit up to an irrelevant constant
with the identification
\begin{equation}
\label{beta-g-relation}
\beta=\frac{1}{(ga)^{2}} \ .
\end{equation}

In the present 2D U(1) theory,
the topological charge can be defined as
\begin{equation}
Q_{\mathrm{log}}=\frac{1}{4\pi}\sum_{n}a^{2}\epsilon^{\mu\nu}F_{n,\mu\nu}
=-\frac{i}{2\pi}\sum_{n}\log P_{n} \ ,
\label{log-Q-def}
\end{equation}
which gives an integer value even at finite $a$.
This can be proved easily by noting that $\prod _n P_{n} = 1$
since each link variable appears twice in this product with 
opposite directions.
We call this definition (\ref{log-Q-def}) the ``log definition''.
As an alternative definition, we consider
\begin{align}
Q_{\mathrm{sin}} & 
=-\frac{i}{4\pi}\sum_{n}\left(P_{n}-P_{n}^{-1}\right)
=\frac{1}{2\pi}\sum_{n}\sin\left(a^{2}F_{n,12}\right) \ ,
\label{simp-Q-def}
\end{align}
which approaches (\ref{def-Q}) in the continuum limit
recalling (\ref{P-F-rel}).
Note, however, that the topological charge defined on the lattice
in this way can take non-integer values
in general before taking the continuum limit.
We call this definition (\ref{simp-Q-def})
the ``sine definition''.
Thus the lattice theory is given by
\begin{equation}
S=S_g+S_\theta \ , 
\label{action-decompose}
\end{equation}
where $S_g$ is given by
(\ref{eq:S_beta}) and $S_\theta$ is given by
(\ref{def-theta-term-2d}) with $Q$ 
defined either by (\ref{log-Q-def}) or by (\ref{simp-Q-def}).

Since this theory is superrenormalizable, we can take 
the continuum limit $a\rightarrow 0$ with fixed $g$,
which is set to unity throughout this paper
without loss of generality.
In this unit, the physical volume of the 2D torus is given by
\begin{align}
V_{\rm phys} = (La)^2  = \frac{L^2}{\beta} \ ,
\label{phys-vol}
\end{align}
where we have used (\ref{beta-g-relation}).

\section{Applying the CLM to the 2D U(1) gauge theory}
\label{sec:CLM}

Since the $\theta$ term
is purely imaginary in general,
it makes Monte Carlo studies of gauge theories extremely difficult
due to the sign problem.
We overcome this problem by using the complex Langevin method 
(CLM) \cite{Klauder:1983sp, Parisi:1984cs, Aarts:2009uq, Aarts:2011ax, Nagata:2015uga, Nagata:2016vkn},
which extends the idea of stochastic quantization to systems
with complex actions.
In this section, we discuss how to apply the CLM 
to 2D U(1) gauge theory with a $\theta$ term and show some results,
which suggest that a naive implementation of the method fails.


\subsection{the complex Langevin equation}
\label{CLMU1} 

The first step of the CLM is to complexify the dynamical variables.
In the present case of U(1) gauge theory, 
we extend the link variables $U_{n,\mu}\in \mathrm{U(1)}$
to $\mathcal{U}_{n,\mu}\in\mathbb{C}\setminus\{0\}$, 
which corresponds to extending the gauge field 
$A_{\mu}(x)\in\mathbb{R}$ to $\mathcal{A}_{\mu}(x)\in\mathbb{C}$
in the continuum theory. 
Then we consider a fictitious time evolution $\mathcal{U}_{n,\mu}(t)$
of the link variables governed by the complex Langevin equation
\begin{equation}
\mathcal{U}_{n,\mu}(t+\Delta t)=\mathcal{U}_{n,\mu}(t)
\exp \Big[i \Big\{ - \Delta t \, D_{n,\mu}S 
+\sqrt{\Delta t} \, \eta_{n,\mu}(t) \Big\} \Big] \ ,
\label{CLE_U1}
\end{equation}
where $\eta_{n,\mu}(t)$ is a real Gaussian noise normalized by
$\langle \eta_{n,\mu}(s) \eta_{k,\nu}(t)\rangle = 2 \delta_{n,k}
\delta_{\mu,\nu} \delta_{s,t}$.
The term $D_{n,\mu}S$ is the drift term defined by
\begin{equation}
D_{n,\mu}S =
\lim_{\epsilon\to0}
\frac{S( e^{i\epsilon} U_{n,\mu})-S(U_{n,\mu})}{\epsilon} \ ,
\label{def-drift-term}
\end{equation}
first for the unitary link variables $U_{n,\mu}(t)$,
and then it is defined for the complexified link variables
$\mathcal{U}_{n,\mu}(t)$ by analytic continuation in order to 
respect holomorphicity.
Using the action (\ref{action-decompose}),
we obtain $D_{n,\mu}S=D_{n,\mu}S_{g}+D_{n,\mu}S_{\theta}$,
where the first term is given as
\begin{alignat}{2}
D_{n,1}S_{g} & = &  & 
-i\frac{\beta}{2}(P_{n}-P_{n}^{-1}-P_{n-\hat{2}}+P_{n-\hat{2}}^{-1}) \ ,
\nn
\\
D_{n,2}S_{g} & = &  & 
-i\frac{\beta}{2}(-P_{n}+P_{n}^{-1}+P_{n-\hat{1}}-P_{n-\hat{1}}^{-1}) \ .
\label{drift-plaquette}
\end{alignat}

The second term $D_{n,\mu}S_{\theta}$
depends on the definition of the topological charge. 
If one uses the log definition (\ref{log-Q-def}),
Eq.~\eqref{def-drift-term} for the $\theta$ term
becomes a $\delta$-function,
which vanishes identically except 
for configurations with $P_{n} = -1$ for some $n$,
reflecting the topological nature of the definition.
Such configurations are precisely 
the ones that appear when 
the topology change occurs within the configuration space of $U_{n , \mu}$.
It is not straightforward to extend such a term to
a holomorphic function of $\mathcal{U}_{n , \mu}$.
%

On the other hand, if one uses the sine definition (\ref{simp-Q-def}),
the drift term becomes
\begin{alignat}{2}
D_{n,1}S_{\theta} & = &  & -i\frac{\theta}{4\pi}
(P_{n}+P_{n}^{-1}-P_{n-\hat{2}}-P_{n-\hat{2}}^{-1}) \ ,
\nn
\\
D_{n,2}S_{\theta} & = &  & -i\frac{\theta}{4\pi}
(-P_{n}-P_{n}^{-1}+P_{n-\hat{1}}+P_{n-\hat{1}}^{-1}) \ ,
\label{drift-simple-Q}
\end{alignat}
which
may be viewed as an approximation of
the $\delta$-function mentioned above.
Moreover, it can be readily extended to
a holomorphic function of $\mathcal{U}_{n , \mu}$.
For this reason, 
we use the sine definition for the non-punctured model.

The criterion \cite{Nagata:2016vkn}
for the validity of the CLM states
that the histogram of the drift term
should fall off exponentially or faster.
There are two cases in which this criterion cannot be met.
The first case occurs when
the configuration comes close to the poles
of the drift terms
(\ref{drift-plaquette}), (\ref{drift-simple-Q}),
which correspond to configurations with $P_{n} = 0$ for some $n$.
If this happens
during the Langevin process,
there is a possibility of violating the criterion.
This problem is called the singular-drift
problem \cite{Nishimura:2015pba,Aarts:2017vrv},
which was found first in simple 
models \cite{Mollgaard:2013qra,Greensite:2014cxa}.
In the present model, the same problem is caused also by
approaching configurations with $|P_{n}| = \infty$ for some $n$,
which are related to the poles by the parity transformation.

The second case occurs when
the dynamical variables make large excursions in the imaginary 
directions \cite{Aarts:2009uq}. 
This problem is called the excursion problem.
In the present model, this corresponds to the situation in which
the link variables have absolute values $|\mathcal{U}_{n,\mu}|$ 
far from unity.

Both the singular-drift problem and the excursion problem
can occur because the link variables $\mathcal{U}_{n,\mu}$
are not restricted to be unitary in the CLM.
In order to avoid these problems,
it is important to perform the gauge cooling, which we
explain in the next section.


\subsection{gauge cooling}
\label{gauge_cooling} 

The idea of gauge cooling \cite{Seiler:2012wz} 
is to reduce the non-unitarity of link variables
as much as possible by making gauge transformations corresponding
to the complexified Lie group after each step (\ref{CLE_U1})
of the Langevin process.
This procedure can be added without affecting the argument for
justifying the CLM as demonstrated explicitly 
in Refs.~\cite{Nagata:2015uga, Nagata:2016vkn}.
Recently, the mechanism of the gauge cooling for stabilizing 
the complex Langevin simulation has been investigated \cite{Cai:2019vmt}.

The deviation of the link variables from U(1) can be defined
by the unitarity norm
\begin{equation}
\mathcal{N}=
\frac{1}{2L^2}\sum_{n,\mu}
\Big\{\mathcal{U}_{n,\mu}^{*}\mathcal{U}_{n,\mu}+
(\mathcal{U}_{n,\mu}^{*}\mathcal{U}_{n,\mu})^{-1}-2 \Big\} \ .
\label{def-unitarity-norm}
\end{equation}
The gauge cooling reduces this quantity by
a complexified gauge transformation,
which
is determined as follows.

First we consider an infinitesimal gauge transformation
\begin{equation}
  \delta \mathcal{U}_{n,\mu}
  = ( \epsilon_{n} - \epsilon_{n+\hat{\mu}}) \, \mathcal{U}_{n,\mu} \ , 
\end{equation}
where $\epsilon_{n}\in\mathbb{R}$.
The change of the unitarity norm due to the transformation is given by
\begin{alignat}{1}
\delta{\mathcal{N}} & =\frac{1}{2L^2}
\sum_{n,\mu} \Big\{2(\epsilon_{n}-\epsilon_{n+\hat{\mu}})
\mathcal{U}_{n,\mu}^{*}\mathcal{U}_{n,\mu}-
2(\epsilon_{n}-\epsilon_{n+\hat{\mu}})
(\mathcal{U}_{n,\mu}^{*}\mathcal{U}_{n,\mu})^{-1} \Big\}
\nonumber \\
 & =\frac{1}{2L^2}\sum_{n}2\epsilon_{n}G_{n} \ ,
\end{alignat}
where $G_{n}$ is defined as 
\begin{equation}
G_{n}=\sum_{\mu}\Big\{\mathcal{U}_{n,\mu}^{*}
\mathcal{U}_{n,\mu}-\mathcal{U}_{n-\hat{\mu},\mu}^{*}
\mathcal{U}_{n-\hat{\mu},\mu}-(\mathcal{U}_{n,\mu}^{*}\mathcal{U}_{n,\mu})^{-1}
+(\mathcal{U}_{n-\hat{\mu},\mu}^{*}\mathcal{U}_{n-\hat{\mu},\mu})^{-1}\Big\} \ .
\end{equation}
Therefore, we find that
the unitarity norm is reduced most efficiently
by choosing $\epsilon_{n}\propto-G_{n}$.

Using this result, we consider a finite gauge transformation
\begin{equation}
  \mathcal{U}_{n,\mu} \mapsto
  g_{n} \, \mathcal{U}_{n,\mu} \, g_{n+\hat{\mu}}^{-1} \ ; \quad \quad
  g_{n}=e^{-\alpha G_{n}} \ ,
\label{g.c.}
\end{equation}
which makes the unitarity norm 
\begin{align}
\mathcal{N(\alpha)}&=
\frac{1}{2L^2}\sum_{n,\mu}
\Big\{\mathcal{U}_{n,\mu}^{*}\mathcal{U}_{n,\mu}e^{-2\alpha(G_{n}-G_{n+\hat{\mu}})}
+(\mathcal{U}_{n,\mu}^{*}\mathcal{U}_{n,\mu})^{-1}e^{2\alpha(G_{n}-G_{n+\hat{\mu}})}-2 \Big\}
\ ,
\label{unitarity-norm-alpha}
\end{align}
depending on $\alpha$ in (\ref{g.c.}).
We search for an optimal
$\alpha$ that minimizes (\ref{unitarity-norm-alpha}).
Note here that it is typically a small number
since the gauge cooling is performed after each step of the Langevin
process.
We therefore expand Eq.~(\ref{unitarity-norm-alpha})
with respect to $\alpha$ up to the second order and
obtain the value of $\alpha$ that minimizes it as
\begin{equation}
\begin{split}
\alpha
&=\frac{1}{2}\frac{\sum_{n}G_n^2}
{\sum_{n,\mu}[(G_{n}-G_{n+\hat{\mu}})^{2}\{\mathcal{U}_{n,\mu}^{*}
\mathcal{U}_{n,\mu}+(\mathcal{U}_{n,\mu}^{*}\mathcal{U}_{n,\mu})^{-1}\}]} \ .
\end{split}
\end{equation}
We repeat this procedure until the unitarity norm
changes by a fraction less than $10^{-5}$.
%

\subsection{adaptive stepsize}
\label{adaptive_step_size} 

When we solve the complex Langevin equation in its discretized version
(\ref{CLE_U1}), it occasionally 
happens that the drift term becomes extremely large,
in particular during the thermalization process.
This causes a large discretization error, which
either makes the thermalization slow or destabilizes the simulation.
We can avoid this problem by using
a small stepsize $\Delta t$,
but the computational cost for a fixed Langevin time increases 
proportionally to $(\Delta t)^{-1}$ and 
the calculation becomes easily unfeasible.
The adaptive stepsize \cite{Aarts:2009dg}
is a useful technique, which amounts to
reducing the stepsize only when the drift term becomes large. 

In our simulation, we measure 
the magnitude of the drift term defined as
\begin{equation}
\label{def-u}
u = \max_{n,\mu} |D_{n,\mu}S|
\end{equation}
at each step, and choose the Langevin stepsize $\Delta t$
in (\ref{CLE_U1}) as
\begin{equation}
\Delta t=\left\{\begin{array}{ll}
\Delta t_{0} &{\rm for\ } u<v_{0} \ ,\\
\displaystyle\frac{v_{0}}{u}\Delta t_{0} 
& {\rm otherwise} \ ,
\end{array}\right.
\end{equation}
where $\Delta t_{0}$ is the default stepsize, and 
$v_{0}$ is the threshold for the magnitude of drift term. 
In the present work,
the default stepsize is set to $\Delta t_{0} = 10^{-5}$,
and the threshold is set to $v_{0}=2\beta$,
considering a bound $u\le2\beta$ for $\theta=0$,
where the CLM reduces to the real Langevin method.
The measurement of the observables should be made 
with the same interval in terms of the Langevin time
but not in terms of the number of steps.

\begin{figure}
\centering
$\!\!\!\!\!\!$
{\includegraphics[scale=0.6]{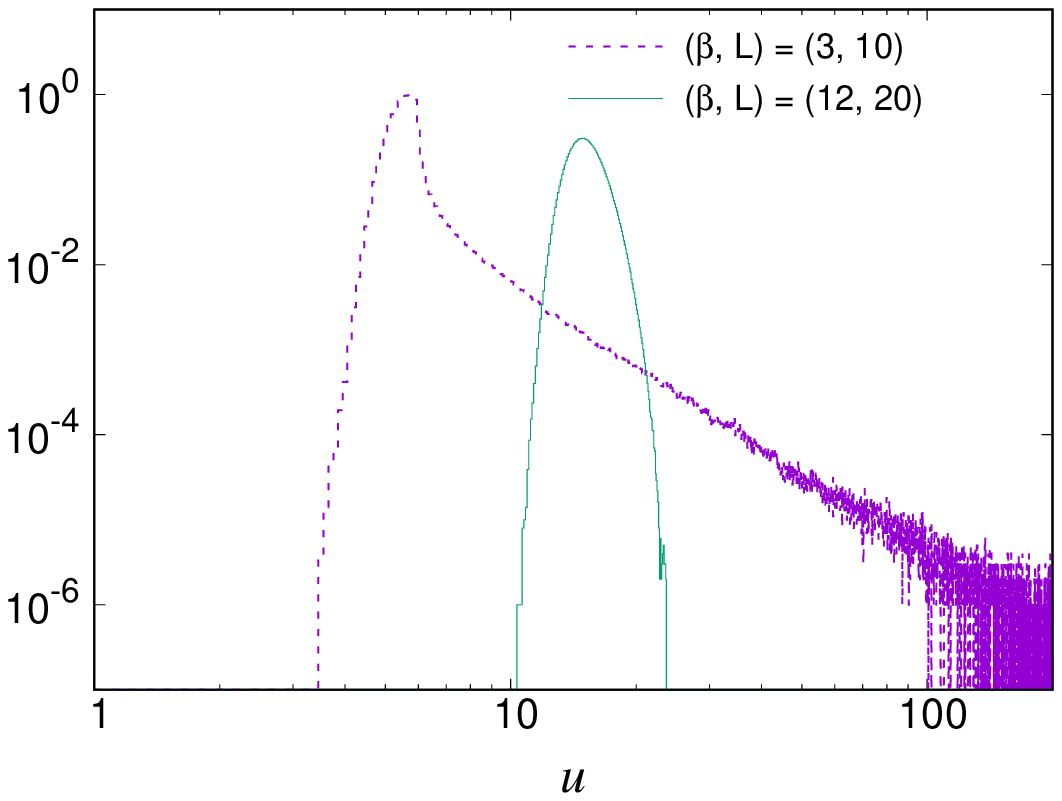}}
$\!\!\!\!\!\!$
{\includegraphics[scale=0.6]{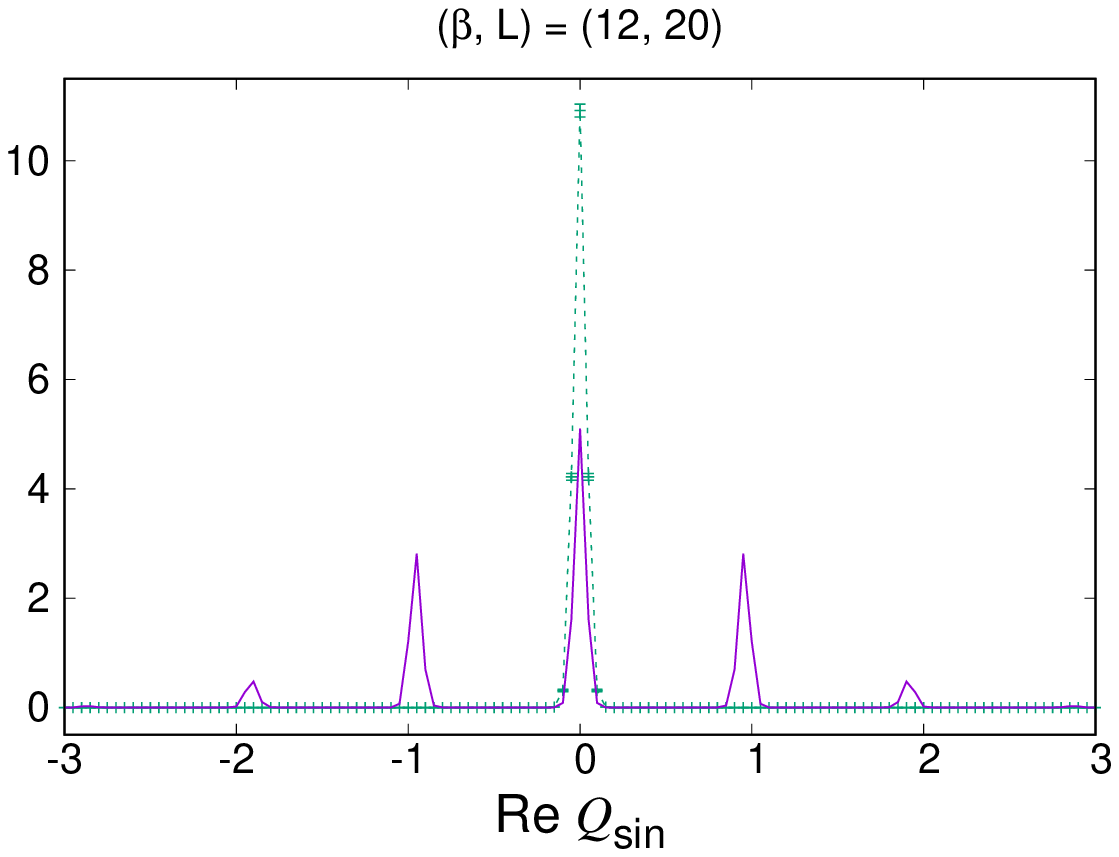}}
\caption{The results obtained by the CLM
  for the non-punctured model using the sine definition $Q_{\mathrm{sin}}$
of the topological charge.
(Left) The histogram of the magnitude $u$ of the drift term 
defined by (\ref{def-u})
is shown
for $(\beta, L)=(3,10),~(12,20)$ with $\theta=\pi$. 
(Right) The histogram of $\mathrm{Re} \, Q_{\mathrm{sin}}$
is shown
for $(\beta, L)=(12,20)$ with $\theta=\pi$.
The exact result
obtained for $(\beta, L)=(12,20)$ with $\theta=0$
is shown by the solid line for comparison.
}
\label{fig:result_NP}
\end{figure}

\subsection{results with the naive implementation}
\label{sec:result_nonpunctured}

In this section, we present our results obtained by the CLM,
which is implemented naively
using the non-punctured model explained above
as opposed to the punctured model, which we use later.
As for the definition of the topological charge,
we adopt the sine definition (\ref{simp-Q-def})
for the reason given in Section \ref{CLMU1}.

We have performed simulations
at various $\theta$
for $(\beta, L)=(3,10),~(12,20)$ corresponding to
a fixed physical volume 
$V_{\rm phys} \equiv L^2/\beta = 10^2/3$.
Below we show our results only for $\theta=\pi$,
where the sign problem becomes severest,
but the situation is the same for all values of $\theta$.

In Fig.~\ref{fig:result_NP}(Left),
we show the histogram of the magnitude $u$ of the drift term.
The distribution
falls off rapidly for $(\beta,L)=(12,20)$,
but it decays slowly with a power law for $(\beta,L)=(3,10)$.
Thus the criterion for correct convergence is satisfied
for $(\beta,L)=(12,20)$ but not for $(\beta,L)=(3,10)$
due to the large drifts.

In Fig.~\ref{fig:result_NP}(Right),
we plot the histogram of $\mathrm{Re} \, Q_{\mathrm{sin}}$
obtained by the CLM for $(\beta, L)=(12,20)$ with $\theta=\pi$,
which has a sharp peak at $\mathrm{Re} \, Q_{\mathrm{sin}} \sim 0$.
In the same figure, we also plot the exact result
for $(\beta, L)=(12,20)$ with $\theta=0$ for comparison,
which exhibits a few sharp peaks at integer values
within the range $-2 \lesssim \mathrm{Re} \, Q_\text{sin}\lesssim 2$.
From these two plots, we conclude that
the transitions between different topological sectors are highly
suppressed in the simulation, which causes a problem with the ergodicity.

This occurs also at $\theta=0$ for large $\beta$,
and it is called the ``topology freezing problem'' in the literature.
In fact, the results one obtains
by simulations suffering from this problem
correspond to the expectation values restricted to the topological
sector specified by the initial configuration.
This is true for both $\theta=0$ and $\theta \neq 0$.
In this case, however, the effect of the $\theta$ term cancels
between the numerator and the denominator of the expectation
values, and the calculation essentially reduces to that of the
real Langevin method at $\theta=0$.

For $(\beta,L)=(3,10)$ with $\theta=\pi$, on the other hand,
the histogram of $\mathrm{Re} \, Q_{\mathrm{sin}}$
obtained by the CLM has broad peaks
that overlap with each other,
%
which looks similar to the exact result 
for $(\beta,L)=(3,10)$ with $\theta=0$.
This implies that the topology freezing problem
does not occur for $(\beta,L)=(3,10)$.
See also Fig.~\ref{fig:Q-drift_history}.



\begin{figure}
\centering
{\includegraphics[scale=0.6]{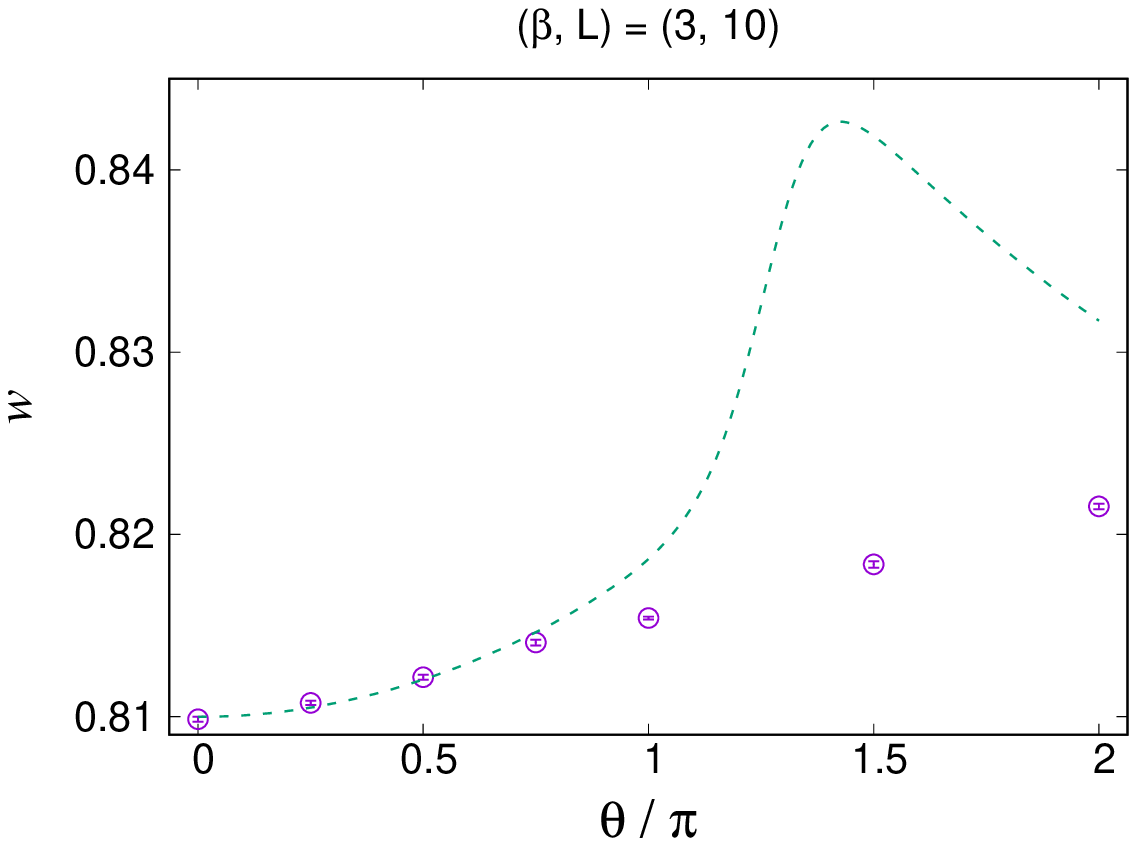}}
{\includegraphics[scale=0.6]{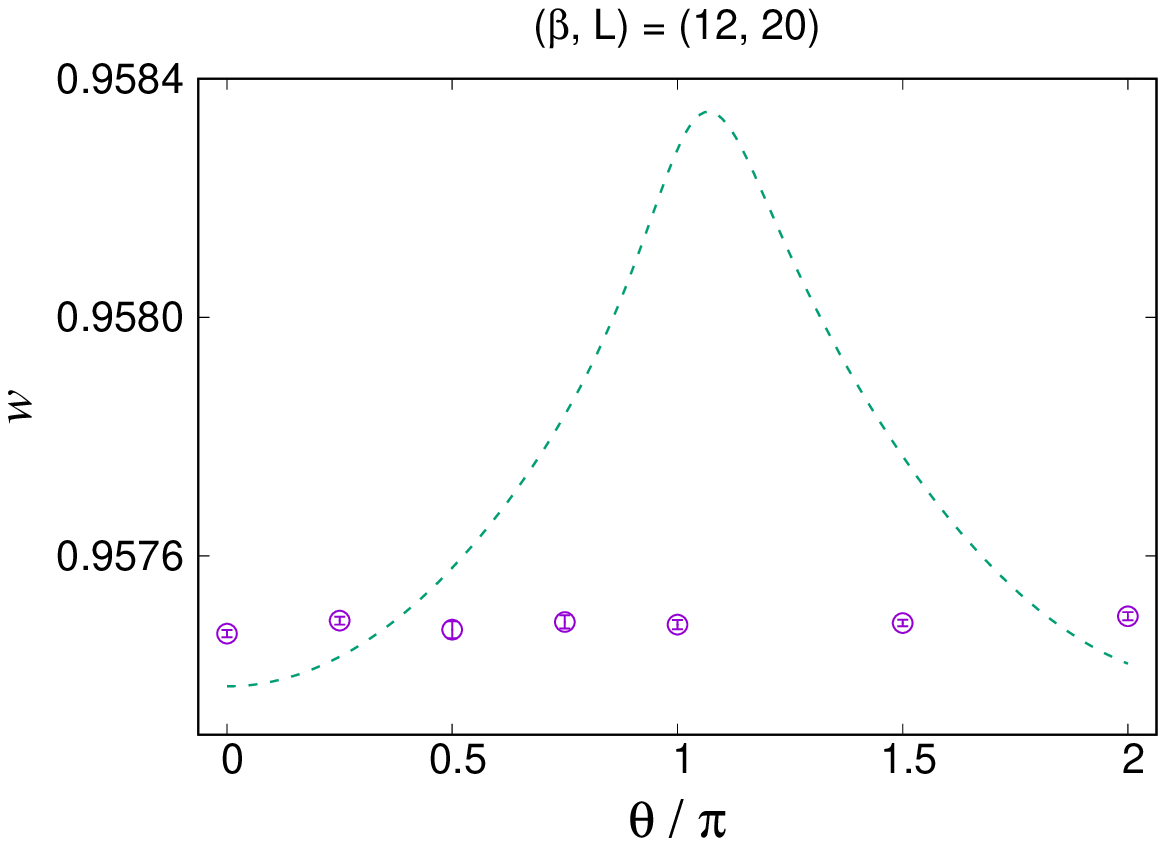}}\\
\vspace{10pt}
{\includegraphics[scale=0.6]{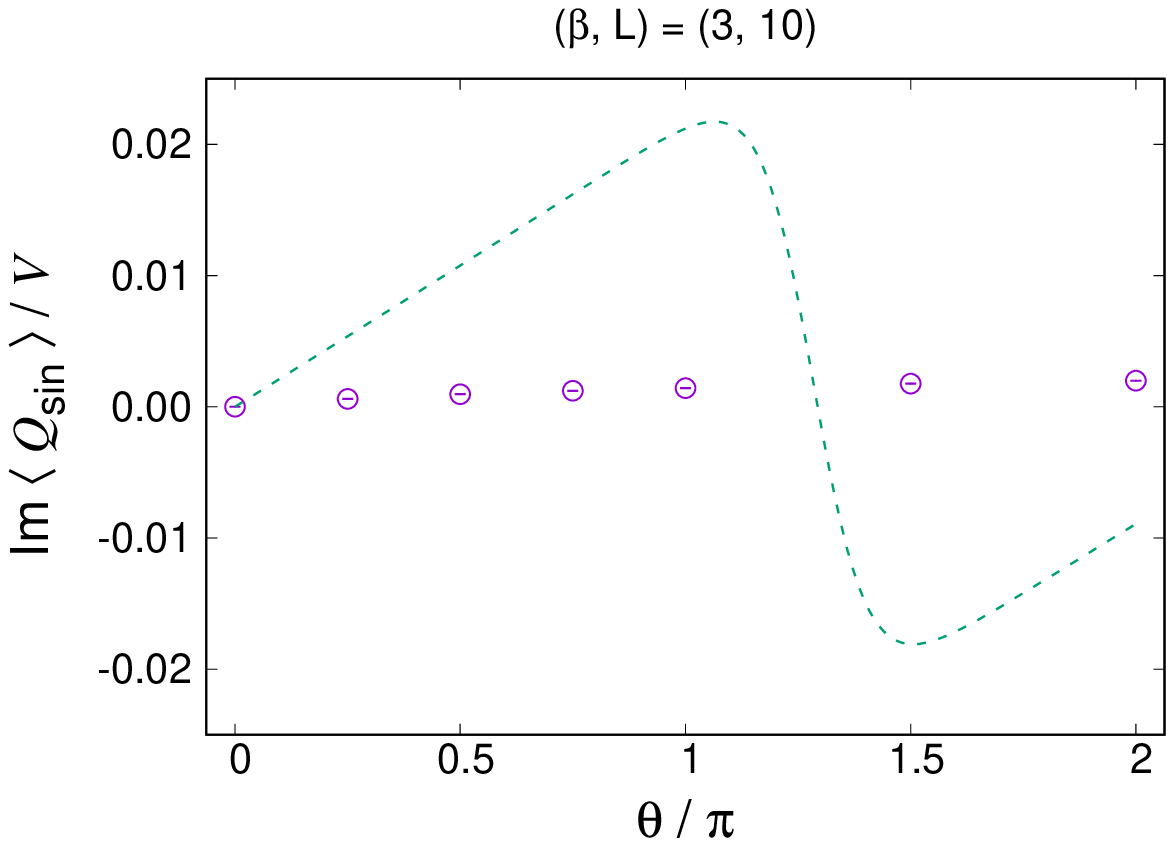}}
{\includegraphics[scale=0.6]{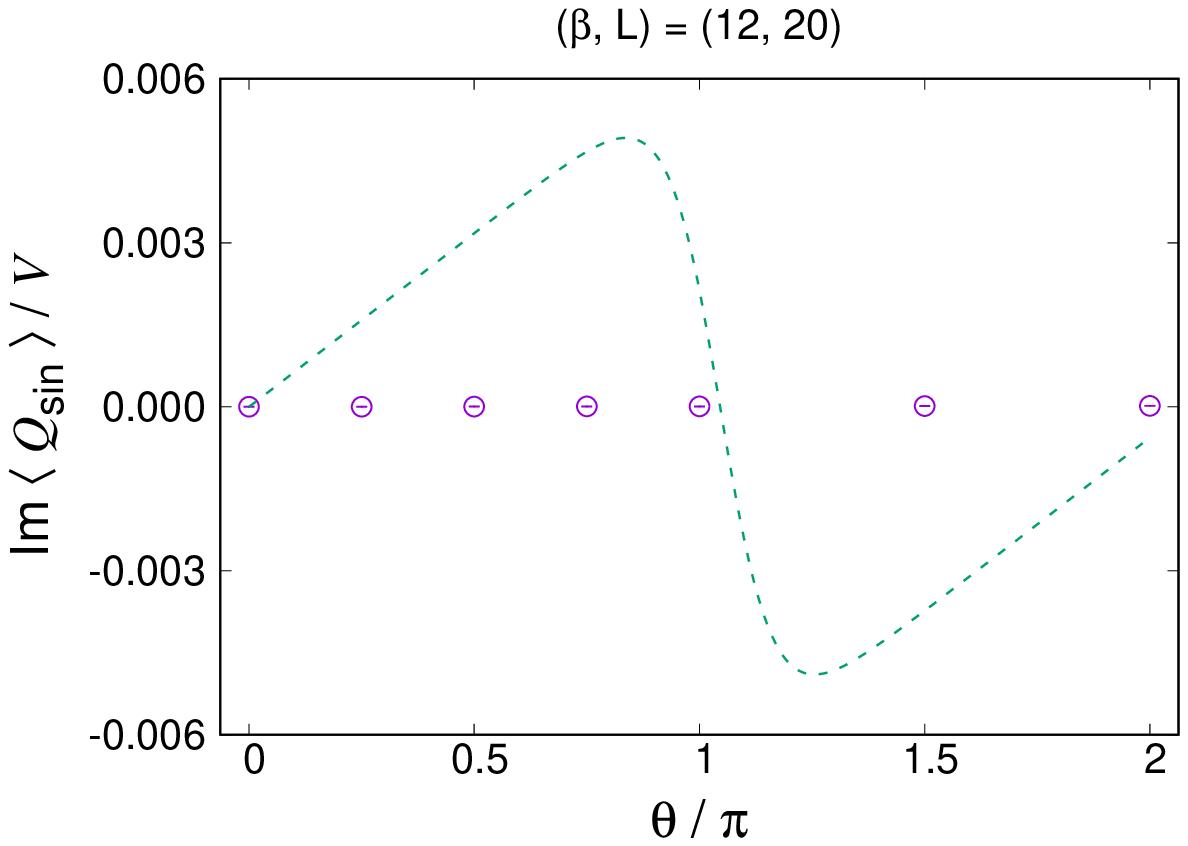}}\\
\vspace{10pt}
{\includegraphics[scale=0.6]{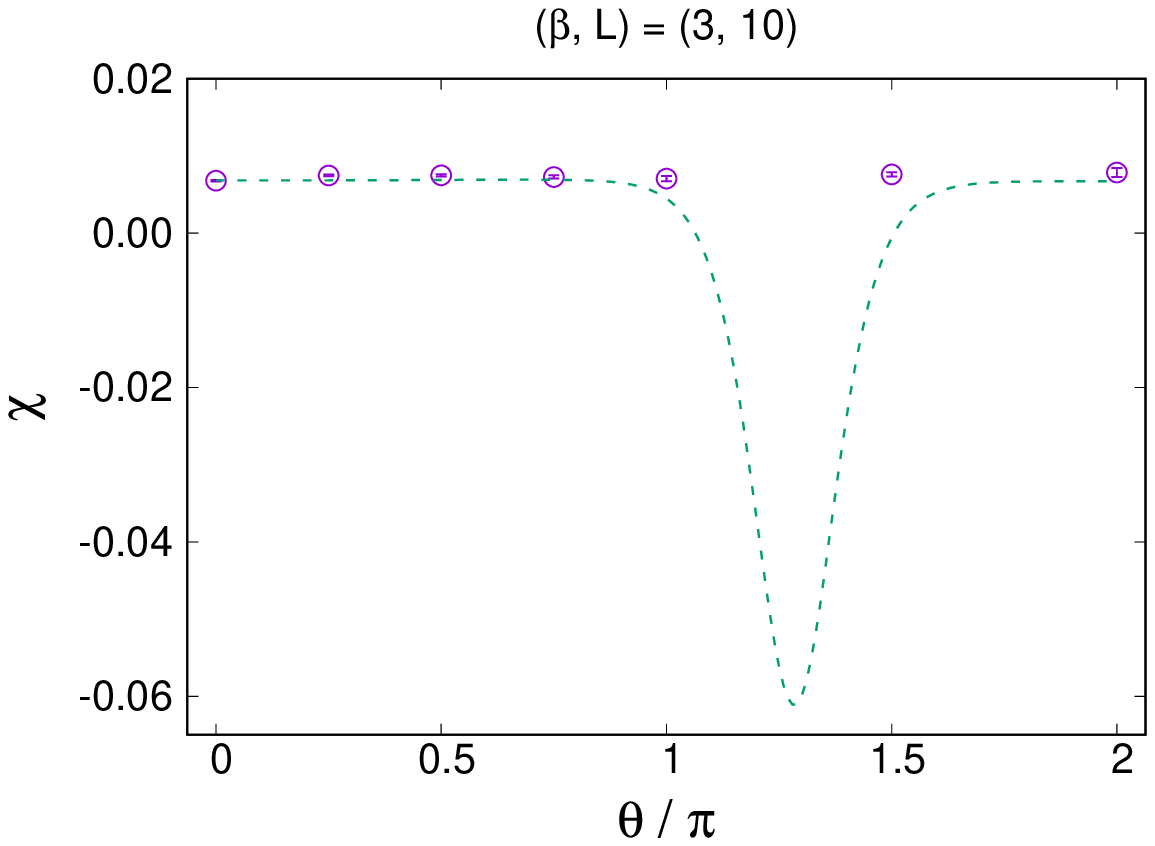}}
{\includegraphics[scale=0.6]{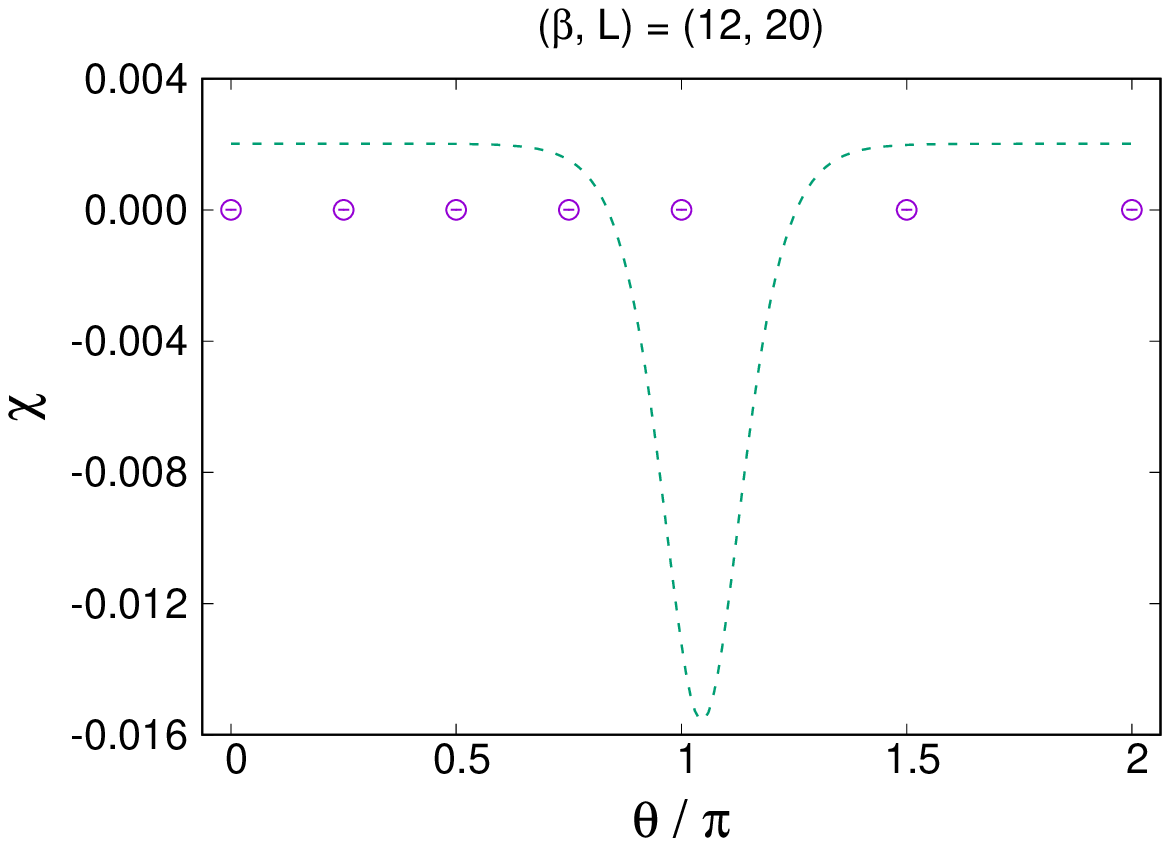}}
\caption{The results for various observables obtained by the CLM 
  for the non-punctured model with the sine definition $Q_{\mathrm{sin}}$.
The average plaquette (Top),
the imaginary part of the topological charge density (Middle),
the topological susceptibility (Bottom)
are plotted against $\theta$
for $(\beta, L)=(3,10)$ (Left) and $(12,20)$ (Right).
The exact results for the same $(\beta, L)$ are shown
by the dashed lines for comparison.}
\label{fig:resultes_naive}
\end{figure}

Below we define the observables we investigate in this paper.
First, we define the average plaquette by
\begin{equation}
w=\frac{1}{V}\frac{\partial}{\partial\beta} \log Z \ .
\label{eq:def_aveP}
\end{equation}
Hereafter, $V$ denotes the number of plaquettes in the
action,
which is $V=L^2$ for the non-punctured model and 
$V=L^2-1$ for the punctured model we define in 
Section \ref{punctured_model}.
The topological charge density is defined by
\begin{equation}
\frac{1}{V} \langle Q\rangle = 
-i \frac{1}{V} \frac{\partial }{\partial\theta}\log Z \ ,
\label{eq:def_ImQ}
\end{equation}
which is zero at $\theta=0$ and purely imaginary for $\theta\neq0$.
Finally, the topological susceptibility is defined by
\begin{equation}
\chi=\frac{1}{V}\left(\langle Q^{2}\rangle-\langle Q\rangle^{2}\right) 
= - \frac{1}{V} \frac{\partial^2 }{\partial\theta ^2}\log Z \ ,
\label{eq:def_chi}
\end{equation}
which is real for all $\theta$.
In fact, the topological susceptibility $\chi$
is related to the topological charge density (\ref{eq:def_ImQ})
through
\begin{equation}
\chi = - i \frac{1}{V} \frac{\partial}{\partial \theta} \langle Q \rangle  \ .
\label{Q-chi-rel}
\end{equation}
Note, however, that this relation can be violated if the CLM
fails to calculate the expectation values correctly.

In Fig.~\ref{fig:resultes_naive}, 
we show the results obtained
by the CLM for the non-punctured model.
We also plot the exact results for comparison,
which are derived in Appendix \ref{sec:analytic_obs}.
%
In the left column, we present our results for $(\beta,L)=(3,10)$, 
which suffer from the incorrect convergence,
whereas in the right column, we present our results for $(\beta,L)=(12,20)$, 
which suffer from the topology freezing problem.
In either case, our results do not reproduce the exact results
as anticipated.
Note that our results at $\theta=0$ agree with the exact results
for $(\beta,L)=(3,10)$ but not for $(\beta,L)=(12,20)$.
This is because the topology freezing problem occurs
for large $\beta$ even at $\theta=0$, where the sign problem is absent.

Thus we find that the CLM with the naive implementation
fails for both $(\beta,L)=(3,10)$ and $(\beta,L)=(12,20)$
for different reasons.
For $(\beta,L)=(3,10)$,
the topology change occurs 
but the criterion for correct convergence is not satisfied
due to the large drifts.
For $(\beta,L)=(12,20)$, 
the criterion for correct convergence is satisfied, but 
the ergodicity is violated due to the
topology freezing problem.
We have searched for a parameter region in which neither of the problems
occur, but we could not find one. 
In fact, we will see in the next section
that these problems are related to each other
at least in the present model.

\subsection{the appearance of large drifts and the topology change}
\label{whypunctureworks}

In this section,
we provide more in-depth discussions
on the relationship between
the appearance of large drifts and the topology change
in the non-punctured model. 
Let us first recall that 
the drift terms are given by
\eqref{drift-plaquette} and \eqref{drift-simple-Q},
which depend on $P_n$.
When $\beta$ is large, the gauge action $S_{g}$
favors configurations with $P_n \sim 1$ for all $n$,
which implies that the drift terms are small.

\begin{figure}
\centering
\includegraphics[scale=0.8]{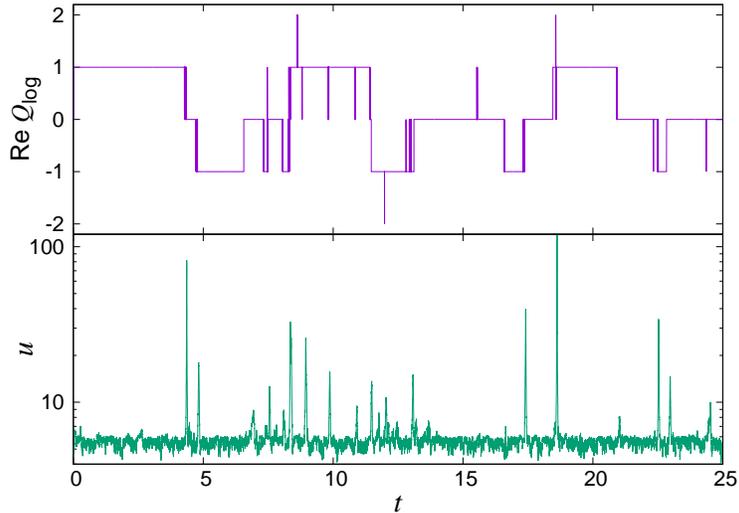}
\caption{The results obtained by the CLM
  for the non-punctured model with the sine definition $Q_{\mathrm{sin}}$
for $(\beta,L)=(3,10)$ with $\theta=\pi$.
The upper plot shows 
the history of the topological charge $Q_{\mathrm{log}}$
with the log definition,
whereas the lower plot shows
the history of the magnitude $u$ of the drift term 
in the log scale.
}
\label{fig:Q-drift_history}
\end{figure}

On the other hand, 
the notion of topological sectors can be defined
by the real part of (\ref{log-Q-def}), which takes 
integer values, even for 
complexified configurations that are generated in the CLM.
In order for a transition between different topological sectors
to occur, one of the plaquettes 
has to cross the branch cut;
namely the phase of the plaquette has to jump
from $-\pi$ to $\pi$ or vice versa.
When this occurs, 
large drift terms can appear 
as can be seen from
Fig.~\ref{fig:Q-drift_history}, where
we plot the histories
of
$\mathrm{Re} \, Q_{\mathrm{log}}$
and the magnitude of the drift term (\ref{def-u}).
We observe clear correlation 
between the large drift term and the topology change.
We have also confirmed
that the large drift term appears 
for the link variables composing the plaquette
that
crosses the branch cut.

\begin{figure}
\centering
{\includegraphics[scale=0.5]{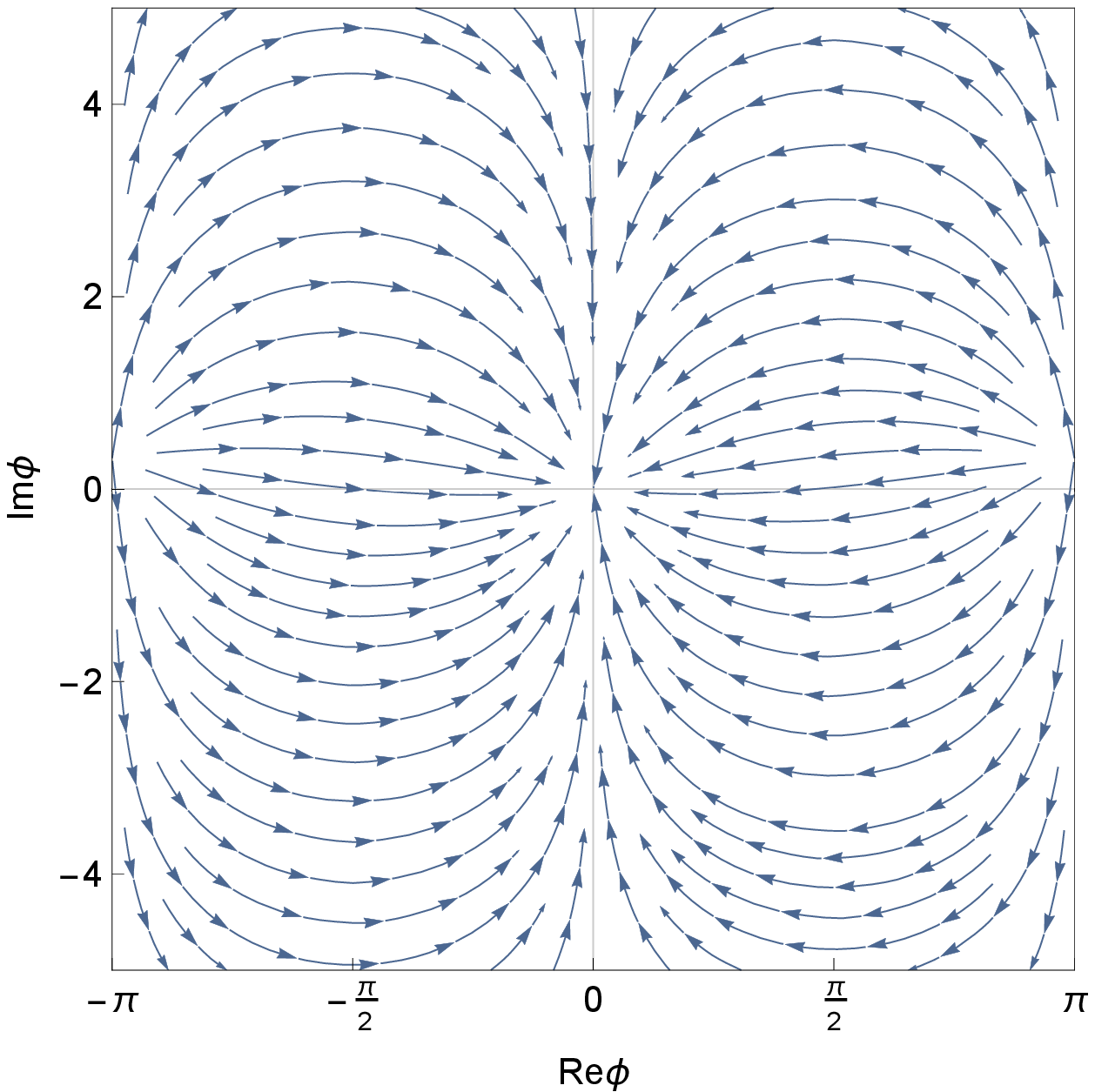}}
{\includegraphics[scale=0.6]{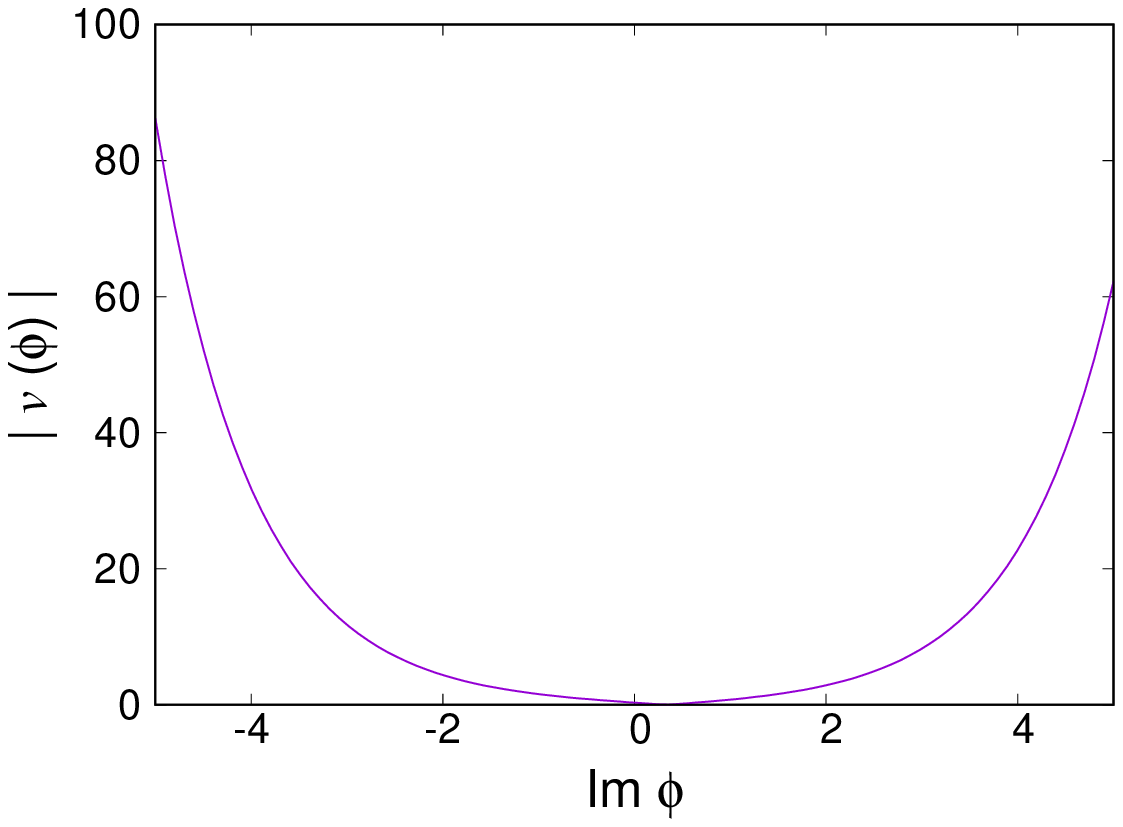}}
\caption{(Left) A flow diagram representing 
$-v$ defined by (\ref{drift-at-phi})
is shown as a function of $\phi$ for $\beta=\theta=1$. 
(Right) The absolute value $|v(\phi)|$ is
plotted against $\mathrm{Im} \, \phi$ for $\mathrm{Re} \, \phi=\pi$.
}
\label{fig:DS}
\end{figure}

In order to understand this observation better,
we focus on a particular link variable $\mathcal{U}_{k,1}$,
and consider the corresponding drift term, which depends on 
the plaquettes $P_{k}$ and $P_{k-\hat{2}}$ sharing the link.
For simplicity, we set $P_{k-\hat{2}}=1$ and 
consider the drift term $v$ as a function of $P_{k}$
\begin{equation}
v=\beta\sin\phi-i\frac{\theta}{2\pi}\left(\cos\phi-1\right) \ ,
\label{drift-at-phi}
\end{equation}
where we have defined a complex parameter $\phi$ by $\phi = -i \log P_{k}$.
A large drift appears when $|\mathrm{Im} \, \phi| \rightarrow \infty$.
In Fig.~\ref{fig:DS}(Left), we plot the drift term as a flow diagram
for $\beta=\theta=1$. 
Considering that the contribution of the drift term $v$
to the change of $\phi$ at a Langevin step is given by
$\Delta \phi   = -  v \Delta t$,
we actually plot $(-v)$ in the complex $\phi$ plane.

In what follows we assume that $\beta > \theta/2\pi$.
Then we find from Eq.~(\ref{drift-at-phi})
that there are two fixed points
corresponding to $v=0$. One is $\phi=0$
and the other is
$\phi=i\log[(\theta/2\pi+\beta)/(\theta/2\pi-\beta)]$,
which is close to $\pm \pi$ for $\beta \gg  \theta/2\pi$.
As one can see from Fig.~\ref{fig:DS}(Left),
the fixed point $\phi=0$ is attractive, which confirms that
$P_{k}$ tends to become unity when $\beta$ is large.
The other fixed point $\phi \sim \pm \pi$ is repulsive,
and the magnitude $|v|$ grows exponentially as one flows away
in the imaginary direction; See Fig.\ref{fig:DS}(Right).
As we mentioned above, when the transition between topological
sectors occurs, one of the plaquettes crosses the branch cut,
which corresponds to
${\rm Re} \, \phi = \pm \pi$ in the flow diagram.
When this happens, the configuration can flow in the imaginary 
direction, which causes a large drift.

\section{Introducing a puncture on the 2D torus}
\label{sec:introduce-puncture}

Since the problem we encounter in the previous section
occurs due to the topological nature of the $\theta$ term,
a simple remedy would be to change the topology of the
base manifold to a noncompact one. 
Here we consider introducing a puncture on the 2D torus.
Once we introduce a puncture, the drift term
$D_{n,\mu}S_{\theta}$ with the log definition of the topological charge
has nonzero contributions for the link variables surrounding the puncture,
which enable us to 
include the effect of the $\theta$ term
correctly in the CLM
as we will see in Section \ref{sec:results}.
Therefore, for the rest of this paper,
we basically use the log definition to simplify our discussions.
Unlike the non-punctured model,
the topological charge is no more restricted to integer values,
and it can be changed freely.

Since the puncture affects the theory only locally, its effect is 
expected to die out in the infinite volume limit for $|\theta|<\pi$
as we demonstrate
explicitly in this section using the exact results.
Thus unless we are interested in a theory with a finite volume,
the punctured model is as good as the original model, the difference
simply being a different choice of ``boundary conditions''.
In fact, we will see that the non-punctured model has slow convergence
to the infinite volume limit for $\theta \sim \pi$,
which is not the case in the punctured model.

\subsection{defining the punctured model on the lattice}
\label{punctured_model}

There are various ways to introduce a puncture
on the periodic lattice.
Here we consider removing a plaquette as a simple choice.
More precisely,
we define
the punctured model
by removing one plaquette, let say $P_{\punc}$,
from the sum appearing in
(\ref{eq:S_beta}) and (\ref{log-Q-def})
when we define the action (\ref{action-decompose}).

As an alternative method,
we have also tried introducing a slit at a particular link,
which amounts to duplicating the corresponding 
link variable and including each of them in the plaquettes
that share the link. The results turn out to be qualitatively the same
as the ones obtained by removing a plaquette.
There are, of course, many others, but in any case,
one can obtain exact results for a finite lattice 
as we explain in Appendix \ref{sec:analytic_calc},
and using them, one can demonstrate explicitly that 
the punctured model is equivalent to the original non-punctured model
in the infinite volume limit for $|\theta|<\pi$.

\subsection{equivalence in the infinite volume limit}
\label{equivalence-punctured_model}


In this section, we show the equivalence of the non-punctured model 
and the punctured model in the infinite volume limit.
Here we use the log definition of the topological charge,
but a similar statement holds as far as the same definition
is used for the two models.\footnote{In the case of the sine definition,
  the equivalence of the two models in the infinite volume limit
  holds for $|\theta|<\theta_{\rm c}(\beta)$,
  where $\theta_{\rm c}(\beta) \sim \pi \{ 1 + 1/(2\beta) \}$
  for large $\beta$.}

The partition function for the non-punctured model 
is given by (See Appendix \ref{subsec:Torus}
for derivation.)
\begin{align}
Z_{\mathrm{nonpunc}} & =\sum_{n=-\infty}^{+\infty}
\left[\mathcal{I}(n,\theta,\beta)\right]^{V}
\label{part-nonpunc-finiteV}
\end{align}
for finite $V=L^2$,
where the function $\mathcal{I}(n,\theta,\beta)$ is defined by
\begin{align}
\mathcal{I}(n,\theta,\beta) & =
\frac{1}{2\pi}
\int_{-\pi}^{\pi}d\phi\,
e^{\beta\cos\phi+i\left(\frac{\theta}{2\pi}-n\right)\phi}  \ .
\label{I-def}
\end{align}
Let us take the infinite volume limit $V\rightarrow\infty$,
in which 
the sum over $n$ in \eqref{part-nonpunc-finiteV}
is dominated by the term that gives the largest absolute value 
$|\mathcal{I}(n,\theta,\beta)|$.
This corresponds to the $n$ that minimizes
$|\frac{\theta}{2\pi}-n|$.
Thus in the infinite volume limit,
the free energy is obtained as
\begin{equation}
\lim_{V\rightarrow \infty} \frac{1}{V} \log Z_{\mathrm{nonpunc}} = 
\log \mathcal{I}(0,\tilde{\theta},\beta)
\ ,
\label{part-fn-nonpunc-exact}
\end{equation}
where $\tilde{\theta}$ is defined by $\tilde{\theta}= \theta - 2\pi k$
with the integer $k$ chosen 
so that $-\pi < \tilde{\theta} \le \pi$.
%



\begin{figure}
\centering
{\includegraphics[scale=0.6]{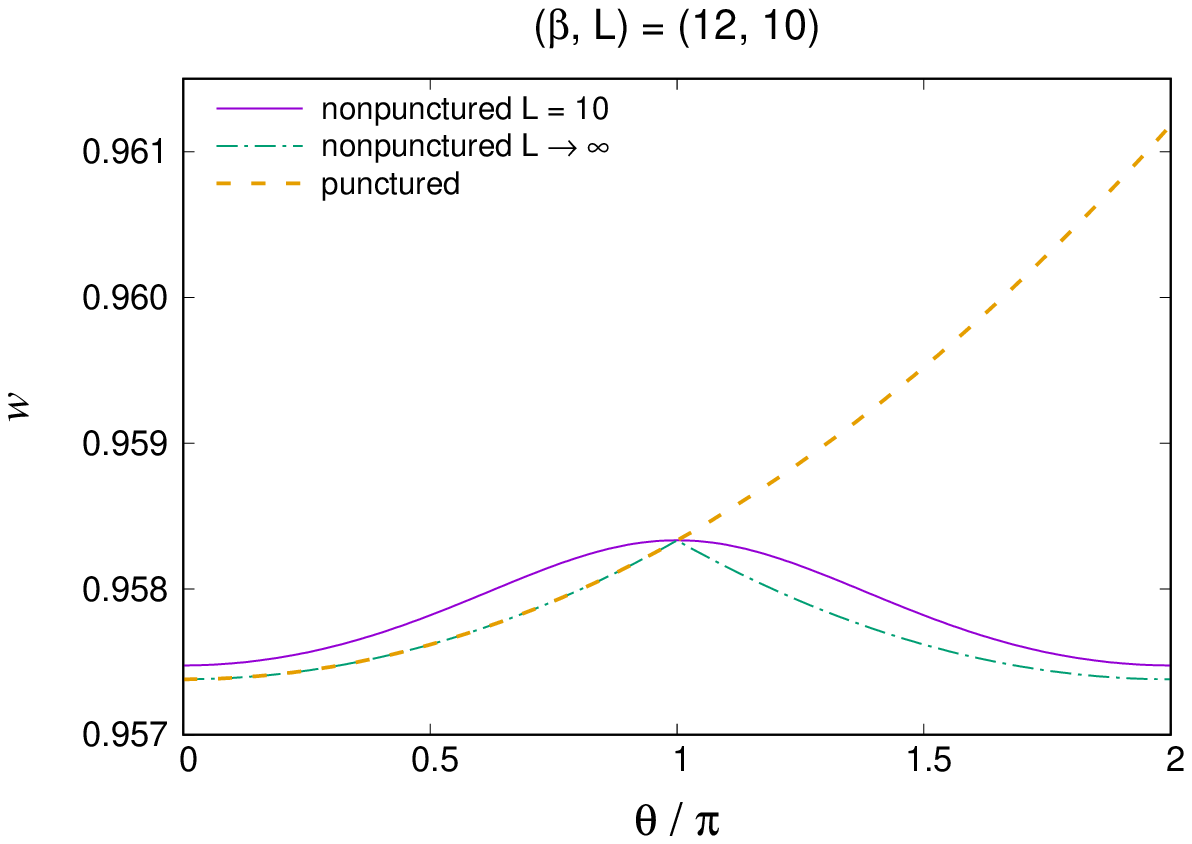}}
{\includegraphics[scale=0.6]{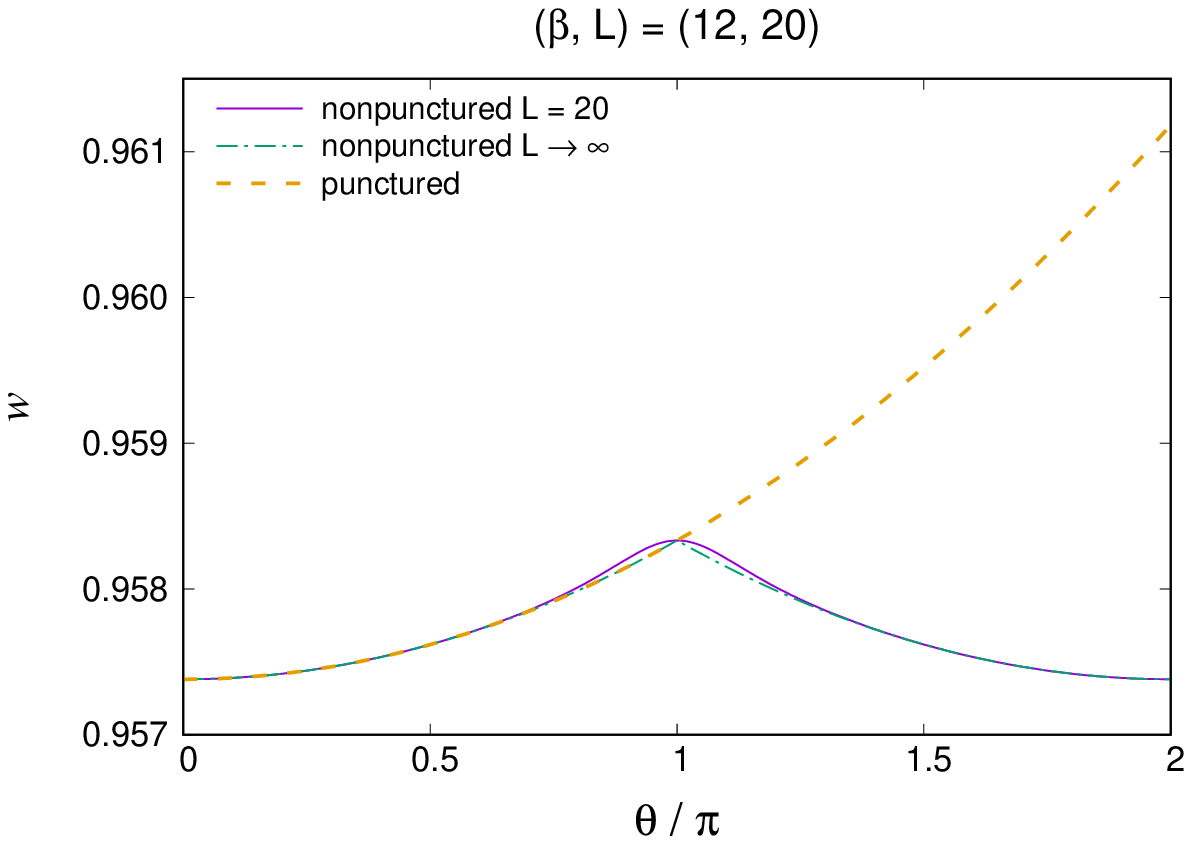}}\\
\vspace{10pt}
{\includegraphics[scale=0.6]{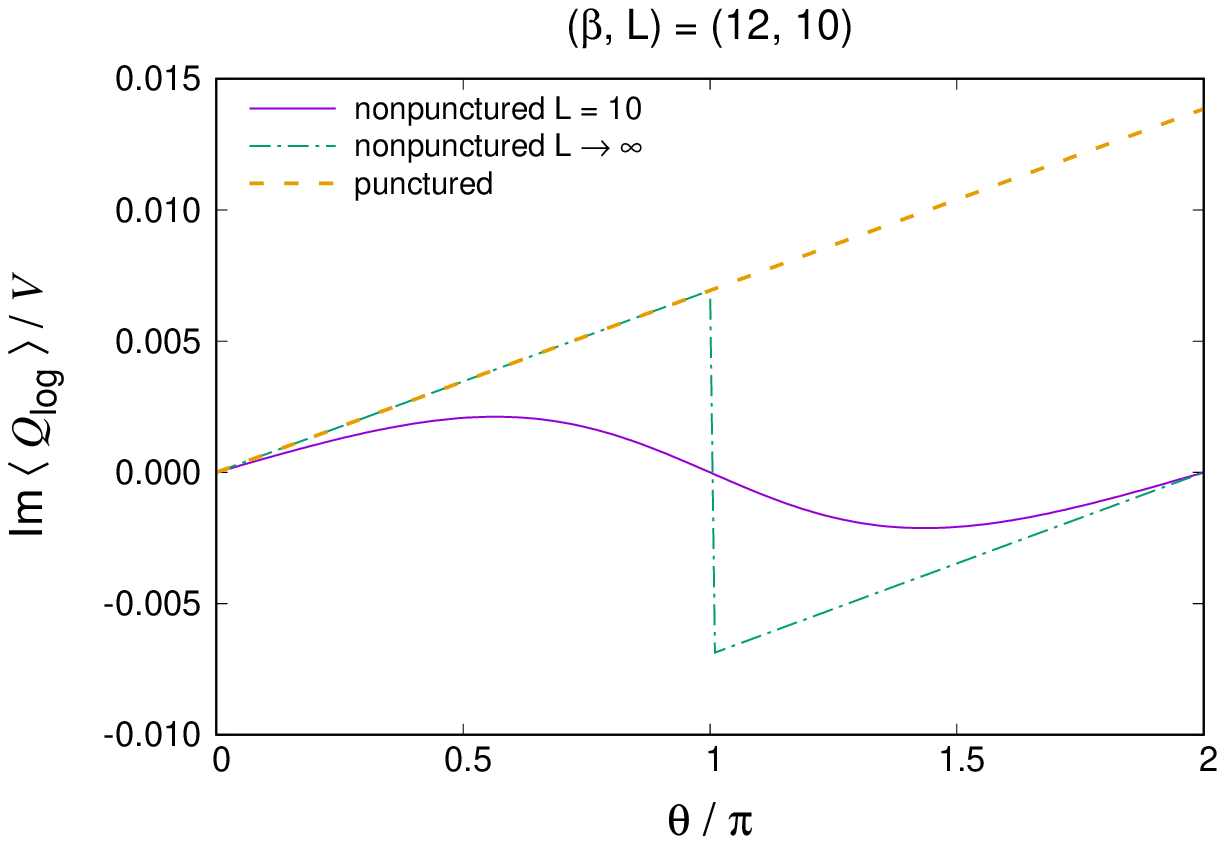}}
{\includegraphics[scale=0.6]{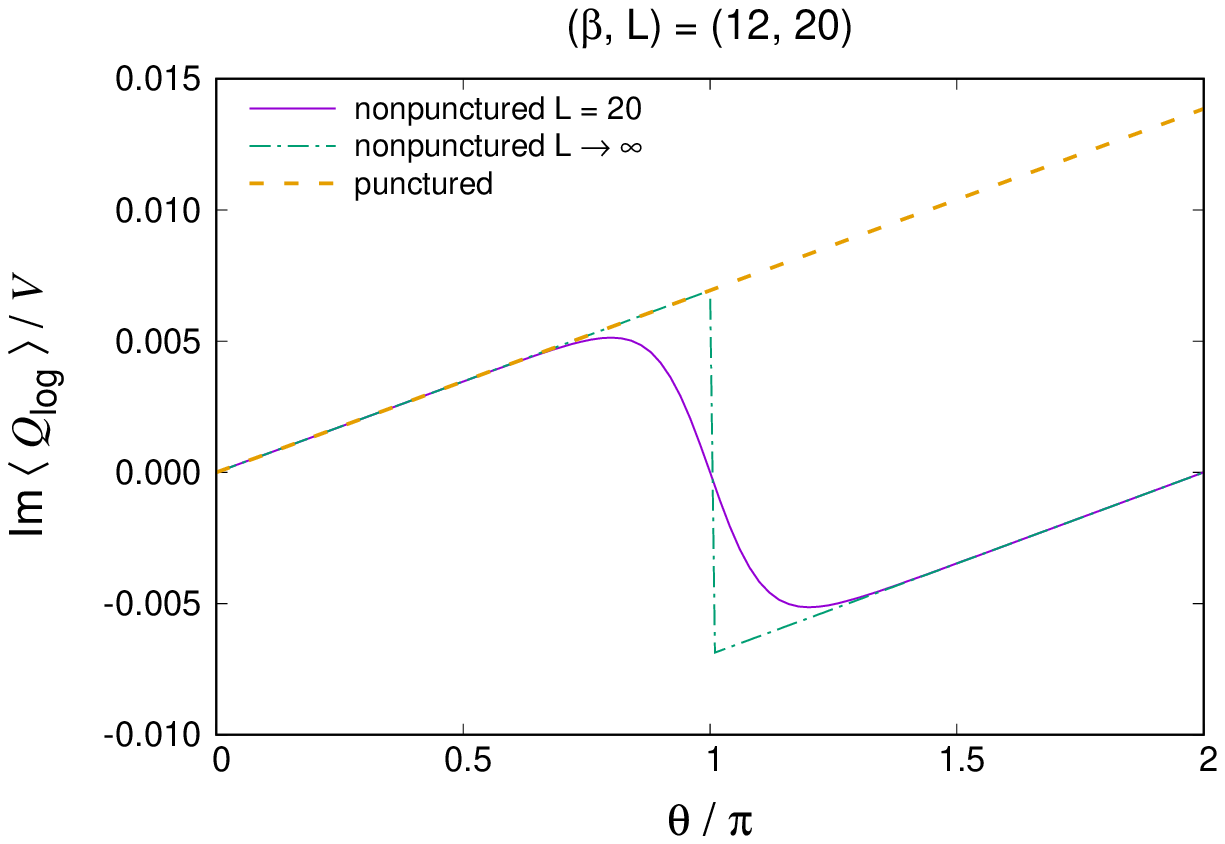}}\\
\vspace{10pt}
{\includegraphics[scale=0.6]{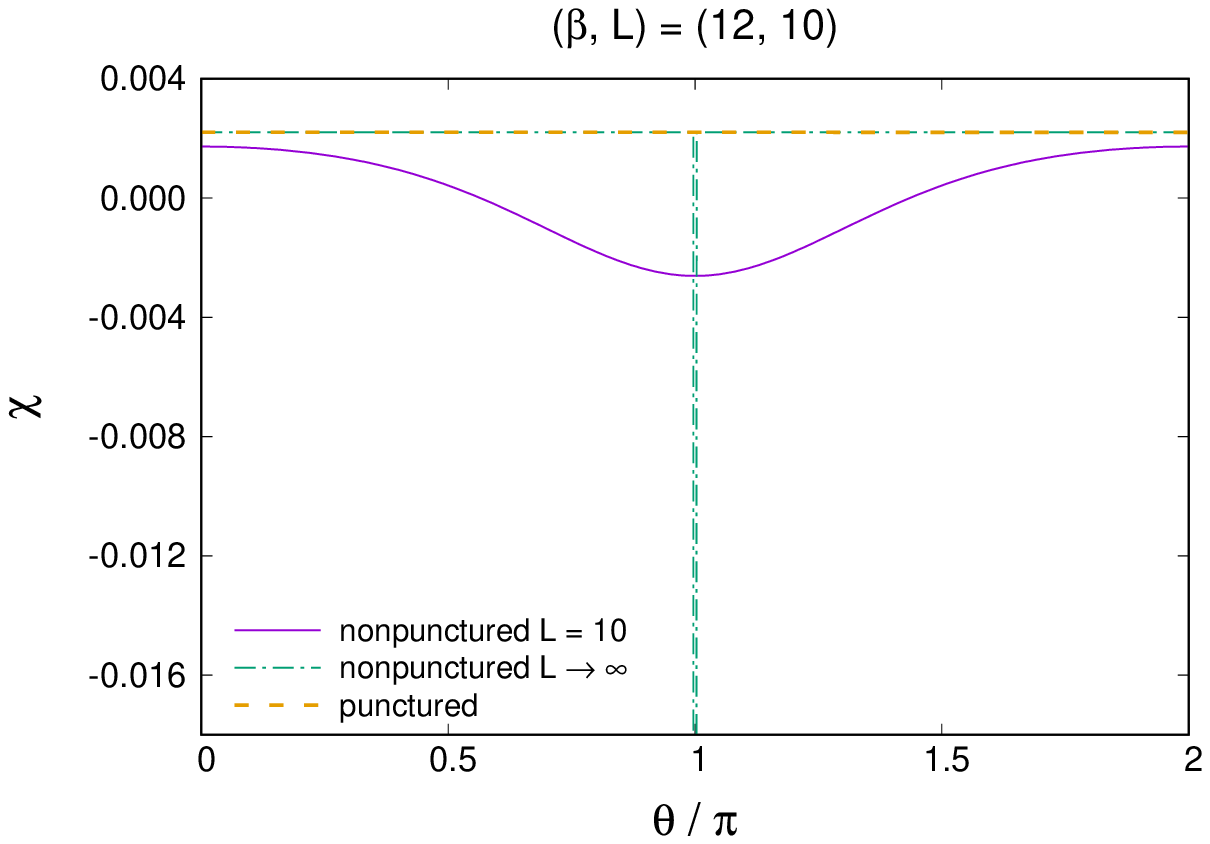}}
{\includegraphics[scale=0.6]{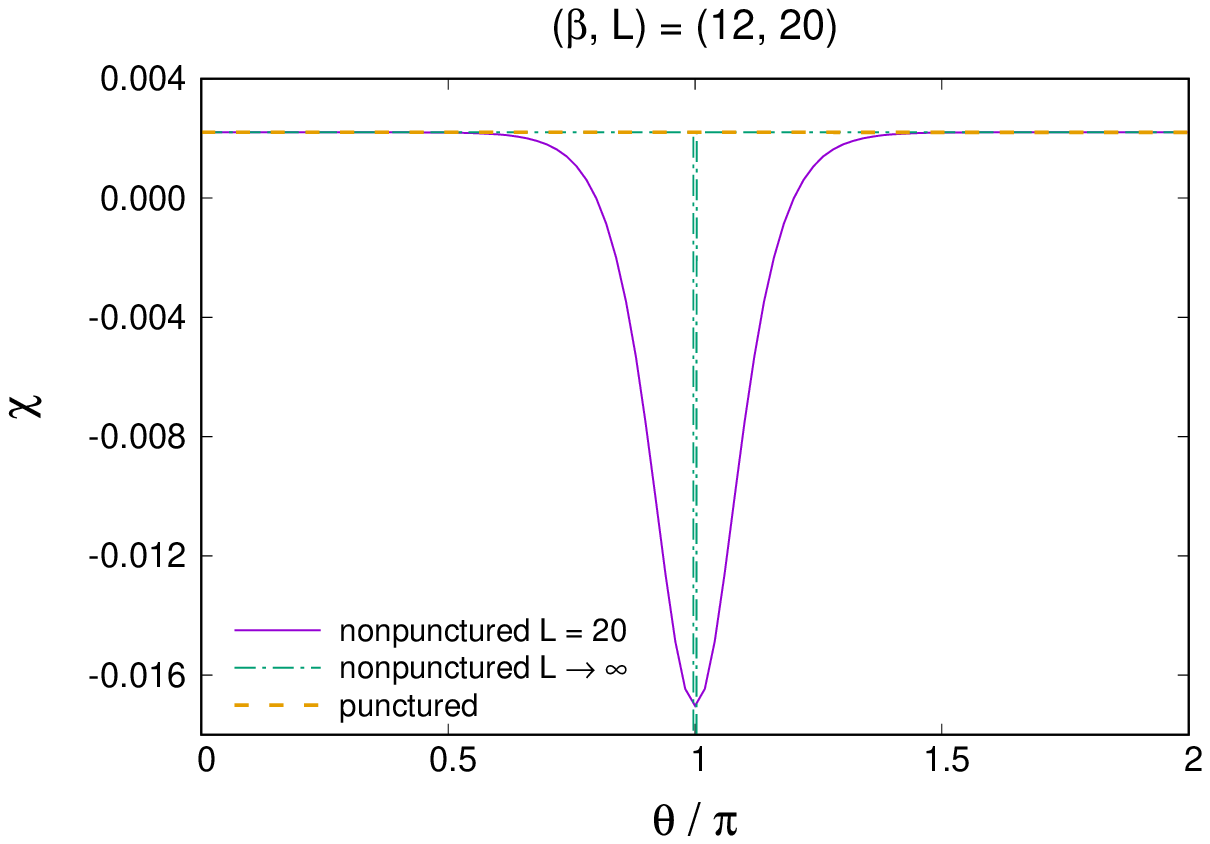}}
\caption{
The exact results for various observables obtained
by using the log definition $Q_{\mathrm{log}}$
of the topological charge.
The average plaquette (Top),
the imaginary part of the topological charge density (Middle),
the topological susceptibility (Bottom)
obtained for the non-punctured (solid line) and 
punctured (dashed line) models
are plotted against $\theta$
for $L=10$ (Left) and $L=20$ (Right) with the same $\beta=12$.
Note that the results for the punctured model are actually 
independent of $L$.
For the non-punctured model,
we also plot the results
in the infinite volume limit $L\rightarrow \infty$
with $\beta=12$ by the dash-dotted lines for comparison.
}
\label{fig:analytic_obs}
\end{figure}

On the other hand, the partition function for the punctured
model is given by
(See Appendix \ref{sec:part-func-punc} for derivation.)
\begin{equation}
 Z_{\mathrm{punc}} = 
\left[\mathcal{I}\left(0,\theta,\beta\right)\right]^V 
\label{part-fn-punc-exact}
\end{equation}
for finite $V=L^2 -1$,
which implies that the free energy
\begin{equation}
\frac{1}{V} \log Z_{\mathrm{punc}} = 
\log \left[\mathcal{I}\left(0,\theta,\beta\right)\right]
\label{part-fn-punc-exact-log}
\end{equation}
is actually $V$ independent.
Hence all the observables that can be derived from it
has no finite size effects.
%
Note also that this model
does not have
the $2\pi$ periodicity in $\theta$.
By comparing (\ref{part-fn-nonpunc-exact}) and 
(\ref{part-fn-punc-exact-log}),
one can see that the two models are equivalent
in the infinite volume limit for $|\theta|<\pi$.

The observables defined in Section \ref{sec:result_nonpunctured}
can be calculated for the two models using
(\ref{part-nonpunc-finiteV}) and
(\ref{part-fn-punc-exact})
by numerical integration (See Appendix \ref{sec:analytic_obs}
for the details.). 
In Fig.~\ref{fig:analytic_obs},
we plot the average plaquette (Top) defined by \eqref{eq:def_aveP},
the imaginary part of the topological charge 
density (Middle)
defined by \eqref{eq:def_ImQ}
and the topological susceptibility (Bottom)
defined by \eqref{eq:def_chi}
for $L=10$ (Left) and $L=20$ (Right), respectively, 
with the same $\beta=12$.
The results for the two models tend to agree
as $L$ increases for $|\theta| < \pi$.



We can evaluate the free energy
(\ref{part-fn-punc-exact-log})
for the punctured model
more explicitly for large $\beta$,
which is relevant in the continuum limit.
By integrating over $\phi$ in Eq.~(\ref{I-def}) as
\begin{align}
\mathcal{I}(n,\theta,\beta) & \simeq
\frac{1}{\sqrt{2\pi\beta}}
e^{\beta - \frac{1}{2\beta} \left( \frac{\theta}{2\pi} -n \right)^2}  \ ,
\end{align}
we get
\begin{equation}
\frac{1}{V} \log Z_{\mathrm{punc}}
\simeq
\beta
-\frac{1}{2}\log2\pi\beta
-\frac{\theta^{2}}{8\pi^{2}\beta}
\ .
\end{equation}
From this, we can obtain various observables for the punctured model as
\begin{align}
 w & \simeq 
1 -\frac{1}{2\beta} + \frac{\theta^{2}}{8\pi^{2}\beta^{2}} \ ,\\
\frac{\langle Q\rangle}{V}
&\simeq\frac{i\theta}{4\pi^{2}\beta} \ ,
\label{imag-Q-exact}
\\
\chi& \simeq \frac{1}{4\pi^{2}\beta} 
\end{align}
for finite $V$, which explains the $\theta$ dependence observed 
in Fig.~\ref{fig:analytic_obs}.

From Fig.~\ref{fig:analytic_obs},
we also find that the results for the non-punctured model have
sizable finite volume effects, in particular 
around $\theta \sim \pi$, which is absent 
in the punctured model.
While the volume independence of the punctured model
may well be peculiar to the present 2D gauge theory case, 
the advantage of the punctured model 
compared with the non-punctured model from the viewpoint
of finite volume effects may 
hold more generally.
%

\section{Application of the CLM to the punctured model}
\label{sec:results}

In this section, we apply the CLM to the punctured model
using the log definition $Q_{\mathrm{log}}$
of the topological charge.
Our results reproduce the exact results discussed 
in the previous section 
as long as we are close enough to the continuum limit.
We also show that the topology freezing problem is circumvented
without causing large drifts thanks to the puncture.

\subsection{the drift terms for the punctured model}
\label{sec:drift_puncture}

We have discussed the drift terms in the non-punctured model
in Section \ref{CLMU1}.
For the punctured model,
we only have to modify
the drift terms 
for the four link variables surrounding the puncture;
i.e., $\mathcal{U}_{\punc,1}$, $\mathcal{U}_{\punc + \hat{2},1}$,
$\mathcal{U}_{\punc,2}$ and $\mathcal{U}_{\punc + \hat{1},2}$.
Thus we obtain
\begin{alignat}{2}
D_{n,1}S & = 
\left\{
\begin{array}{ll}
 -i\frac{\beta}{2}(P_{n}-P_{n}^{-1}-P_{n-\hat{2}}+P_{n-\hat{2}}^{-1})  & 
\mbox{~for~}n \neq \punc , \ \punc+\hat{2}  \ , \\
 -i\frac{\beta}{2}(-P_{\punc-\hat{2}}+P_{\punc-\hat{2}}^{-1}) 
 +i\frac{\theta}{2\pi}
 & 
\mbox{~for~}n = \punc   \ , \\
 -i\frac{\beta}{2}(P_{\punc +\hat{2}}-P_{\punc +\hat{2}}^{-1})
 - i\frac{\theta}{2\pi}  & \mbox{~for~}n = \punc+\hat{2}   \ ,
\end{array}
\right. 
\label{eq:drift_theta_smp_1}\\
D_{n,2}S & = 
\left\{
\begin{array}{ll}
 -i\frac{\beta}{2}(-P_{n}+P_{n}^{-1}+P_{n-\hat{1}}-P_{n-\hat{1}}^{-1})  & 
\mbox{~for~}n \neq \punc , \ \punc+\hat{1}  \ , \\
 -i\frac{\beta}{2}(P_{\punc-\hat{1}}-P_{\punc-\hat{1}}^{-1}) 
- i\frac{\theta}{2\pi} & 
\mbox{~for~}n = \punc   \ , \\
 -i\frac{\beta}{2}(-P_{\punc + \hat{1}}+P_{\punc + \hat{1}}^{-1}) 
+ i\frac{\theta}{2\pi}  & 
\mbox{~for~}n = \punc+\hat{1}  \ ,
\end{array}
\right.
\label{eq:drift_theta_smp_2}
\end{alignat}
where
we have ignored
the issue of $\delta$-function
discussed in Section \ref{CLMU1}.
This is justified
if all the plaquettes in the action
never cross the branch cut; i.e., $|\mathrm{Im} \log P_n | \le \pi - \epsilon$
for $\forall n \neq K$
with a strictly positive $\epsilon$ during the Langevin simulation.
We will see that this assumption is justified at sufficiently large $\beta$
in Section \ref{subsec:validity}. 

Note that the drift term from the $\theta$ term 
appears only for the link variables surrounding the puncture,
and it is actually a constant independent of the configuration.
While these properties are peculiar to the log definition $Q_{\rm log}$,
similar properties hold also for the sine definition $Q_{\rm sin}$
at large $\beta$, where all the plaquettes $P_n$ approach unity
except for $P_K$, which corresponds to the puncture.
We discuss the case
with the sine definition in Appendix \ref{sec:results-punctured},
where we see that the obtained results 
are qualitatively the same as those obtained 
with the log definition.

\subsection{the $\theta$ dependence of the partition function}
\label{subsec:chargedist}

As we have seen in Section \ref{equivalence-punctured_model},
the punctured model is equivalent to the non-punctured model
in the infinite volume limit for $|\theta|<\pi$, beyond which
the equivalence ceases to hold.
In particular, the punctured model does not have the
$2\pi$ periodicity in $\theta$, which exists
in the non-punctured model.

\begin{figure}
\centering
{\includegraphics[scale=0.6]{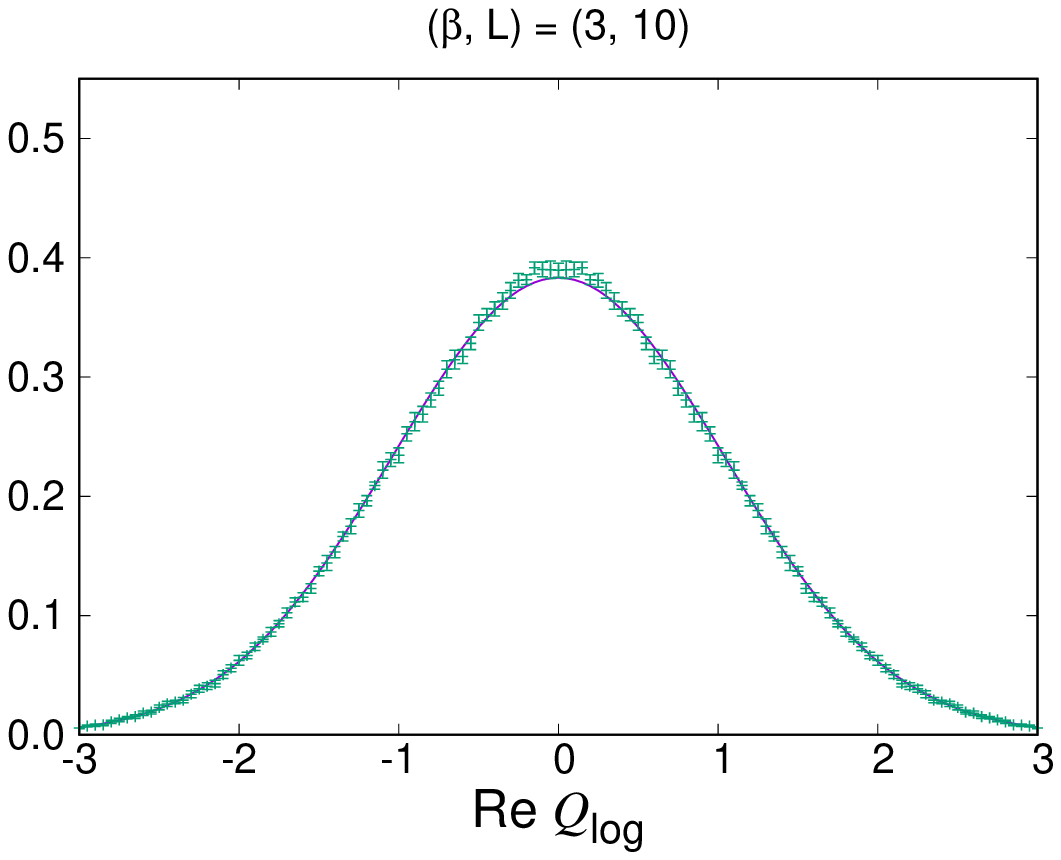}}
{\includegraphics[scale=0.6]{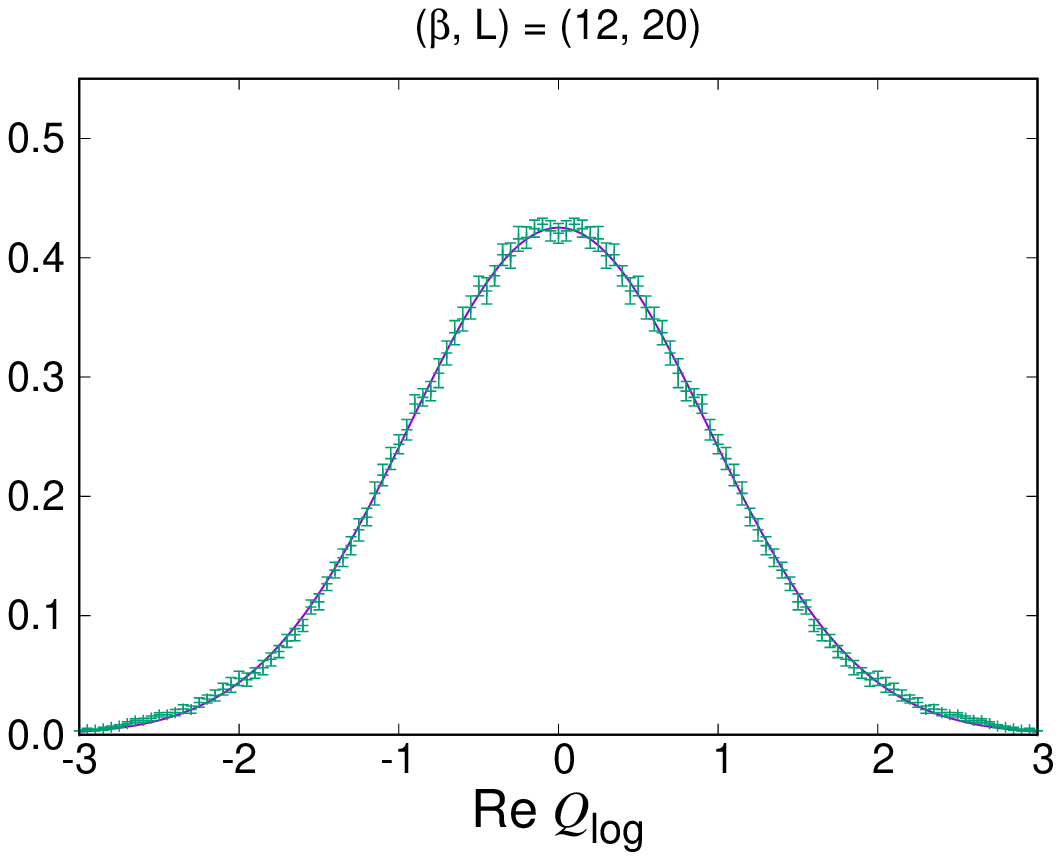}}
\caption{The topological charge distribution for $\theta=0$
obtained by the CLM for the punctured model
using the log definition $Q_{\rm log}$
is plotted for $(\beta,L)=(3,10)$ (Left) and $(\beta,L)=(12,20)$ (Right).
The solid lines represent 
the exact results obtained by evaluating
\eqref{eq:charge-partition-inverse-transform}
using the partition function (\ref{part-fn-punc-exact}).
}
\label{fig:partitiontheta0}
\end{figure}

In order to understand this point better,
we discuss the $\theta$ dependence of the partition function
in this section.
Let us first note that the partition function 
for arbitrary $\theta$
is related 
to the topological charge distribution $\rho(q)$ for $\theta=0$
through Fourier transformation as
\begin{align}
    Z(\theta) &= \int dU e^{-S_g[U]+i\theta Q[U]}\nonumber\\
    &=\int  dU e^{-S_g[U]}\int dq \,
    e^{i\theta q}\delta(Q[U]-q)
\nonumber\\
    &=Z(0)\int dq \, e^{i\theta q}\rho(q) \ .
\label{eq:charge-partition-transform}
\end{align}
Therefore, the absence of the $2\pi$ periodicity in $\theta$
in the punctured model is directly related
to its property that the topological charge can take 
non-integer values
even if we use the log definition $Q_{\rm log}$.
Going beyond the fundamental region $ - \pi < \theta \le \pi$
simply amounts to probing
the fine structure of the topological charge distribution $\rho(q)$,
which is irrelevant in the infinite volume limit.

By making an inverse Fourier transform,
we can obtain 
the topological charge distribution $\rho(q)$ for $\theta=0$ as
\begin{equation}
    \rho(q)=\frac{1}{Z(0)}\int_{-\infty}^{\infty} \frac{d\theta}{2\pi}
    Z(\theta)\, e^{-i\theta q} \ .
    \label{eq:charge-partition-inverse-transform}
\end{equation}
We calculate
this quantity
for the punctured model by the CLM for $\theta=0$.
In Fig.~\ref{fig:partitiontheta0}, we show the results
for $(\beta,L)=(3,10)$ (Left) and $(\beta,L)=(12,20)$ (Right),
which agree well with the exact results
obtained by evaluating (\ref{eq:charge-partition-inverse-transform})
using the partition function (\ref{part-fn-punc-exact}).
Note that the calculation actually 
reduces to that of the real Langevin method
due to the absence of the sign problem for $\theta=0$.
We therefore have no concerns 
about the criterion for correct convergence here.

While the sign problem is absent for $\theta=0$,
the topology freezing problem can still be an issue for large $\beta$.
The agreement we see
for $(\beta,L)=(12,20)$
confirms that this problem is resolved in the punctured model
at least for $\theta = 0$.


\subsection{validity of the CLM}
\label{subsec:validity} 

\begin{figure}
\centering
\includegraphics[scale=0.6]{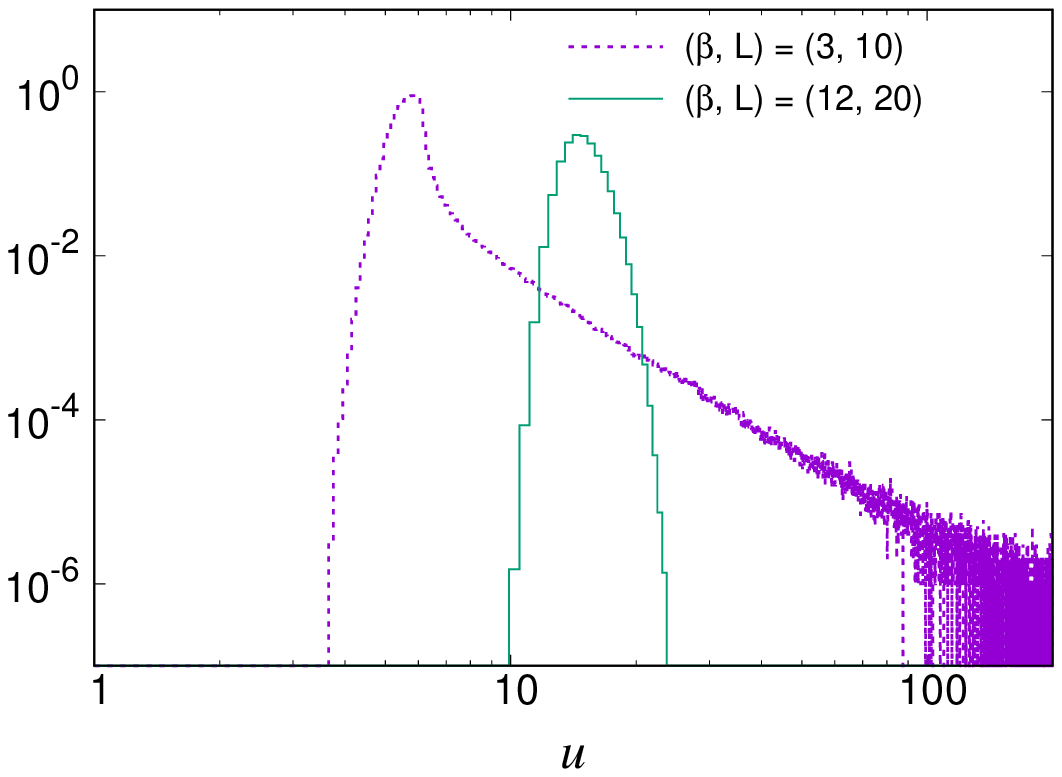}
\includegraphics[scale=0.6]{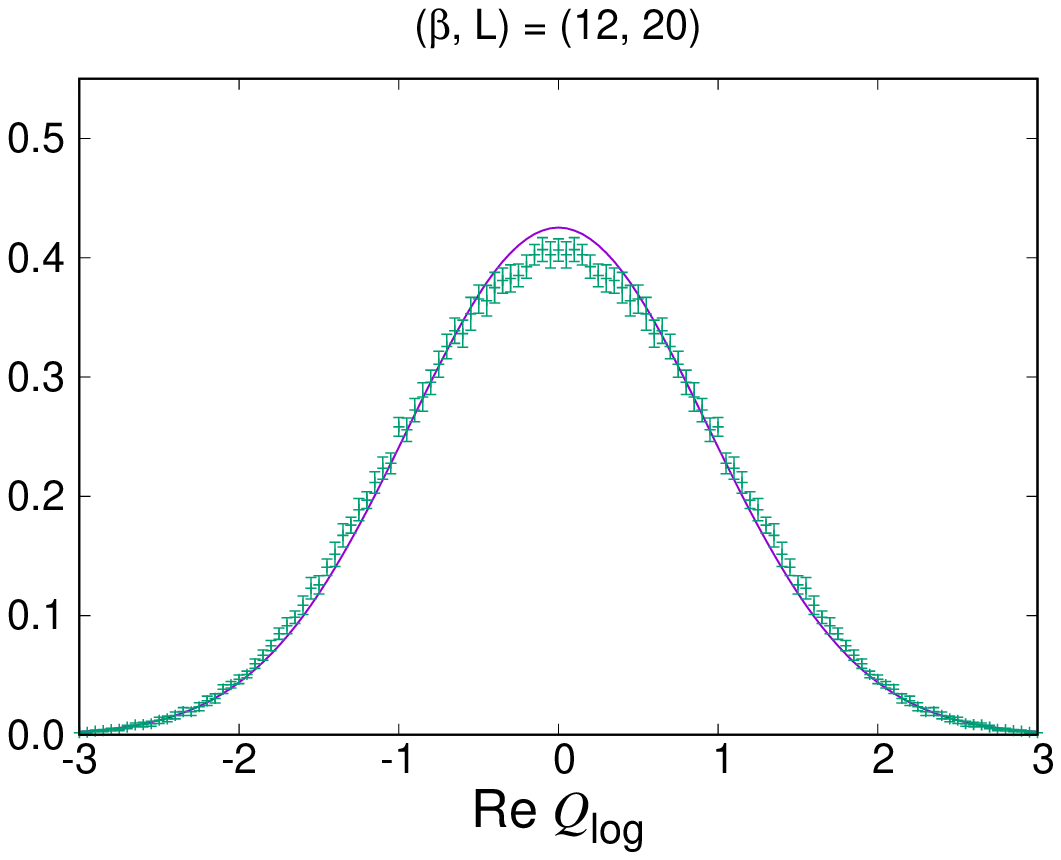}
\caption{The results obtained by the CLM 
  for the punctured model using the log definition $Q_{\mathrm{log}}$
of the topological charge.
(Left) The histogram of the magnitude $u$ of the drift term 
defined by (\ref{def-u})
is shown for $(\beta,L)=(3,10)$ and $(12,20)$ with $\theta=\pi$.
(Right) The histogram of $\mathrm{Re} \, Q_{\mathrm{log}}$
is shown
for $(\beta, L)=(12,20)$ with $\theta=\pi$.
The exact result
obtained for $(\beta, L)=(12,20)$ with $\theta=0$
is shown by the solid line for comparison.
}
\label{fig:compare_drift}
\end{figure}

In this section, 
we discuss the validity of the CLM for the punctured model.
Fig.~\ref{fig:compare_drift}(Left) shows
the histogram of the drift term 
for $(\beta,L)=(3,10)$ and $(\beta,L)=(12,20)$
with $\theta=\pi$, which are the parameters used in
Section \ref{sec:result_nonpunctured} for the non-punctured model.
We find that the criterion is satisfied for $(\beta,L)=(12,20)$
but not for $(\beta,L)=(3,10)$, similarly to the situation
in the non-punctured model.
The difference from the non-punctured model is seen, however,
in Fig.~\ref{fig:compare_drift}(Right),
where we show
the histogram of $\mathrm{Re} \, Q_{\mathrm{log}}$ 
obtained by the CLM for $(\beta, L)=(12,20)$ with $\theta=\pi$.
(The result for $(\beta,L)=(3,10)$ looks quite similar to this plot.)
It is widely distributed
within the range $-3 \lesssim \mathrm{Re} \, Q_{\mathrm{log}}\lesssim 3$,
which is in sharp contrast to the plot in Fig.~\ref{fig:result_NP}(Right)
for the same $(\beta,L)=(12,20)$ in the case of the non-punctured model.
In fact, it turns out to be close\footnote{Note,
  however, that precise agreement is not expected here
since the histogram of $\mathrm{Re} \, Q_{\mathrm{log}}$
is a non-holomorphic quantity, for which the CLM does not allow
a clear interpretation.}
to the exact result obtained
for the same $(\beta,L)=(12,20)$ with $\theta=0$, which is
plotted in the same figure.
Thus we find that the topology freezing problem at large $\beta$
is circumvented in the punctured model and yet the CLM remains valid.


Next we discuss the reason why the punctured model can avoid the 
topology freezing problem without causing large drifts.
The difference from the non-punctured model is that one
of the plaquettes, $P_{K}$,
is removed from the action.
Note that the topological charge $Q_{\mathrm{log}}$ for the punctured
model is given by
\begin{equation}
Q_{\mathrm{log}}
= - \frac{i}{2\pi}\sum_{n}\log P_{n}
  +\frac{i}{2\pi}\log P_{K} \ ,
\label{top-punctured}
\end{equation}
where the first term is
nothing but the topological charge
defined for the non-punctured model,
whose real part takes integer values.
The second term has a real part which 
lies within the interval $[-\frac{1}{2},\frac{1}{2})$.
Therefore it makes sense to define the ``topology change''
in the punctured model
as the situation in which the real part of the first term 
changes by $\pm 1$.
As we discussed in Section \ref{whypunctureworks} for the
non-punctured model,
one of the plaquettes
should inevitably cross the branch cut
in order for the topology change to occur in the above sense.
When $\beta$ is large, this process is highly suppressed
for all the plaquettes that are included in the action.
In the non-punctured model, the topology freezing problem
occurs precisely for this reason.
However, in the punctured model, the particular
plaquette $P_K$ is removed from the action,
and therefore it can change freely even for large $\beta$.

\begin{figure}
\centering
\includegraphics[scale=0.6]{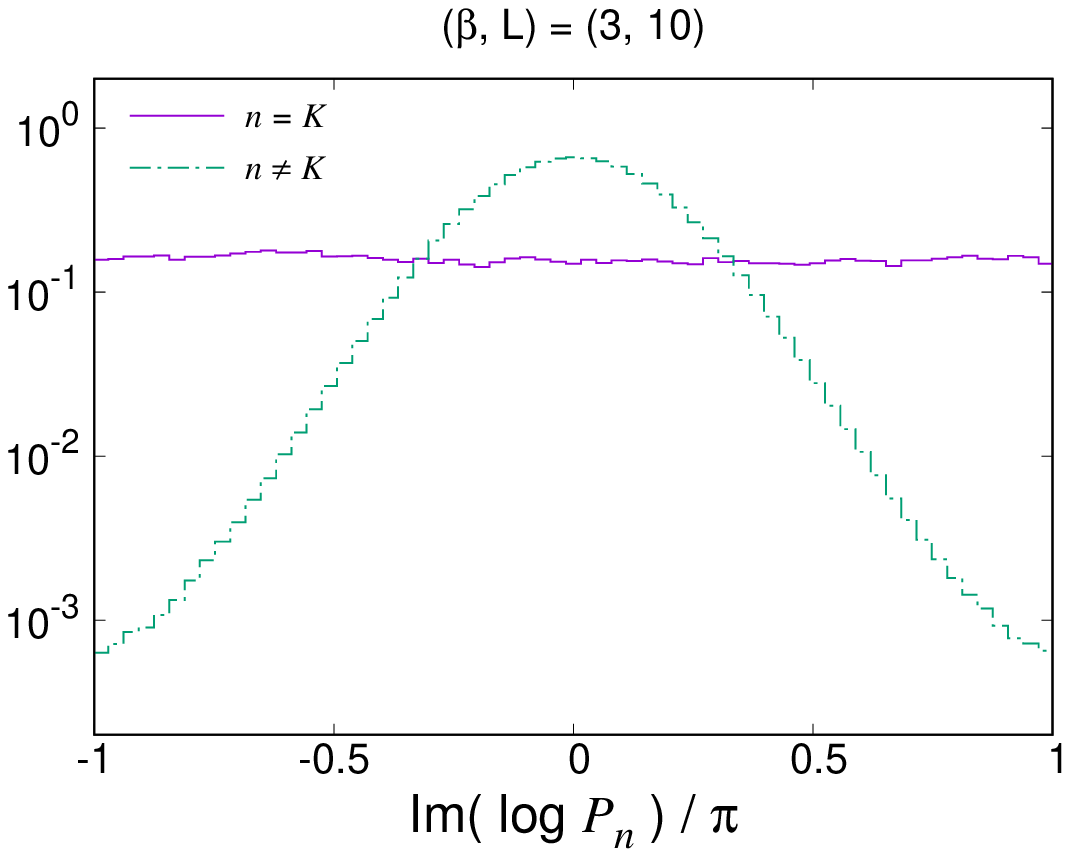}
\includegraphics[scale=0.6]{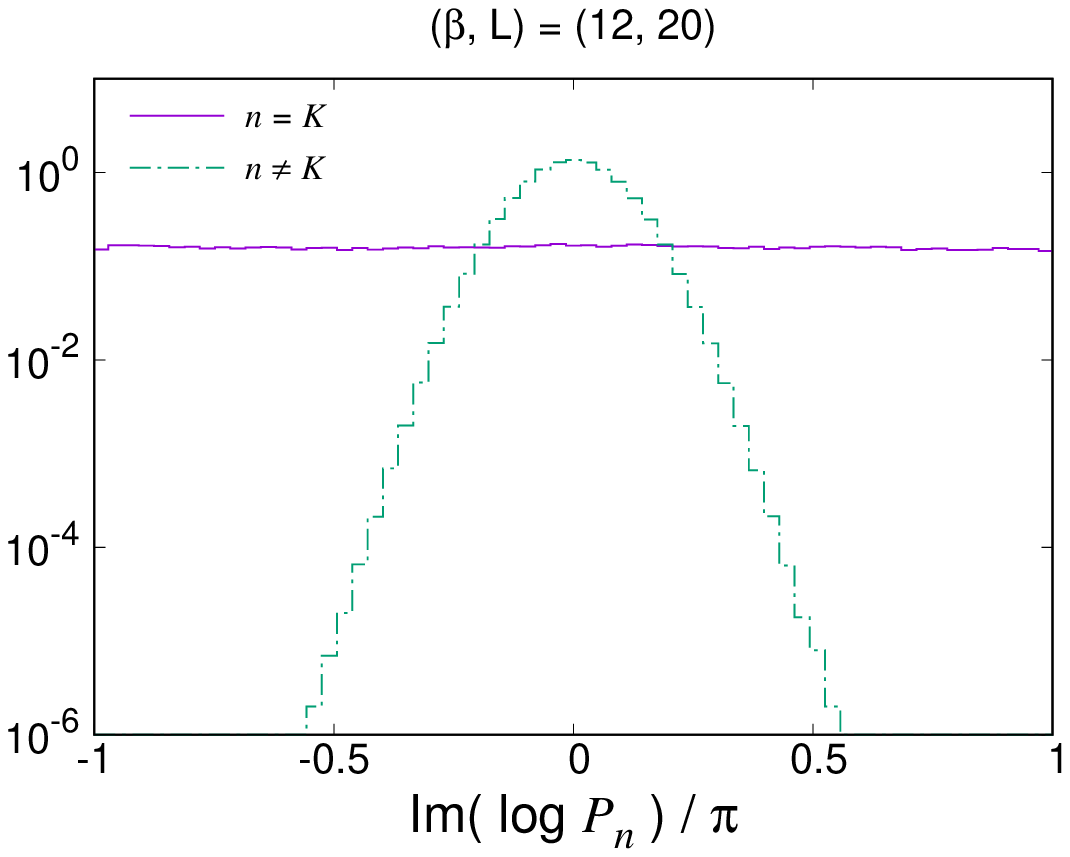}
\caption{The distribution of the phase of the plaquettes
is plotted for the punctured model 
with the log definition (\ref{log-Q-def}) of the topological charge
for $(\beta,L)=(3,10)$ (Left) and
$(\beta,L)=(12,20)$ (Right) with $\theta=\pi$.
We show the results for the plaquette ($n=K$) removed from the action
and those for all the other plaquettes ($n\neq K$) separately.
}
\label{fig:phase_dist}
\end{figure}

This is demonstrated in Fig.~\ref{fig:phase_dist},
where we plot the probability 
distribution of the phase of the plaquette $P_{K}$
as well as that of the other plaquettes $P_{n}\mbox{~($n\neq K$)}$
for $(\beta,L)=(3,10)$ (Left) and $(\beta,L)=(12,20)$ (Right).
We find that the phase of the removed plaquette $P_{K}$ is almost
uniformly distributed for both $(\beta,L)$.
On the other hand, the distribution of 
the phase of the other plaquettes depends on $(\beta,L)$.
It has a compact support for $(\beta,L)=(12,20)$ but
not for $(\beta,L)=(3,10)$.
In the former case, there is no distribution 
at the branch cut, which implies that the branch cut crossing of the
plaquettes $P_{n}\mbox{~($n\neq K$)}$ does not occur at all. 
In the latter case, there is a small but finite distribution 
at the branch cut, which means that the value of $\beta$
is not large enough to suppress the branch cut crossing of the
plaquettes $P_{n}\mbox{~($n\neq K$)}$ completely.

This is consistent with the fact that the histogram of the 
drift term has fast fall-off for $(\beta,L)=(12,20)$ but not 
for $(\beta,L)=(3,10)$
considering the discussion given in Section \ref{whypunctureworks}.
While the flow diagram in Fig.~\ref{fig:DS}(Left)
is obtained for the sine definition of the topological charge,
it looks similar for the log definition,
which simply corresponds to setting $\theta=0$ in
\eqref{drift-at-phi}.
Therefore,
large drifts can appear when
one of the plaquettes $P_{n}\mbox{~($n\neq K$)}$ crosses the
branch cut, which indeed occurs for $(\beta,L)=(3,10)$ also for the
punctured model.
For $(\beta,L)=(12,20)$, on the other hand,
the topology change is made possible
by allowing the removed plaquette $P_K$ to cross the branch cut freely,
but all the plaquettes that are included in the action
are forced to stay close to unity because of large $\beta$.
This justifies our assumption that the issue of $\delta$-function
can be neglected in deriving the drift terms
(\ref{eq:drift_theta_smp_1})
and (\ref{eq:drift_theta_smp_2}).
Since the plaquette $P_K$ does not appear in the drift terms,
it does not cause large drifts even if it crosses the branch cut.
This makes it possible for the punctured model
to avoid the topology freezing problem
without causing large drifts.


Let us next discuss how the unitarity norm (\ref{def-unitarity-norm})
behaves in our complex Langevin simulations.
As we can see from (\ref{eq:drift_theta_smp_1})
and (\ref{eq:drift_theta_smp_2}),
the link variables surrounding the puncture have
a drift term in the imaginary direction coming from the $\theta$ term.
At each Langevin step, two of the link variables are
multiplied by $e^{\theta \Delta t /2 \pi }$ and the other two
are multiplied by $e^{-\theta \Delta t /2 \pi }$
so that the removed plaquette is multiplied by
$e^{2 \theta \Delta t / \pi }$ due to this drift term.
Therefore, there is a danger that
the magnitude of
these four link variables increases or decreases exponentially
and hence
the unitarity norm (\ref{def-unitarity-norm}) 
grows exponentially with the Langevin time.

\begin{figure}
\centering
{\includegraphics[scale=0.8]{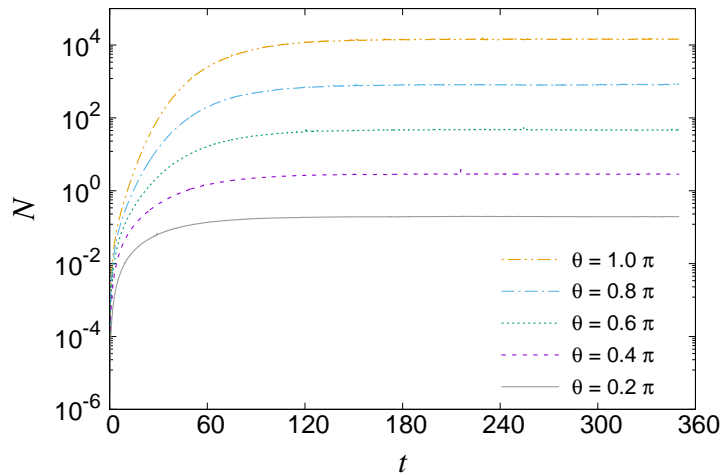}}
\caption{The history of the unitarity norm $\mathcal{N}$ 
is plotted for the punctured model 
with the log definition (\ref{log-Q-def}) of the topological charge
for various $\theta$ with $(\beta,L)=(5,16)$.}
\label{fig:uninormsaturation}
\end{figure}

In Fig.~\ref{fig:uninormsaturation}, we plot the history of
the unitarity norm (\ref{def-unitarity-norm}) for various $\theta$
with $(\beta , L)=(5,16)$.
Similar results are obtained for other $(\beta , L)$.
(Here and for the rest of this subsection, we restrict ourselves
to the parameter sets, for which the histogram of the drift term
has fast fall-off.)
Indeed we observe an exponential growth at early stage, but
the unitarity norm actually saturates
to a constant 
depending on $\theta$ at sufficiently long Langevin time.
This saturation occurs since the non-unitarity of the
four link variables surrounding the puncture
propagates to all the other link variables on the lattice
due to the interaction caused by the gauge action $S_{g}$,
which tries to make each plaquette except the removed one
close to unity.
We find that thermalization of various observables can be achieved
only after the saturation of the unitarity norm.

\begin{figure}
\centering
{\includegraphics[scale=0.6]{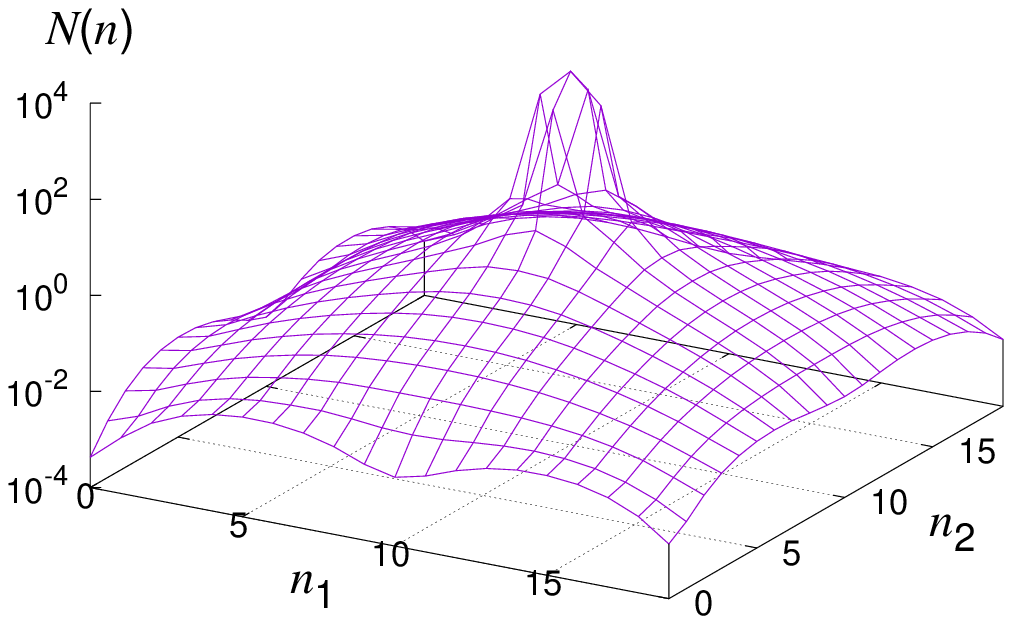}}
{\includegraphics[scale=0.6]{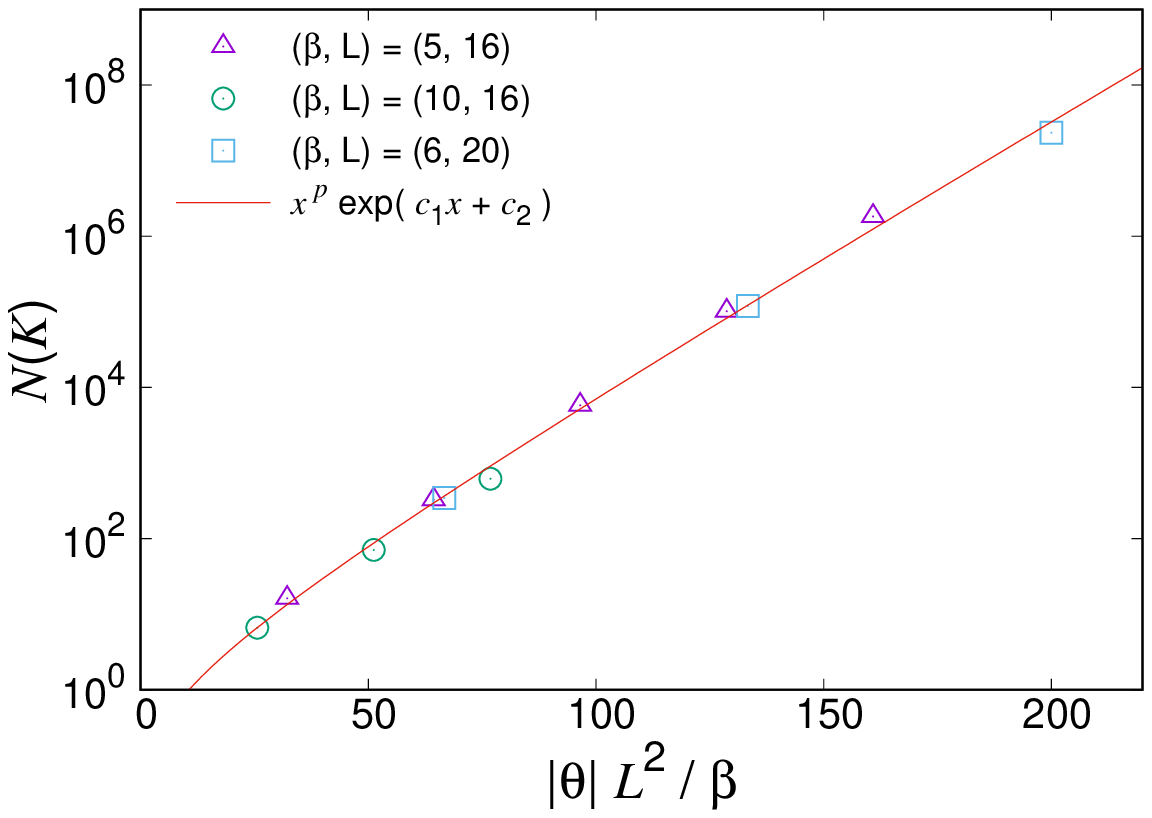}}
\caption{(Left) The local unitarity norm $\mathcal{N}(n)$
for each plaquette $P_n$
defined by (\ref{def-local-unitarity-norm})
is plotted against $n=(n_1,n_2)$
for the punctured model 
with the log definition (\ref{log-Q-def}) of the topological charge
for $(\beta,L)=(12,20)$ with $\theta=\pi$.
The removed plaquette corresponds to $n=K=(10,10)$ in this figure.
(Right) The local unitarity norm $\mathcal{N}(K)$
for the removed plaquette $P_K$ 
obtained for various $\beta$, $L$ and $\theta$
is plotted against $x=|\theta| L^{2}/\beta$.
The solid line represents a fit to $x^{p}\exp(c_{1} x + c_{2})$
with $c_{1}=0.079(5)$, $c_{2}=-3(1)$ and $p=0.8(4)$.}
\label{fig:link on lattice}
\end{figure}

In fact, we find that the unitarity norm is not distributed uniformly
on the lattice due to the existence of the puncture,
as is also expected from the above discussion.
In order to see this, we
define the ``local unitarity norm'' by
\begin{equation}
    \mathcal{N}(n)=
\frac{1}{4} \sum_{(k,\mu)\in P_n}
\Big\{ \mathcal{U}_{k,\mu}^*
\mathcal{U}_{k,\mu}+(\mathcal{U}_{k,\mu}^*\mathcal{U}_{k,\mu})^{-1}-2
\Big\} \ ,
\label{def-local-unitarity-norm}
\end{equation}
which is an average of the unitarity norm for the four link variables 
surrounding each plaquette $P_n$.
The unitarity norm defined by
(\ref{def-unitarity-norm}) is simply 
an average of $\mathcal{N}(n)$
over all the plaquettes including the removed one;
namely $\mathcal{N}=\frac{1}{L^2}\sum_n \mathcal{N}(n)$.
In Fig.~\ref{fig:link on lattice}(Left), we
plot this quantity $\mathcal{N}(n)$ against $n=(n_1,n_2)$
for $(\beta,L)=(12,20)$ with $\theta = \pi$,
where the puncture is located at $n=K=(10,10)$.
We observe a sharp peak at the puncture,
which goes up to $\mathcal{N}(K)\sim 6 \times 10^{3}$.
The plaquettes adjacent to the puncture
have a local unitarity norm $\sim 1.5 \times 10^{3}$.
This implies that
the unitarity norm is mostly dominated by the
four link variables surrounding the puncture.

\begin{figure}
\centering
{\includegraphics[scale=0.8]{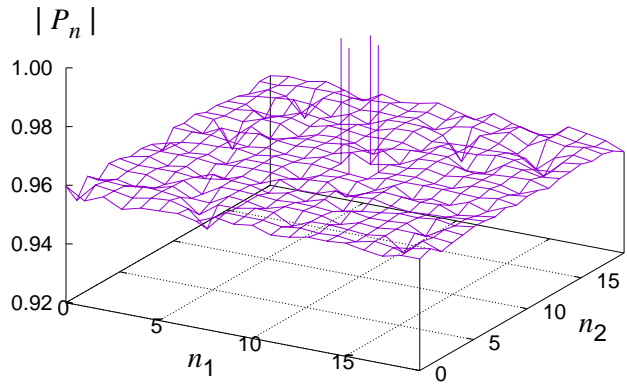}}
\caption{The absolute value of the plaquette $P_n$ is plotted
against $n=(n_1, n_2)$ for the punctured model
with the log definition (\ref{log-Q-def}) of the topological charge
for $(\beta,L)=(12,20)$ with $\theta=\pi$.
The removed plaquette corresponds to $n=K=(10,10)$ in this figure.
}
\label{fig:plaq on lattice}
\end{figure}

The local unitarity norm $\mathcal{N}(K)$ at the puncture
depends not only on $\theta$ but also on $\beta$ and $L$.
In Fig.~\ref{fig:link on lattice}(Right),
we plot this value
against $x=|\theta| V_{\rm phys}=|\theta| L^2/\beta$ 
for various $\theta$, $\beta$ and $L$.
All the data can be
fitted to a single curve $\mathcal{N}(K)=x^{p}\exp(c_{1} x + c_{2})$,
which reveals an exponential behavior at large $x$.

What actually matters for the validity of the CLM is
not so much the local unitarity norm $\mathcal{N}(n)$
as the absolute value of each plaquette $P_n$,
which we plot
in Fig.~\ref{fig:plaq on lattice}
against $n=(n_1,n_2)$
for the same parameters as in Fig.~\ref{fig:link on lattice}(Left).
The absolute value of $P_K$ corresponding to the removed
plaquette
is close to
$(\sqrt{\mathcal{N}(K)})^4 \sim 3.6 \times 10^{7}$,
which implies that $|\mathcal{U}_{K, 1}|$,
$|\mathcal{U}_{K+\hat{1}, 2}|$,
$|\mathcal{U}_{K+\hat{2}, 1}^{-1}|$ and
$|\mathcal{U}_{K, 2}^{-1}|$ are close to $\sqrt{\mathcal{N}(K)}$.
Except for this removed plaquette,
the absolute value of the plaquette
deviates only slightly from unity
due to large $\beta$.

In fact, this deviation of $|P_n|$ from unity for $n \neq K$
has a physical meaning since
${\rm Im} \, \langle Q_{\log}  \rangle
= - \frac{1}{2\pi} \sum_{n\neq K} \langle \log |P_n| \rangle$
as one can see from (\ref{top-punctured}).
From the exact result (\ref{imag-Q-exact})
obtained 
at large $\beta$,
we find that $|P_n| \sim  e ^{-  \theta / (2 \pi \beta)}$ for $n \neq K$,
which is $\sim 0.96$ for $\theta=\pi$ and $\beta=12$
in agreement with the value observed in 
Fig.~\ref{fig:plaq on lattice}.
If we flip the sign of $\theta$, which corresponds 
to the parity transformation,
we find that $|P_n|\mapsto |P_n|^{-1}$ for all $n$.

Note also that $P_{K}$ does not appear in the drift term,
which implies that its absolute value can become large 
without causing large drifts.
We have confirmed that
the criterion for correct convergence is satisfied
for sufficiently large $\beta$,
and indeed the exact results for various observables
can be reproduced correctly as we will see in the next section.
This remains to be the case
even for large $\theta$ and/or large $V_{\rm phys}$,
where the unitarity norm becomes large.\footnote{We find,
  however, that the fluctuation of the local unitarity norm is small even
  for the one $\mathcal{N}(K)$ at the puncture, which implies that
  the distribution of each link variable has fast fall-off.
  Therefore, it is suggested
  that the problem due to the boundary
  terms discussed in
  Refs.~\cite{Aarts:2009uq,Aarts:2011ax,Scherzer:2018hid,Scherzer:2019lrh}
  does not occur.}
Thus the present model provides a counterexample to the 
common wisdom
that the CLM fails when the unitarity norm becomes large.


\subsection{results for the observables}
\label{subsec:observables} 

In this section, we calculate the observables
for the punctured model by the CLM
and compare our results with
the exact results derived in Appendix \ref{sec:analytic_obs}.
Let us recall that,
in the definitions
\eqref{eq:def_aveP}, \eqref{eq:def_ImQ} and \eqref{eq:def_chi},
$V$ denotes the number of plaquettes in the action,
which is $V=L^2-1$
for the punctured model.
In contrast, we define the physical volume $V_{\rm phys}$
by Eq.~\eqref{phys-vol} not only for the non-punctured model but also
for the punctured model,
which simplifies the relationship between
$\beta$ and $L$ for fixed $V_{\rm phys}$.

\begin{figure}{}
\centering
{\includegraphics[scale=0.6]{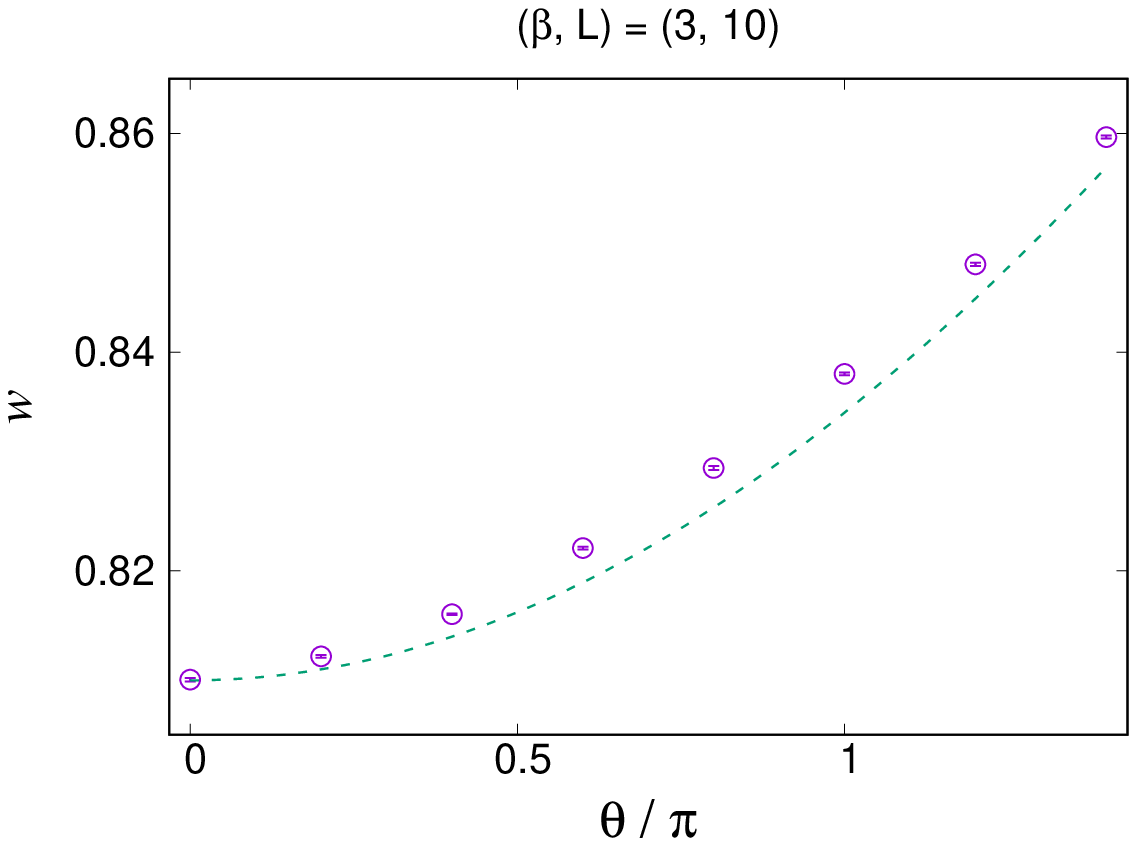}}
{\includegraphics[scale=0.6]{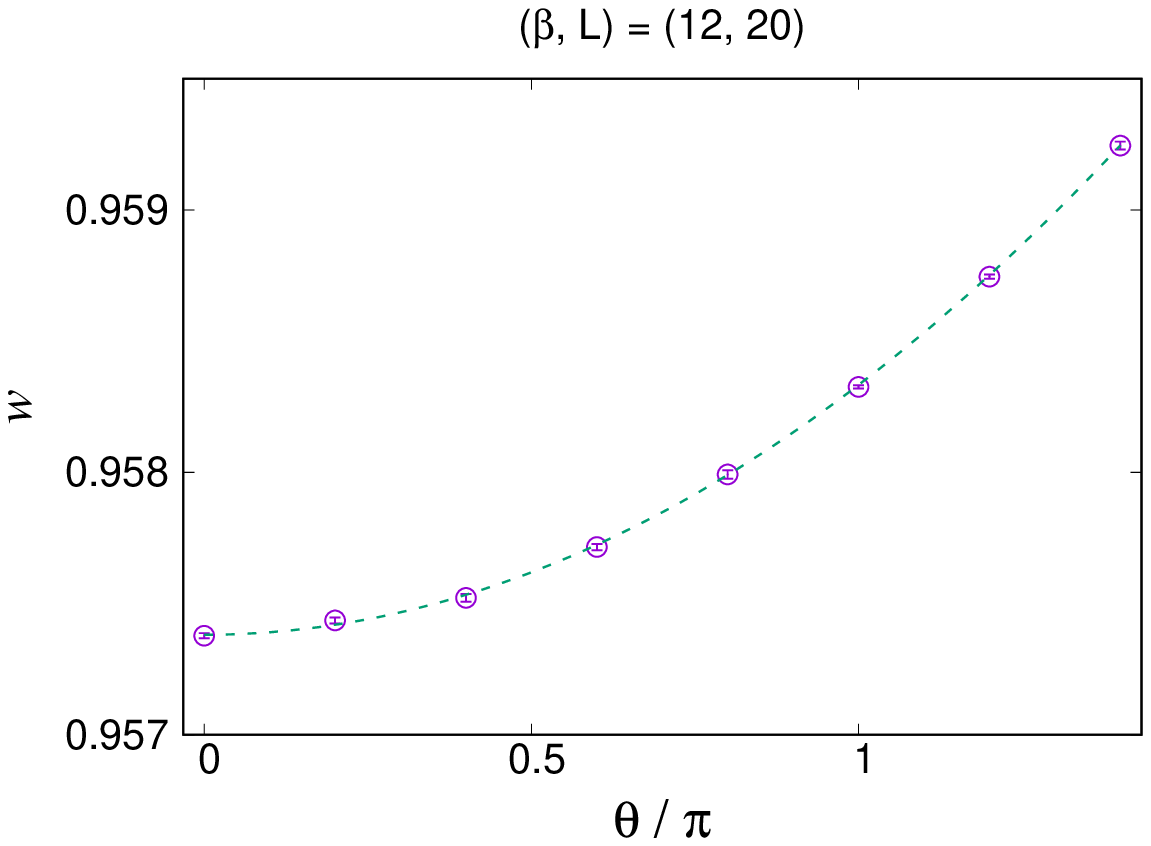}}\\
\vspace{10pt}
{\includegraphics[scale=0.6]{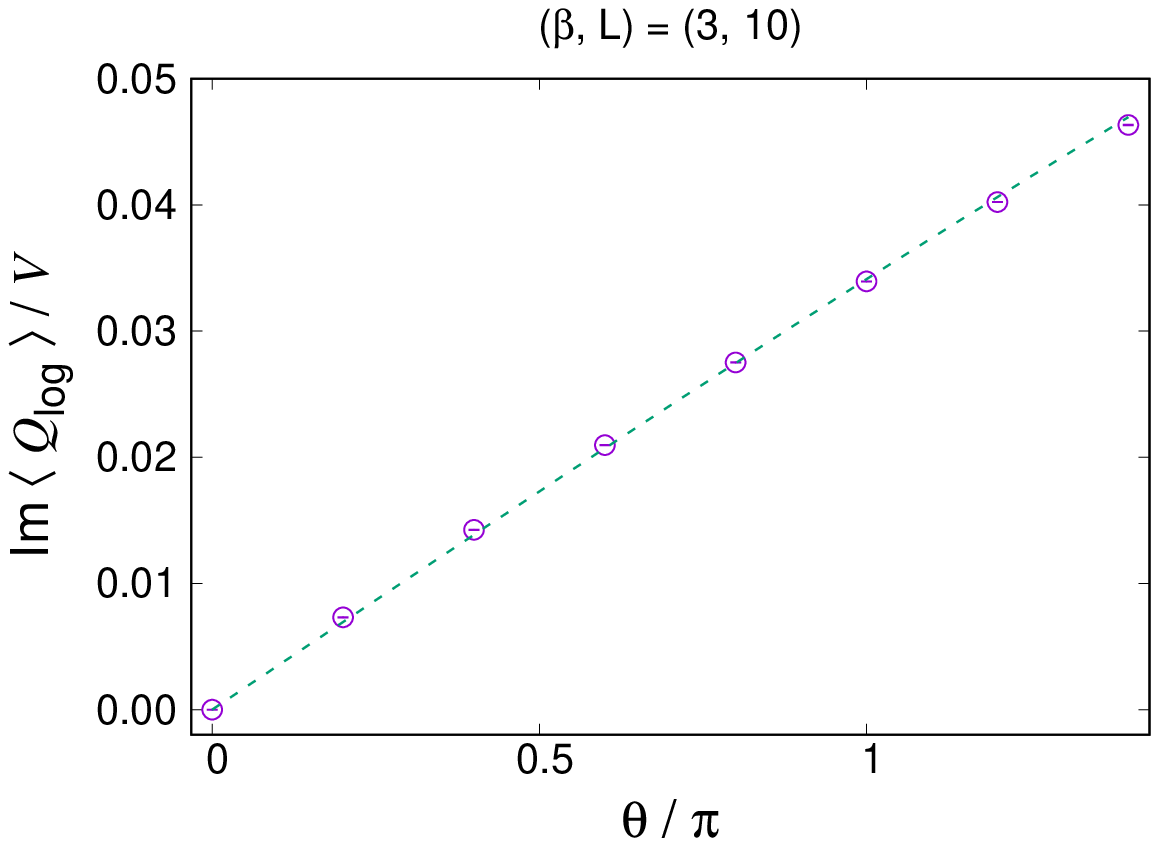}}
{\includegraphics[scale=0.6]{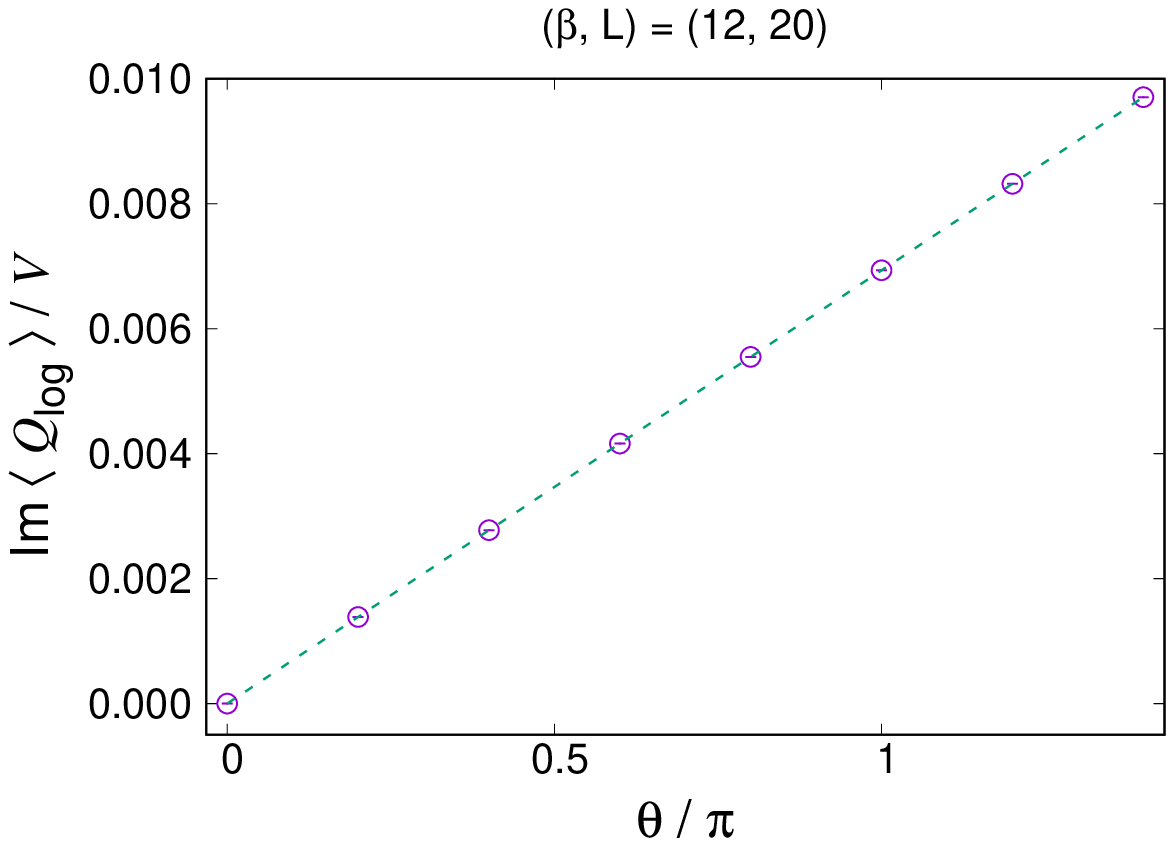}}\\
\vspace{10pt}
{\includegraphics[scale=0.6]{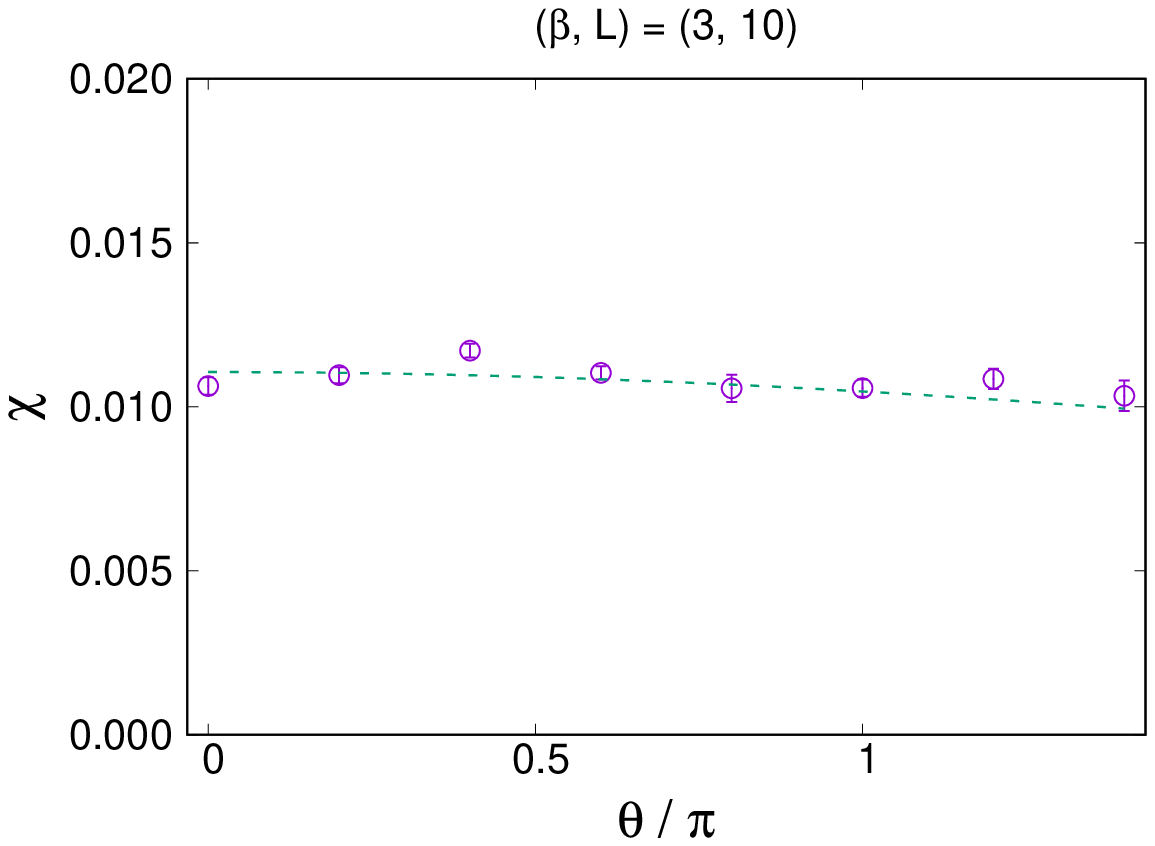}}
{\includegraphics[scale=0.6]{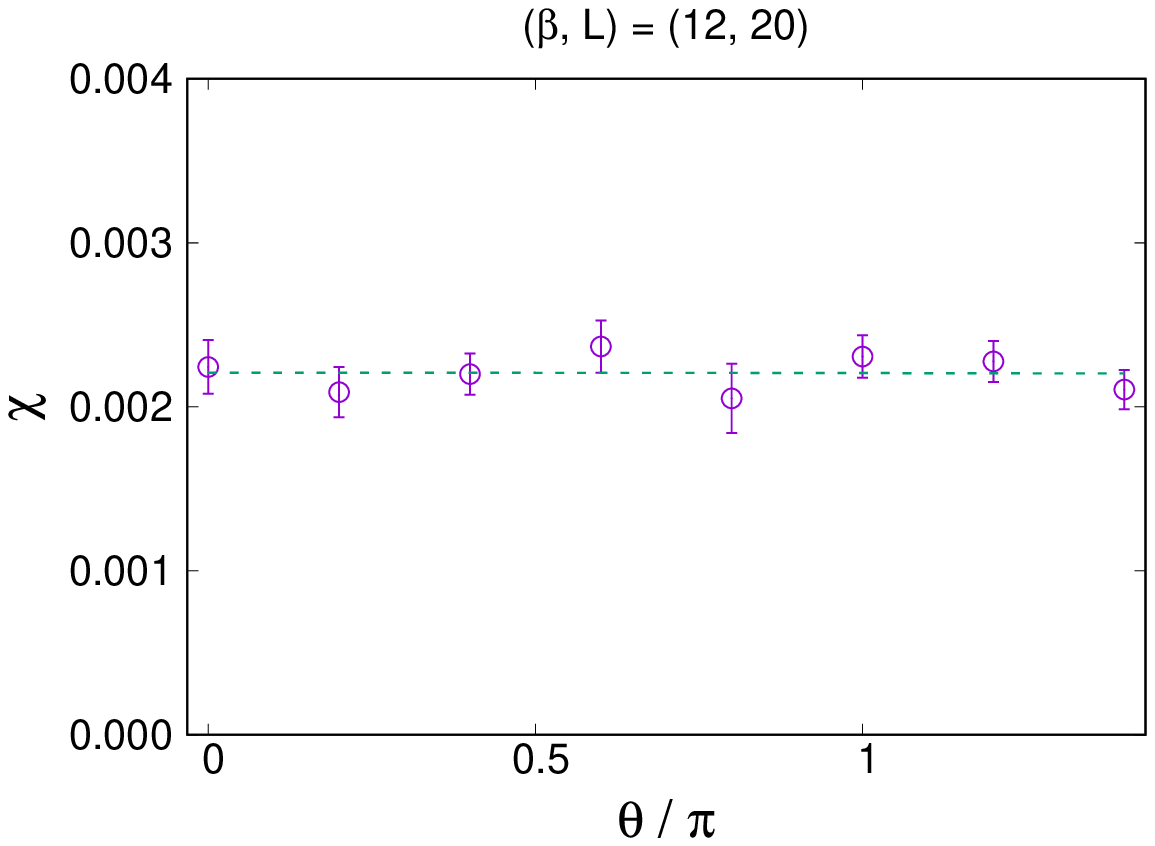}}
\caption{The results for various observables obtained by the CLM 
  for the punctured model with the log definition $Q_{\mathrm{log}}$.
  The average plaquette (Top),
the imaginary part of the topological charge density (Middle),
the topological susceptibility (Bottom)
are plotted against $\theta$
for $(\beta, L)=(3,10)$ (Left) and $(12,20)$ (Right).
The exact results for the same $(\beta, L)$ are shown by the dashed lines
for comparison.}
\label{fig:obs_punc}
\end{figure}

In Fig.~\ref{fig:obs_punc}, we
show our results 
for
the average plaquette $w$ (Top),
the topological charge (Middle) 
and the topological susceptibility (Bottom)
against $\theta$ for $(\beta,L)=(3,10)$ and $(12,20)$
in the left and right columns, respectively,
which correspond to a fixed physical volume 
$V_{\rm phys} \equiv L^2/\beta = 10^2/3$.
The exact results obtained for the same model with the same parameter sets
are also shown for comparison.
We find from our results for the average plaquette that the 
exact results are reproduced for $(\beta,L)=(12,20)$,
but there is slight deviation for $(\beta,L)=(3,10)$.
This is consistent with our observation in Section \ref{subsec:validity} 
that the condition for correct convergence is
met for $(\beta,L)=(12,20)$ but not for $(\beta,L)=(3,10)$.

For the topological charge, we find that
our results reproduce the exact results not only for 
$(\beta,L)=(12,20)$
but also for $(\beta,L)=(3,10)$.
The same holds for the topological susceptibility.
We consider that the agreement observed here
for $(\beta,L)=(3,10)$ is accidental, though, since
the condition for correct convergence is not satisfied.
The fact that the results of the CLM for the punctured model 
with $(\beta,L)=(3,10)$ is not as bad as those for the
non-punctured model with the same $(\beta,L)$
shown in Fig.~\ref{fig:resultes_naive}(Left)
can be understood by considering that
the effect of the $\theta$ term is 
included correctly by the drift terms for the link variables
composing the removed plaquette, but it is only
the infrequent branch cut crossing
of the other plaquettes that spoils the validity of the CLM.


\section{Summary and discussions}
\label{sec:summary}

In this paper, we have made an attempt to apply
the CLM to gauge theories with a $\theta$ term.
As a first step, we applied the CLM 
to the 2D U(1) case, which is exactly solvable on a finite lattice
with various boundary conditions.
We find that a naive implementation of the method fails due to the
topological nature of the $\theta$ term.

While the gauge configurations are complexified
in the CLM, one can still define the notion of 
topological sectors by ${\rm Re} \, Q_{\rm log}\in\mathbb{Z}$.
When a transition between different topological sectors occurs,
one of the plaquettes has to cross the branch cut inevitably,
which causes the appearance of large drift terms.
This indeed happens at small $\beta$, where we find that the 
criterion for correct convergence of the CLM is not satisfied.
Increasing $\beta$ makes all the plaquettes close to unity.
The large drift terms do not appear in this case, and the 
criterion for correct convergence of the CLM is satisfied.
However, the topology change does not occur during the simulation
and the ergodicity is violated.
This is analogous to the topology freezing problem,
which is known to occur for $\theta=0$.
The results obtained in this case correspond to the
expectation values for 
an ensemble restricted to a particular topological sector specified
by the initial configuration.

In order to avoid this problem, we have considered the 
punctured model, which can be obtained
by removing
one plaquette from the action, both from the gauge action
and from the $\theta$ term.
While the quantity ${\rm Re} \, Q_{\rm log}$
is no more restricted to integer values,
we can still formally classify the complexified configurations 
into ``topological sectors''
by adding back the contribution of the removed plaquette to
${\rm Re} \, Q_{\rm log}$.
Even for large $\beta$,
the removed plaquette can cross the branch cut easily,
which results in frequent 
transitions between different ``topological sectors''.
Note also that, as far as $\beta$ is sufficiently large,
all the other plaquettes 
are close to unity,
and hence large drift terms do not appear.
Thus the criterion for correct convergence of the CLM can be satisfied
by simply approaching the continuum limit
without causing the topology freezing problem.
Indeed our results obtained by the CLM for the punctured model
reproduce the exact results even at large $\theta$.

In the case of the punctured model,
the drift term from the $\theta$ term appears only 
for the link variables composing the removed plaquette,
and it is given by $\pm i \frac{\theta}{2\pi}$, which
causes multiplication by a constant factor 
$e^{\mp \Delta t \frac{\theta}{2\pi}}$ to these link variables
at each Langevin step.
The local unitarity norm of these link variables grows exponentially
at early Langevin times, but it saturates at some point
to some constant, which 
increases exponentially for large $|\theta| V_{\rm phys}$.
We have seen that the CLM works perfectly even in this situation
as far as $\beta$ is sufficiently large.
This provides a counterexample to the common wisdom
that the CLM fails when the unitarity norm becomes large.
Thus our results also give us new insights into the method itself.

The punctured model is
actually equivalent
to the non-punctured model
in the infinite volume limit for $|\theta|<\pi$.
In that limit, 
the topological charge can take arbitrarily large values,
so the discretization of $Q$ to integers is no more important.
This equivalence has been confirmed
explicitly by obtaining exact results for the punctured model.
In fact, the exact results also reveal 
the absence of finite volume effects in the punctured model
as opposed to the non-punctured model,
which exhibits sizable finite volume effects around $\theta \sim \pi$.
It is conceivable that the smearing of the topological charge
somehow results in the reduction of finite volume effects.
If so, a similar conclusion should hold more generally.

We are currently working on the application of the CLM
to the 4D gauge theory with a $\theta$ term.
(See Ref.~\cite{Bongiovanni:2014rna} for an earlier attempt.)
Some preliminary results obtained in the SU(2) case look promising.
We plan to investigate, in particular,
the interesting phase structure around $\theta = \pi$ 
predicted in Ref.~\cite{Gaiotto:2017yup}.

\subsection*{Acknowledgements}

We would like to thank H.~Fukaya, S.~Hashimoto, K.~Hatakeyama, 
M.~Honda, Y.~Ito and S.M.~Nishigaki for valuable discussions.
The authors are also grateful to
R.~Kitano and N.~Yamada for carefully reading our manuscript.
The computations were carried out on
the PC clusters in KEK Computing Research Center
and KEK Theory Center.
Part of this work was completed through discussions during 
the workshop ``Discrete Approaches to the Dynamics of Fields and Space-Time 2019'' 
held at Shimane University, Matsue, Japan.

\appendix

\section{Derivation of the exact result}
\label{sec:analytic_calc}

In 2D lattice gauge theory,
we can 
obtain the partition function 
explicitly on any manifold at finite
lattice spacing and finite volume \cite{Bonati:2019ylr},
from which various observables can be obtained.
In this section,
we review the derivation
using the so-called K-functional \cite{Rusakov:1990rs}. 

\subsection{the K-functional}

Let us consider a lattice gauge theory
with a $\theta$ term 
on a 2D lattice manifold $\mathcal{M}$. 
Here we take the gauge group 
to be U($N$), which is a generalization of U(1) considered so far. 
Note that
the topology of the gauge field becomes trivial for SU($N$)
in 2D gauge theories.

\begin{figure}
\centering
\includegraphics[scale=0.8]{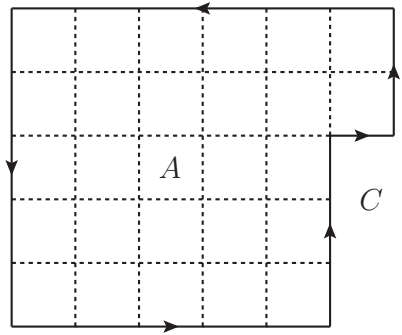}
\caption{An example of the region $A$, which 
has a boundary $C=\partial A$. 
The K-functional for this region is defined by
integrating out the link variables represented by the dashed lines.
The result depends on $\Gamma$ defined by (\ref{def-gamma})
for the loop $C$ represented by the solid line with arrows.}
\label{fig:region-A} 
\end{figure}

As a building block for evaluating the partition function,
we define the K-functional $K_{A}$ for the region
$A\subset\mathcal{M}$
defined by \cite{Rusakov:1990rs}
\begin{equation}
K_{A}(\Gamma) =
\int\left(\prod_{U_{i}\in A\setminus C}dU_{i}\right)e^{-S_{A}} \ ,
\label{def-K-functional-A}
\end{equation}
where the integral goes over the link variables inside $A$
leaving out those on the boundary $C$. (See Fig.~\ref{fig:region-A}.)
The action $S_{A}$ in Eq.~(\ref{def-K-functional-A}) is given by 
\begin{equation}
S_{A}=\sum_{P_{i}\in A}\mathrm{Tr}
\left[-\frac{\beta}{2}
\left(P_{i}+P_{i}^{-1}\right)-\frac{\theta}{2\pi}\log P_{i}\right] \ ,
\end{equation}
where
the sum goes over
the plaquettes $P_{i}$ included in the region $A$. 
Here we use the log definition (\ref{log-Q-def})
of the topological charge,
but the results for the sine definition (\ref{simp-Q-def})
can be obtained in a similar manner as we mention at the end of 
Section \ref{sec:analytic_obs}.


The K-functional depends on the link variables on the boundary 
$C=\partial A$, but due to the gauge invariance,
it actually depends only on
\begin{equation}
\Gamma =\prod_{U_{i}\in C}U_{i} \ ,
\label{def-gamma}
\end{equation}
which is a consecutive product of link variables along the loop $C$.
The choice of the starting point of the loop $C$ does not matter
since a different choice simply corresponds to making
a gauge transformation of $\Gamma$, which leaves the K-functional
invariant.

We can calculate the K-functional for any $A$ by gluing the
K-functional for a single plaquette $P$, which is nothing but
\begin{equation}
K(P)=\exp\mathrm{Tr}
\left[\frac{\beta}{2}\left(P+P^{-1}\right)+
\frac{\theta}{2\pi}\log P\right] \ .
\label{def-K-P}
\end{equation}
Note here that (\ref{def-K-P}) 
is a function of the group element $P\in \mathrm{U}(N)$,
which is invariant under
\begin{equation}
P\rightarrow  g P g^{-1} \ ; \qquad g\in \mathrm{U}(N) \ .
\end{equation}
It is known that any function 
having this property
can be 
expressed by the so-called character expansion
\begin{equation}
K(P)=\sum_{r}\lambda_{r}\chi_{r}(P) \ ,
\label{character-expansion-K}
\end{equation}
which is analogous to the Fourier expansion for periodic functions.
Here $\chi_{r}(P)$ is the group character,
which is defined by the trace of $P$ 
for an irreducible representation $r$, and
it satisfies the orthogonality relation
\begin{equation}
\int dU \, \chi_{r_{1}}(U^{-1})\chi_{r_{2}}(U)
=\delta_{r_{1},r_{2}} \ .
\label{eq:orthogonality}
\end{equation}
Using this relation, the coefficient $\lambda_{r}$ in the expansion
(\ref{character-expansion-K}) can be readily obtained as
\begin{equation}
\lambda_{r}=\int dU \, \chi_{r}(U^{-1})K(U) \ .
\end{equation}

\begin{figure}
\centering
\includegraphics[scale=0.8]{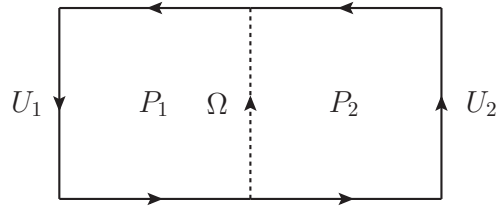}
\caption{The K-functional $K_{2\times1}$ for a $2\times1$ rectangle
is obtained by
considering the K-functional for
the two plaquettes $P_{1}=U_{1}\Omega$ and $P_{2}=\Omega^{-1}U_{2}$,
which are glued together
by integrating out the shared link variable $\Omega$.
}
\label{fig:gluing-plaquettes} 
\end{figure}

As an example, 
let us obtain the K-functional $K_{2\times1}$ for a $2\times1$
rectangle by gluing two neighboring plaquettes $P_{1}=U_{1}\Omega$
and $P_{2}=\Omega^{-1}U_{2}$ as shown in Fig.~\ref{fig:gluing-plaquettes}.
The group elements $U_{1}$ and $U_{2}$ are 
the products of three link variables, 
and $\Omega$ represents the link variable shared by $P_{1}$ and $P_{2}$.
Integrating out the shared link variable $\Omega$, we get
\begin{align}
K_{2\times1}(U_{1}U_{2}) & =\int d\Omega \, K(P_{1}) K(P_{2}) \nonumber \\
 & =\sum_{r_{1},r_{2}}\lambda_{r_{1}}\lambda_{r_{2}}\int d\Omega
\, \chi_{r_{1}}(U_{1}\Omega)\chi_{r_{2}}(\Omega^{-1}U_{2})\nonumber \\
 & =\sum_{r}d_{r}\left(\frac{\lambda_{r}}{d_{r}}\right)^{2}
\chi_{r}(U_{1}U_{2}) \ ,
\end{align}
where $d_{r}=\chi_{r}(1)$ is the dimension of the representation $r$
and we have used a formula
\begin{equation}
\int d\Omega \,
\chi_{r_{1}}(U_{1}\Omega)\chi_{r_{2}}(\Omega^{-1}U_{2})
=\frac{1}{d_{r_{1}}}\chi_{r_{1}}(U_{1}U_{2})\delta_{r_{1},r_{2}} \ .
\end{equation}
Iterating this procedure, we obtain the K-functional 
for any simply connected region $A$ as
\begin{equation}
K_{A}(\Gamma)=\sum_{r}d_{r}
\left(\frac{\lambda_{r}}{d_{r}}\right)^{\left|A\right|}\chi_{r}(\Gamma) \ ,
\label{eq:K-func_of_A}
\end{equation}
where $\left|A\right|$ is the number of plaquettes in $A$, 
and $\Gamma$ is defined by \eqref{def-gamma}.

In the U(1) case, the representation can be labeled by the charge 
$n\in\mathbb{Z}$,
and the dimension of the representation
is $d_{n}=1$ for $\forall n\in\mathbb{Z}$.
Since the character for the plaquette $P=e^{i \phi}$
is given by $\chi_{n}(P)=e^{in\phi}$,
the K-functional for a single plaquette 
(\ref{character-expansion-K}) reduces to
\begin{equation}
K(P)=\sum_{n=-\infty}^{+\infty}\lambda_{n}e^{in\phi} \ ,
\end{equation}
where the coefficient $\lambda_{n}$ is a function of
$\theta$ and $\beta$ given explicitly as
\begin{align}
\lambda_{n} & \equiv \mathcal{I}(n,\theta,\beta) \nn \\
&= \frac{1}{2\pi}\int_{-\pi}^{\pi}
d\phi \, e^{-in\phi}  K\left(P=e^{i\phi}\right)
\nn \\
&=\frac{1}{2\pi}\int_{-\pi}^{\pi}
d\phi\,
\exp \left[ 
\beta\cos\phi +
i\left(\frac{\theta}{2\pi}-n\right)\phi \right]
\label{eq:integral}
\end{align}
using \eqref{def-K-P} with $P=e^{i \phi}$.
This function reduces to the modified Bessel function of the first kind
for $\theta=0$.

The character expansion in the U($N$) case is more complicated, 
so we only show the end results referring the reader, for instance,
to the appendix of Ref.~\cite{Drouffe:1983fv} for the details.
The representation of the U($N$)
group is labeled by $N$ integers
\begin{equation}
\rho=\left(\rho_{1},\rho_{2},\cdots,\rho_{N}\right)
\in\mathbb{Z}^{N} 
\end{equation}
satisfying $\rho_{i}\geq\rho_{i+1}$,
and the dimension of the representation $\rho$ can be calculated by
\begin{equation}
d_{\rho}=
\chi_{\rho}(1)=\prod_{i>j}^{N}\left(1-\frac{\rho_{i}-\rho_{j}}{i-j}\right) \ .
\end{equation}
The coefficient $\lambda_{\rho}$
in (\ref{character-expansion-K})
that corresponds to the representation $\rho$
is expressed as a
determinant
\begin{equation}
\lambda_{\rho}=\det\mathcal{M}(\rho,\theta,\beta) \ ,
\end{equation}
where the matrix $\mathcal{M}(\rho,\theta,\beta)$ is given as
\begin{equation}
\mathcal{M}_{jk} (\rho,\theta,\beta)
=\frac{1}{2\pi}\int_{-\pi}^{\pi}d\phi \,
\exp \left[ 
\beta\cos\phi +
i\left(\frac{\theta}{2\pi}+\rho_{k}+j-k\right)\phi
\right] \ ,
\end{equation}
which may be viewed as a generalization of \eqref{eq:integral}.

\subsection{partition function for the non-punctured model
\label{subsec:Torus}}

\begin{figure}
\centering
\includegraphics[scale=0.8]{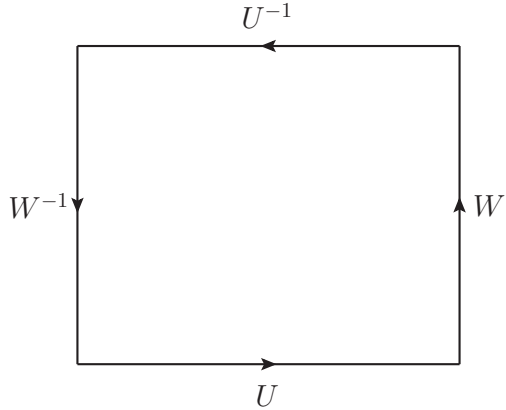}
\caption{The partition function for the 2D U($N$) gauge theory
on a torus is obtained
from the K-functional for the rectangle
by integrating out the group elements $U$ and $W$
corresponding to the identified sides.}
\label{fig:torus} 
\end{figure}

Let us evaluate the partition function 
for the 2D U($N$) lattice gauge theory on a torus.
For that, we first consider the K-functional 
$K_{L_1 \times L_2 }$
for a rectangle
composed of $V = L_{1}\times L_{2}$ plaquettes,
which can be expressed as \eqref{eq:K-func_of_A}. 
As is shown in Fig.~\ref{fig:torus},
we identify
the top and bottom sides represented by $U^{-1}$ and $U$, respectively,
and identify the left and right sides
represented by $W^{-1}$ and $W$, respectively.
Integrating out the group elements $U$ and $W$,
we obtain the partition function for the non-punctured model as
\begin{align}
Z_{\mathrm{nonpunc}} & =\int dUdW \, K_{L_1 \times L_2 }(UWU^{-1}W^{-1})\nonumber \\
 & =\sum_{r}d_{r}
\left(\frac{\lambda_{r}}{d_{r}}\right)^{V}
\int dUdW\chi_{r}(UWU^{-1}W^{-1})\nonumber \\
 & =\sum_{r}\left(\frac{\lambda_{r}}{d_{r}}\right)^{V}
\int dU\chi_{r}(U)\chi_{r}(U^{-1})\nonumber \\
 & =\sum_{r}\left(\frac{\lambda_{r}}{d_{r}}\right)^{V} \ ,
\label{part-fn-torus}
\end{align}
where we have used the orthogonality relation \eqref{eq:orthogonality} 
and a formula
\begin{equation}
\int d\Omega \, \chi_{r}(U\Omega W\Omega^{-1})
=\frac{1}{d_{r}}\chi_{r}(U)\chi_{r}(W) \ .
\end{equation}
In the U(1) case, the partition function (\ref{part-fn-torus}) 
reduces to
\begin{equation}
Z_{\mathrm{nonpunc}} = \sum_{n=-\infty}^{+\infty}
\left[\mathcal{I}(n,\theta,\beta)\right]^{V} \ .
\label{eq:Z_U(1)}
\end{equation}
As one can see from (\ref{eq:integral}),
the integral $\mathcal{I}(n,\theta,\beta)$
has a property $\mathcal{I}(n,\theta+2\pi k,\beta) =
\mathcal{I}(n-k,\theta,\beta)$ for $\forall k\in\mathbb{Z}$, 
which guarantees the $2\pi$ periodicity
of (\ref{eq:Z_U(1)}) in $\theta$.

Let us consider taking the $V\rightarrow\infty$
and $\beta \rightarrow \infty$ limits simultaneously
with fixed $V_{\rm phys} \equiv V/\beta$,
which corresponds to the continuum limit.
In this limit, the integral \eqref{eq:integral}
can be evaluated as
\begin{align}
\mathcal{I}(n,\theta,\beta) & \simeq
\frac{1}{\sqrt{2\pi\beta}}
e^{\beta - \frac{1}{2\beta} \left( \frac{\theta}{2\pi} -n \right)^2}  \ .
\end{align}
Plugging this into \eqref{eq:Z_U(1)}, we obtain
\begin{align}
Z_{\mathrm{nonpunc}}
& \simeq
\left(\frac{e^{\beta}}{\sqrt{2\pi\beta}}\right)^{V}
\sum_{n=-\infty}^{+\infty}\exp\left[-\frac{V}{2\beta}
\left(\frac{\theta}{2\pi}-n\right)^{2}\right]  \nn \\
& \sim 
\sum_{n=-\infty}^{+\infty}\exp\left[
-\frac{1}{2} V_{\mathrm{phys}}
\left(\frac{\theta}{2\pi}-n\right)^{2}\right]  \ ,
\label{eq:Z_U(1)_cont}
\end{align}
omitting the divergent constant factor.

\subsection{partition function for the punctured model}
\label{sec:part-func-punc}

\begin{figure}
\centering
\includegraphics[scale=0.8]{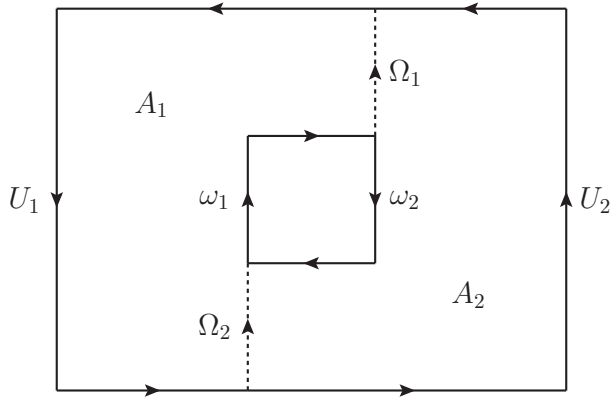}
\caption{The K-functional for 
a rectangle with a hole 
is obtained by gluing 
the two regions $A_{1}$ and $A_{2}$.
From this, the K-functional
for the punctured torus 
is obtained similarly to
what we did in Fig.~\ref{fig:torus}.
Integrating out the link variables surrounding the puncture,
we obtain the partition function for the 2D U($N$) gauge theory
on a punctured torus.
}
\label{fig:punctured-torus} 
\end{figure}

Let us extend the calculation in the previous section
to the punctured model.
First, we calculate
the K-functional for a rectangle with a hole
shown in Fig.~\ref{fig:punctured-torus},
which we divide into two regions $A_{1}$ and $A_{2}$
by cutting along two segments $\Omega_{1}$ and $\Omega_{2}$.
The outer and inner boundaries 
of the rectangle are divided into two segments
$(U_{1} , U_{2})$ and $(\omega_{1} , \omega_{2})$, respectively.
Then, the K-functional for each region is given, respectively, as
\begin{align}
K_{A_{1}}(U_{1}\Omega_{2}\omega_{1}\Omega_{1}) 
&= 
\sum_{r_{1}} d_{r_{1}}
\left(\frac{\lambda_{r_{1}}}{d_{r_{1}}}\right)^{\left|A_{1}\right|}
\chi_{r_{1}}(U_{1}\Omega_{2}\omega_{1}\Omega_{1})  \ ,
\\
K_{A_{2}}(\Omega_{1}^{-1}\omega_{2}\Omega_{2}^{-1}U_{2})
&= 
\sum_{r_{2}} d_{r_{2}}
\left(\frac{\lambda_{r_{2}}}{d_{r_{2}}}\right)^{\left|A_{2}\right|}
\chi_{r_{2}}(\Omega_{1}^{-1}\omega_{2}\Omega_{2}^{-1}U_{2}) \ .
\end{align}
By gluing the two regions $A_{1}$ and $A_{2}$ together
at $\Omega_{1}$ and $\Omega_{2}$,
we obtain the K-functional for the rectangle with a hole as
\begin{align}
K_{A_{1} \cup A_{2}} & =
\int d\Omega_{1}d\Omega_{2} \, 
K_{A_{1}}(U_{1}\Omega_{2}\omega_{1}\Omega_{1})
K_{A_{2}}(\Omega_{1}^{-1}\omega_{2}\Omega_{2}^{-1}U_{2})
\nonumber \\
 & =\sum_{r_{1},r_{2}}d_{r_{1}}d_{r_{2}}
\left(\frac{\lambda_{r_{1}}}{d_{r_{1}}}\right)^{\left|A_{1}\right|}
\left(\frac{\lambda_{r_{2}}}{d_{r_{2}}}\right)^{\left|A_{2}\right|}
\int d\Omega_{1}d\Omega_{2}\chi_{r_{1}}(U_{1}\Omega_{2}\omega_{1}\Omega_{1})
\chi_{r_{2}}(\Omega_{1}^{-1}\omega_{2}\Omega_{2}^{-1}U_{2})\nonumber \\
 & =\sum_{r}d_{r}\left(\frac{\lambda_{r}}{d_{r}}\right)^{\left|A_{1}\right|}
\left(\frac{\lambda_{r}}{d_{r}}\right)^{\left|A_{2}\right|}
\int d\Omega_{2}\chi_{r}(U_{1}\Omega_{2}\omega_{1}\omega_{2}\Omega_{2}^{-1}U_{2})
\nonumber \\
 & =\sum_{r}\left(\frac{\lambda_{r}}{d_{r}}\right)^{V}
\chi_{r}(U_{1}U_{2})\chi_{r}(\omega_{1}\omega_{2}) \ ,
\end{align}
where we have defined $V=|A_{1}\cup A_{2}|=|A_{1}|+|A_{2}|$.

Let us introduce the group elements $U$ and $W$ 
for the outer boundary as we did in Fig.~\ref{fig:torus}
so that $U_{1}U_{2} = UWU^{-1}W^{-1}$,
and define $\omega=\omega_{1}\omega_{2}$ for the inner boundary.
Integrating out the group elements $U$ and $W$,
we obtain the K-functional for the punctured torus as
\begin{align}
K_{\mathrm{punc}}(\omega) 
& =\sum_{r}\left(\frac{\lambda_{r}}{d_{r}}\right)^{V}
\chi_{r}(\omega)\int dUdW\chi_{r}(UWU^{-1}W^{-1})
\nonumber \\
 & =\sum_{r}\left(\frac{\lambda_{r}}{d_{r}}\right)^{V}
\chi_{r}(\omega)\int dU\frac{1}{d_{r}}\chi_{r}(U)\chi_{r}(U^{-1})
\nonumber \\
 & =\sum_{r} \frac{1}{d_{r}} \left(\frac{\lambda_{r}}{d_{r}}\right)^{V}
\chi_{r}(\omega) \  .
\end{align}
Finally, 
we integrate out the link variables surrounding the puncture
to get the partition function for the punctured model as
\begin{align}
Z_{\mathrm{punc}} 
& =\int d\omega \, K_{\mathrm{punc}}(\omega) 
 =\sum_{r} \frac{1}{d_{r}} \left(\frac{\lambda_{r}}{d_{r}}\right)^{V}
\delta_{r,0}
 =\left(\lambda_{0}\right)^{V} \ ,
\end{align}
where $r=0$ corresponds to the trivial representation, 
which has $d_0 = 1$.

In the U(1) case, the partition function reduces to
\begin{equation}
Z_{\mathrm{punc}}
=\left[\mathcal{I}(0,\theta,\beta)\right]^{V} \ ,
\label{eq:Z_Tx_U(1)}
\end{equation}
which does not have the $2\pi$ periodicity in $\theta$.

Let us consider taking the $V\rightarrow\infty$
and $\beta \rightarrow \infty$ limits simultaneously
with fixed $V_{\rm phys} \equiv V/\beta$,
which corresponds to the continuum limit.
Similarly to the case of the non-punctured model discussed
in Section \ref{subsec:Torus},
we obtain
\begin{align}
Z_{\mathrm{punc}} 
& \simeq\left(\frac{e^{\beta}}{\sqrt{2\pi\beta}}\right)^{V}
\exp\left[-\frac{V}{2\beta}\left(\frac{\theta}{2\pi}\right)^{2}\right]
\nonumber \\
 & \sim\exp\left[-\frac{1}{2} V_{\mathrm{phys}}
\left(\frac{\theta}{2\pi}\right)^{2}\right] \ ,
\end{align}
omitting the divergent constant factor.
This coincides with \eqref{eq:Z_U(1)_cont} in the 
$V_{\mathrm{phys}}\rightarrow\infty$ limit for $|\theta|<\pi$.
Note, however, that the equivalence between the punctured and
non-punctured models does not hold for finite $V_{\mathrm{phys}}$.


\subsection{evaluation of the observables}
\label{sec:analytic_obs}

We can evaluate the expectation values of various observables
defined in Section \ref{sec:result_nonpunctured}
from the partition function derived above,
namely \eqref{eq:Z_U(1)} for the non-punctured model
and \eqref{eq:Z_Tx_U(1)} for the punctured model.
Since the latter case is easier
due to the absence of an infinite sum,
we only discuss the former case in what follows.

The average plaquette $w$ defined by (\ref{eq:def_aveP})
is given as
%
\begin{align}
  w
  & =\frac{1}{Z_{\mathrm{nonpunc}}}
  \sum_{n=-\infty}^{+\infty}\mathcal{A}(n,\theta,\beta)
  \left[\mathcal{I}(n,\theta,\beta)\right]^{V} \ ,
  \label{eq:analytic_w}
\end{align}
where we have defined
\begin{align}
  \mathcal{A}(n,\theta,\beta)
   & =\frac{\partial}{\partial\beta}
   \log\mathcal{I}(n,\theta,\beta)\nonumber \\
  & =\frac{1}{\mathcal{I}(n,\theta,\beta)}
  \frac{1}{2\pi}\int_{-\pi}^{\pi}
  d\phi  \, \cos\phi 
\exp \left[ 
\beta\cos\phi +
i\left(\frac{\theta}{2\pi}-n\right)\phi \right]
\nonumber \\
 & =\frac{\mathcal{I}(n-1,\theta,\beta)+
    \mathcal{I}(n+1,\theta,\beta)}{2\mathcal{I}(n,\theta,\beta)} \ .
  \label{def-calA}
\end{align}
Similarly,
the topological charge density defined by (\ref{eq:def_ImQ})
can be obtained from
\begin{align}
  \langle Q\rangle
  & =-i\frac{V}{Z_{\mathrm{nonpunc}}}
  \sum_{n=-\infty}^{+\infty}\mathcal{B}(n,\theta,\beta)
  \left[\mathcal{I}(n,\theta,\beta)\right]^{V} \ ,
  \label{eq:analytic_q}  
\end{align}
where we have defined
\begin{align}
  \mathcal{B}(n,\theta,\beta) &
  =\frac{1}{\mathcal{I}(n,\theta,\beta)}
  \frac{\partial}{\partial\theta}\mathcal{I}(n,\theta,\beta)\nonumber \\
  & =\frac{i}{\mathcal{I}(n,\theta,\beta)}
  \frac{1}{4\pi^{2}}\int_{-\pi}^{\pi}
    d\phi \, \phi 
\exp \left[ 
\beta\cos\phi +
i\left(\frac{\theta}{2\pi}-n\right)\phi \right]
\ .
\label{def-calB}
\end{align}
Finally, the topological susceptibility defined by
\eqref{eq:def_chi} can be obtained from
\begin{align}
  \langle Q^{2}\rangle
  & =-\frac{V}{Z_{\mathrm{nonpunc}}}
  \sum_{n=-\infty}^{+\infty}\left[\mathcal{C}(n,\theta,\beta)+
    (V-1)\mathcal{B}(n,\theta,\beta)^{2}\right]
  \left[\mathcal{I}(n,\theta,\beta)\right]^{V} \ ,
  \label{eq:analytic_chi}
\end{align}
where we have defined
\begin{align}
  \mathcal{C}(n,\theta,\beta) &
   =\frac{1}{\mathcal{I}(n,\theta,\beta)}
   \frac{\partial^{2}}{\partial\theta^{2}}\mathcal{I}(n,\theta,\beta)\nonumber \\
  & = -  \frac{1}{\mathcal{I}(n,\theta,\beta)}
  \frac{1}{8\pi^{3}}\int_{-\pi}^{\pi}
  d\phi \, \phi^{2} 
\exp \left[ 
\beta\cos\phi +
i\left(\frac{\theta}{2\pi}-n\right)\phi \right]
\ .
  \label{def-calC}
\end{align}

Note that $\mathcal{I}(n,\theta,\beta)$ and the functions
\eqref{def-calA}, \eqref{def-calB} and \eqref{def-calC}
derived from it are all real-valued,
and we can calculate them by numerical integration with sufficient precision.
Also, when we evaluate the infinite sum
in the expressions \eqref{eq:analytic_w}, \eqref{eq:analytic_q}
and \eqref{eq:analytic_chi},
we have to truncate it at some $n$.
Note here that $|\mathcal{I}(n,\theta,\beta)|$
vanishes
quickly as $|\theta/2\pi-n|$ increases.
We can therefore evaluate the infinite sum
with
sufficient precision
by keeping only a few terms
when the lattice volume $V$ is sufficiently large.


In this section, we have derived the exact results for
the log definition (\ref{log-Q-def}) of the topological charge.
As is clear from the derivation,
we can obtain the exact
results for the sine definition (\ref{simp-Q-def})
by simply replacing $\mathcal{I}(n,\theta,\beta)$ with
\begin{equation}
  \tilde{\mathcal{I}}(n,\theta,\beta)
  =\frac{1}{2\pi}\int_{-\pi}^{\pi}d\phi
\exp \left[ 
\beta\cos\phi +
i\frac{\theta}{2\pi} \sin \phi
- i n \phi \right]
\ .
\end{equation}

\section{The punctured model with the sine definition $Q_{\rm sin}$}
\label{sec:results-punctured}

In Sections \ref{sec:introduce-puncture} and \ref{sec:results},
we have discussed the punctured model
with the log definition (\ref{log-Q-def}) of the topological charge
for simplicity.
In fact, we can also use
the sine definition
\eqref{simp-Q-def}
in the punctured model.
Here we discuss what happens in this case.




The drift terms for the sine definition
are given already for the non-punctured model in Section \ref{CLMU1}.
When we consider the punctured model,
the only modification from the non-punctured model appears
in the drift terms for the four link variables surrounding the puncture;
i.e., $\mathcal{U}_{\punc,1}$, $\mathcal{U}_{\punc + \hat{2},1}$,
$\mathcal{U}_{\punc,2}$ and $\mathcal{U}_{\punc + \hat{1},2}$.
Thus the drift terms are given as
\begin{alignat}{2}
D_{n,1}S & =
\left\{
\begin{array}{ll}
  -i\frac{\beta}{2}(P_{n}-P_{n}^{-1}-P_{n-\hat{2}}+P_{n-\hat{2}}^{-1})
  -i\frac{\theta}{4\pi}(P_{n}+P_{n}^{-1}-P_{n-\hat{2}}-P_{n-\hat{2}}^{-1})\\
 &\hspace{-80pt}\mbox{~for~}n \neq \punc , \ \punc+\hat{2}  \ , \\
 -i\frac{\beta}{2}(-P_{\punc-\hat{2}}+P_{\punc-\hat{2}}^{-1})
 +i\frac{\theta}{4\pi}(P_{\punc-\hat{2}}+P_{\punc-\hat{2}}^{-1}) &
 \hspace{-80pt}\mbox{~for~}n = \punc   \ , \\
 -i\frac{\beta}{2}(P_{\punc+\hat{2}}-P_{\punc+\hat{2}}^{-1})
 -i\frac{\theta}{4\pi}(P_{\punc+\hat{2}}+P_{\punc+\hat{2}}^{-1})  &
 \hspace{-80pt}\mbox{~for~}n = \punc+\hat{2}   \ ,
\end{array}
\right.
\\[2.5mm]
D_{n,2}S & =
\left\{
\begin{array}{ll}
  -i\frac{\beta}{2}(-P_{n}+P_{n}^{-1}+P_{n-\hat{1}}-P_{n-\hat{1}}^{-1})
  -i\frac{\theta}{4\pi}(-P_{n}-P_{n}^{-1}+P_{n-\hat{1}}+P_{n-\hat{1}}^{-1})\\
 &\hspace{-98pt}\mbox{~for~}n \neq \punc , \ \punc+\hat{1}  \ , \\
 -i\frac{\beta}{2}(P_{\punc-\hat{1}}-P_{\punc-\hat{1}}^{-1})
 -i\frac{\theta}{4\pi}(P_{\punc-\hat{1}}+P_{\punc-\hat{1}}^{-1}) &
 \hspace{-98pt}\mbox{~for~}n = \punc   \ , \\
 -i\frac{\beta}{2}(-P_{\punc+\hat{1}}+P_{\punc+\hat{1}}^{-1})
 +i\frac{\theta}{4\pi}(P_{\punc+\hat{1}}+P_{\punc+\hat{1}}^{-1})  &
\hspace{-98pt}\mbox{~for~}n = \punc+\hat{1}  \ .
\end{array}
\right.
\end{alignat}
At large $\beta$,
all the plaquettes except $P_{\punc}$, namely the one that is removed,
approach unity.
The drift term from the $\theta$ term therefore vanishes for all the
link variables except for those surrounding the puncture,
which have
constant drifts $\pm i\frac{\theta}{2\pi}$.
Thus in the continuum limit, the drift terms for the sine definition
agree with
those for the log definition given
by \eqref{eq:drift_theta_smp_1} and \eqref{eq:drift_theta_smp_2}.
This connection makes it easier to understand why we can safely
ignore the issue of $\delta$-function
in the drift term
for the log definition described in Section \ref{CLMU1}.



\begin{figure}
\centering
\includegraphics[scale=0.6]{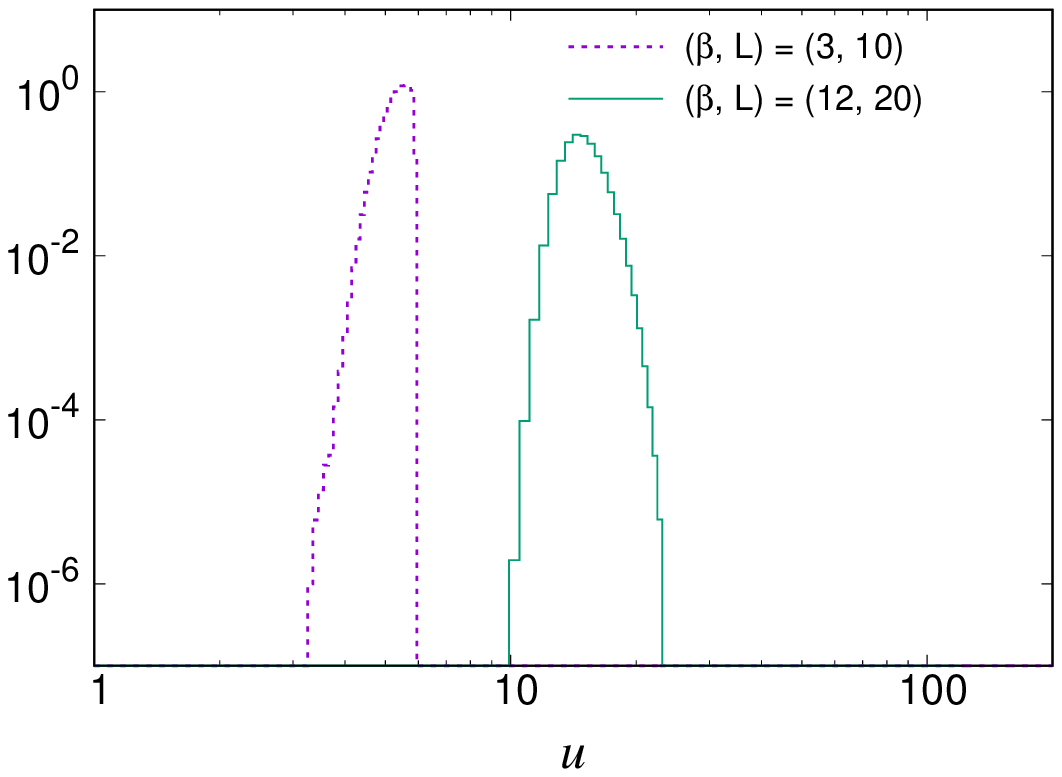}
\includegraphics[scale=0.6]{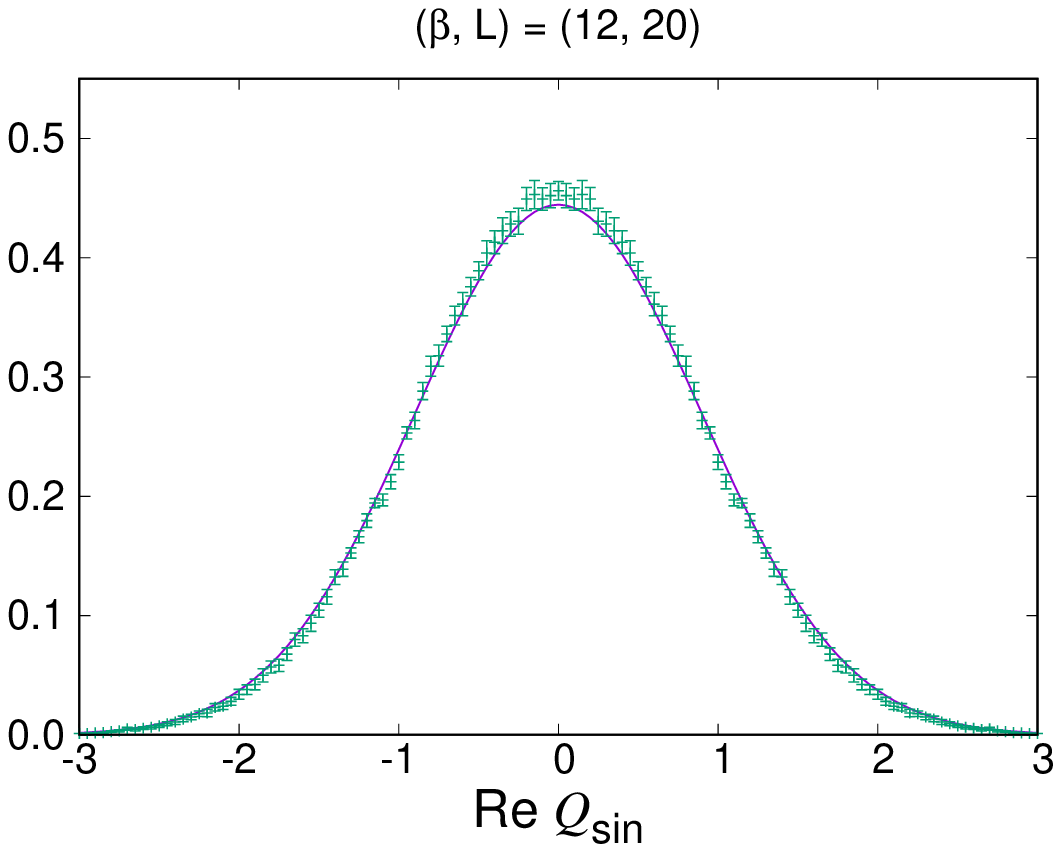}
\caption{The results obtained by the CLM
  for the punctured model using the sine definition of
the topological charge.
(Left) The histogram of the magnitude $u$ of the drift term 
is shown for $(\beta,L)=(3,10)$ and $(12,20)$ with $\theta=\pi$.
(Right) The histogram of $\mathrm{Re} \, Q_{\mathrm{sin}}$
for the punctured model 
is shown for $(\beta, L)=(12,20)$ with $\theta=\pi$.
The exact result
obtained for $(\beta, L)=(12,20)$ with $\theta=0$
is shown by the solid line for comparison.}
\label{fig:compare_drift_simp}
\end{figure}

It is therefore expected that
the results of the CLM for the sine definition
are essentially the same as those for the log definition
for large $\beta$.
In Fig.~\ref{fig:compare_drift_simp}, we show
our results for the punctured model with the sine definition
for the same $(\beta,L)$ as those
in Fig.~\ref{fig:compare_drift} with the log definition.
For $(\beta,L)=(12,20)$,
we find that the histogram of the magnitude $u$ of the drift term
falls off rapidly, and that
the histogram of $\mathrm{Re} \, Q_\text{sin}$ obtained by the CLM
is widely distributed
within the range $-3 \lesssim \mathrm{Re} \, Q_\text{sin}\lesssim 3$.
Hence the topology freezing problem is circumvented without
causing large drifts similarly to the situation with the log definition.

\begin{figure}
\centering
{\includegraphics[scale=0.6]{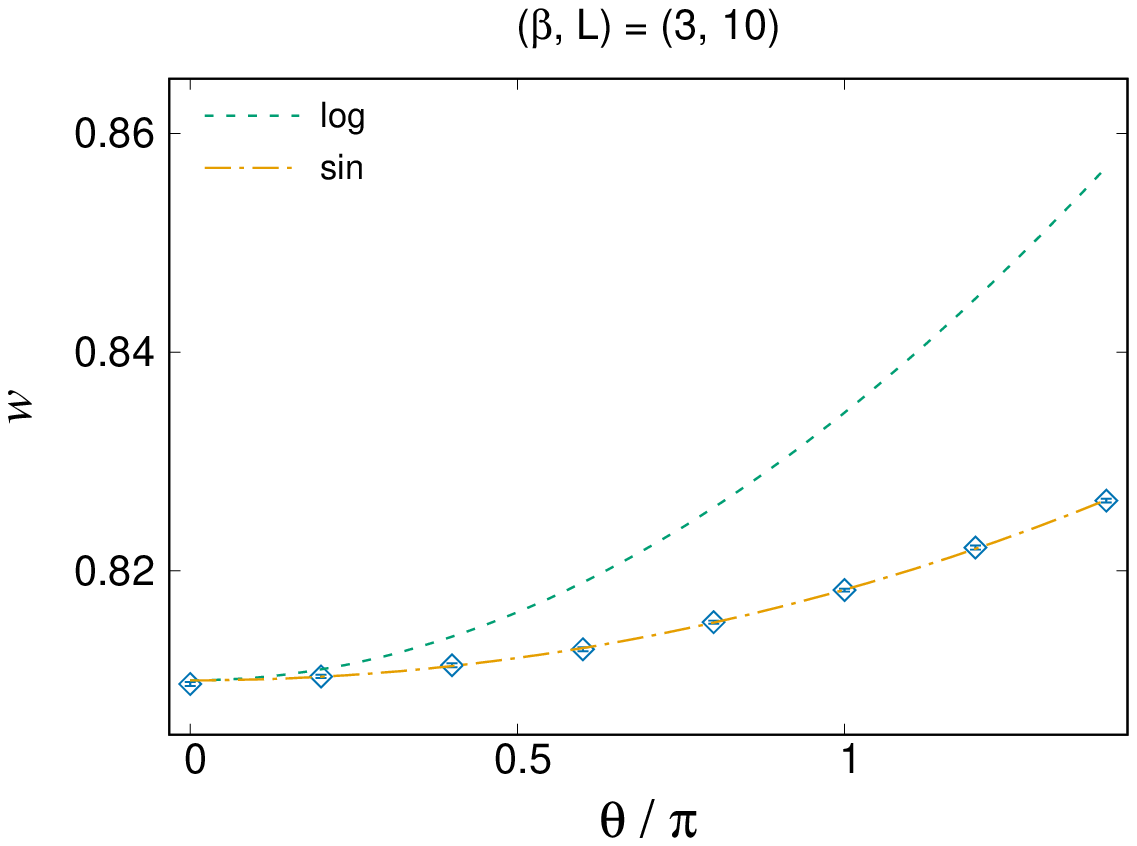}}
{\includegraphics[scale=0.6]{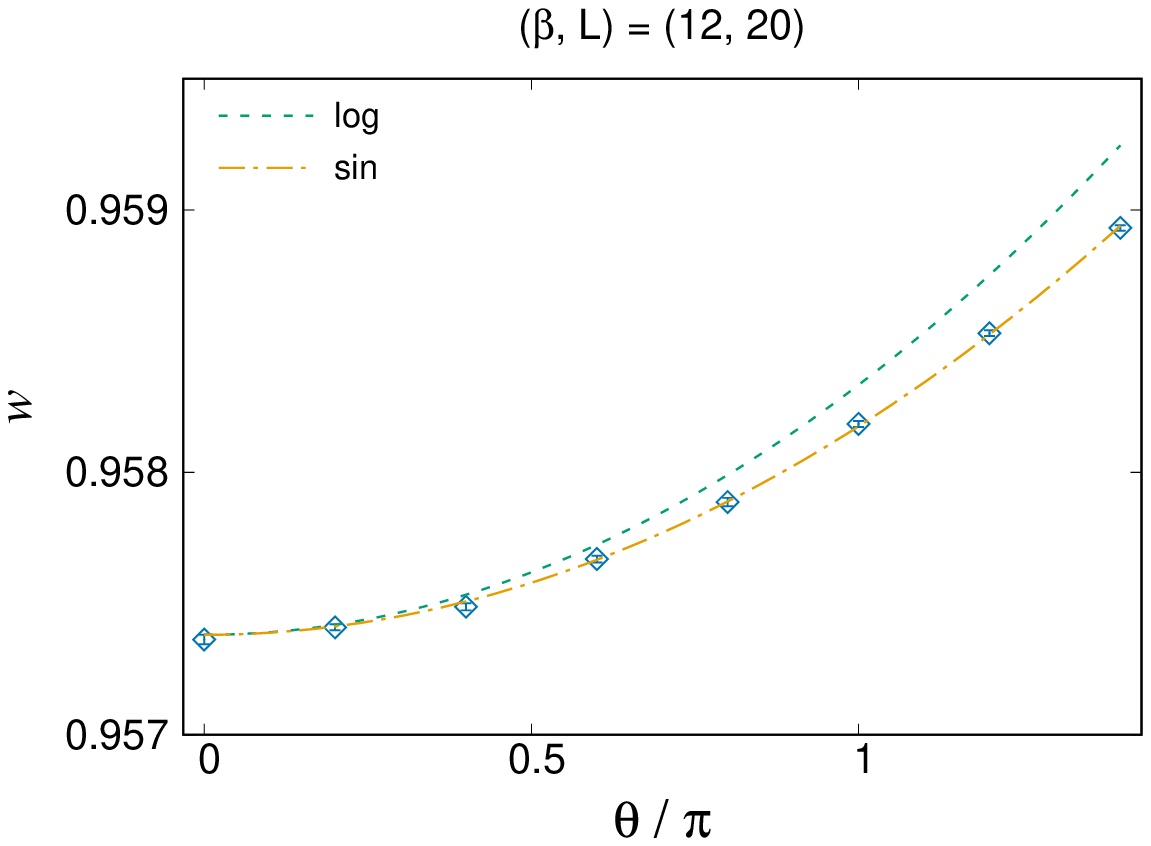}}\\
\vspace{10pt}
{\includegraphics[scale=0.6]{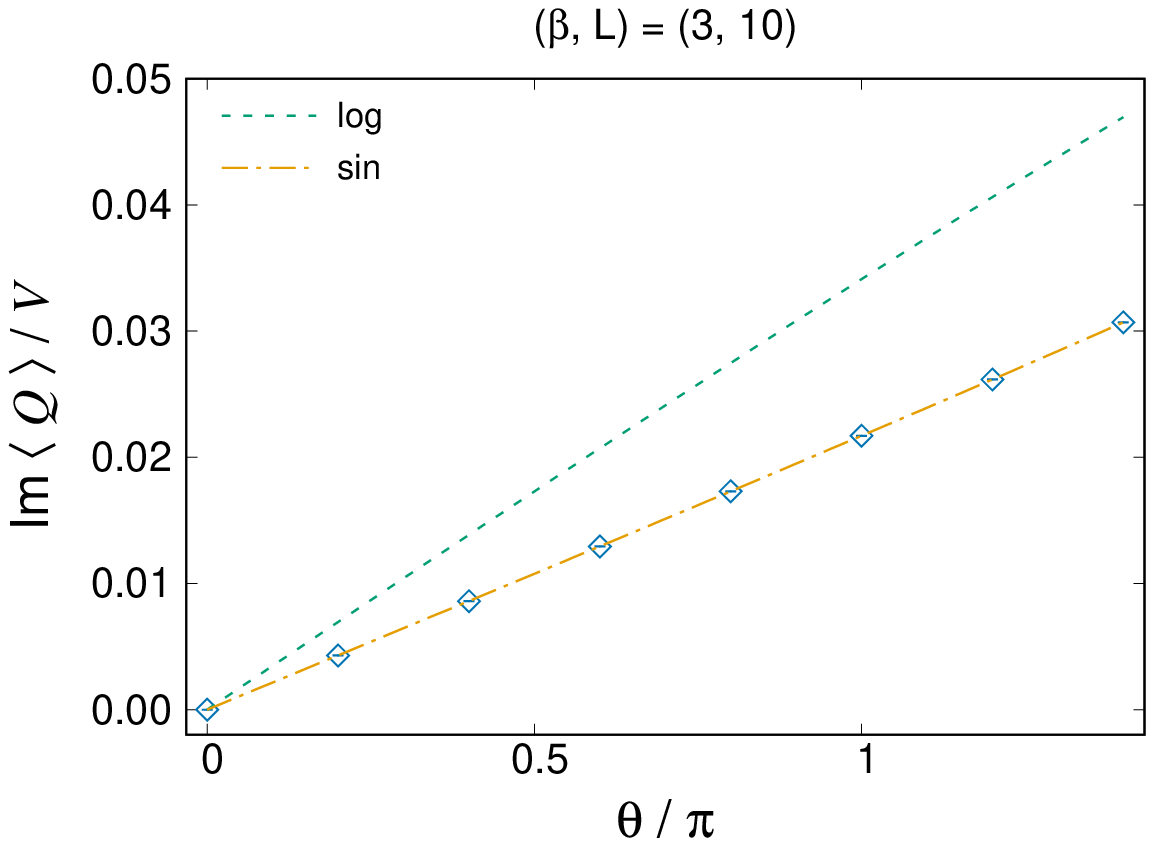}}
{\includegraphics[scale=0.6]{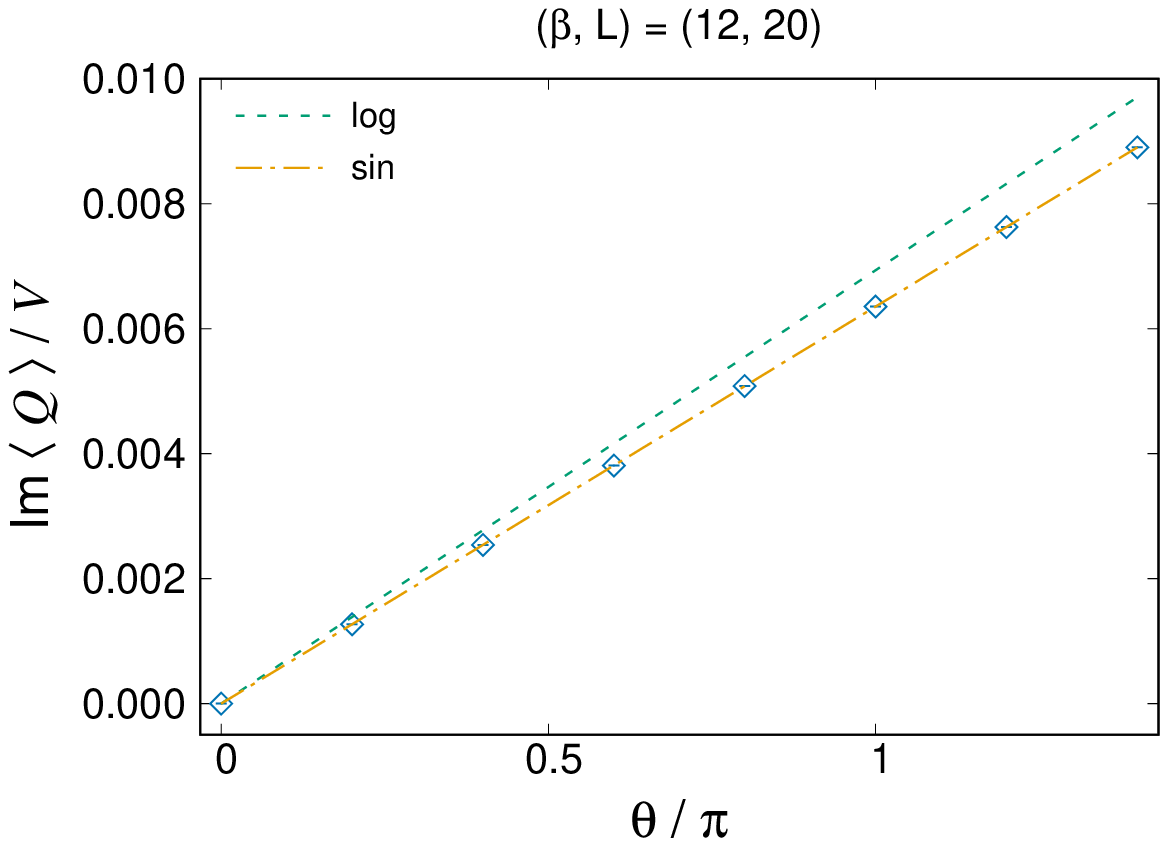}}\\
\vspace{10pt}
{\includegraphics[scale=0.6]{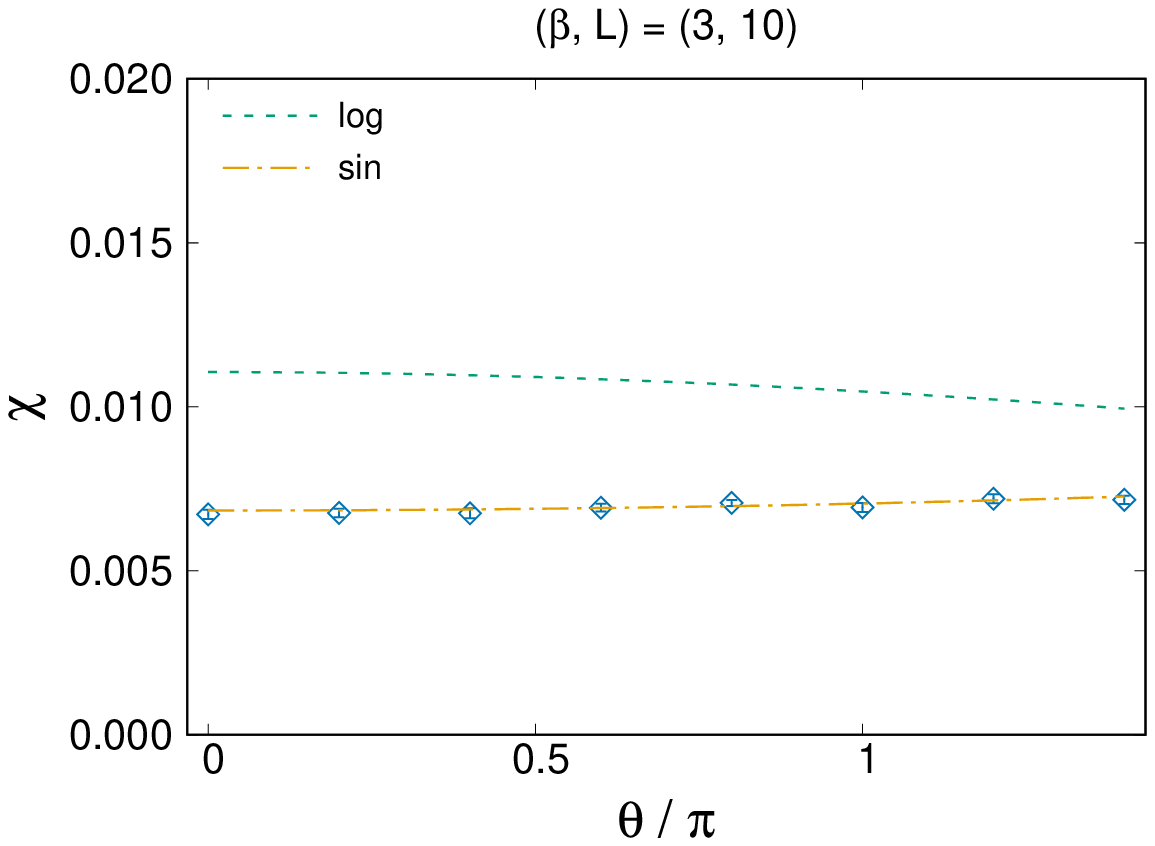}}
{\includegraphics[scale=0.6]{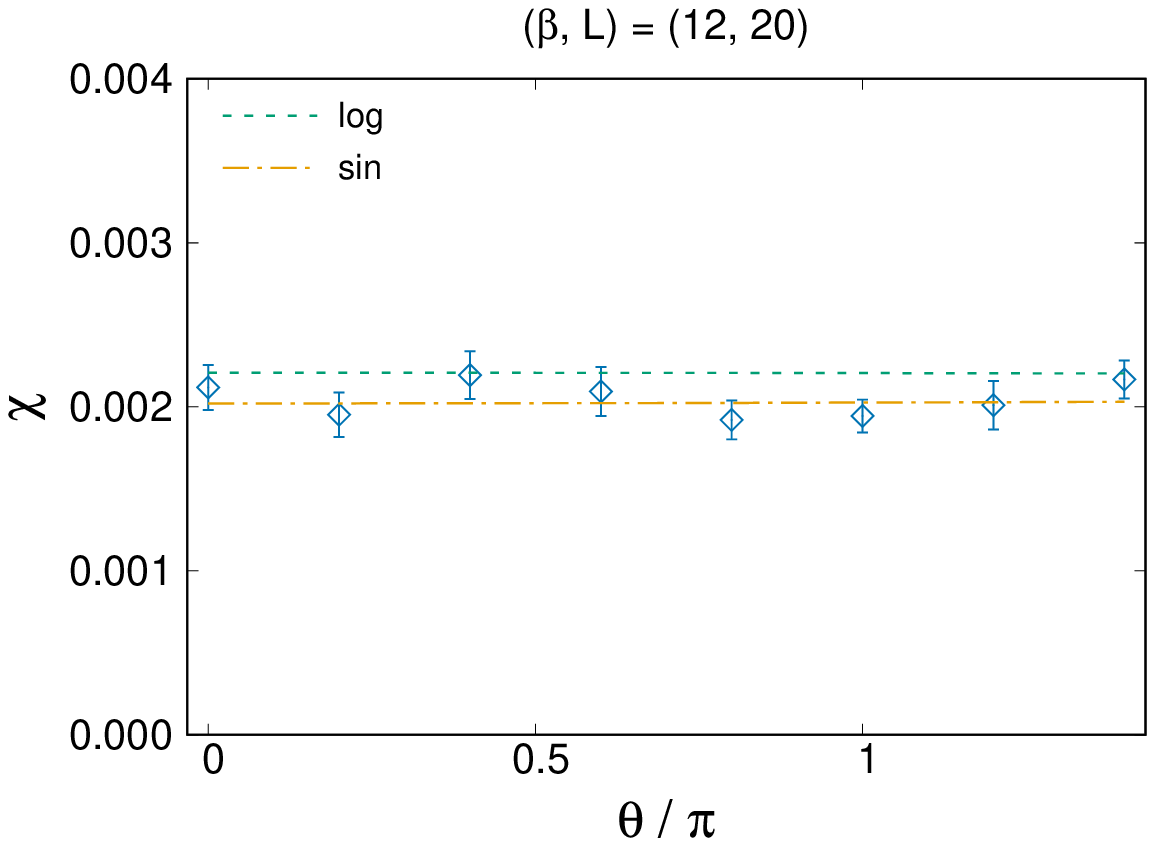}}
\caption{The results for various observables obtained by the CLM 
  for the punctured model with the sine definition $Q_{\mathrm{sin}}$.
  The average plaquette (Top),
the imaginary part of the topological charge density (Middle),
the topological susceptibility (Bottom)
are plotted against $\theta$
for $(\beta, L)=(3,10)$ (Left) and $(12,20)$ (Right).
The exact results for the punctured model
with the log and sine definitions
are shown for the same $(\beta, L)$
by the dashed lines and the dash-dotted lines, respectively,
for comparison.}
\label{fig:obs_punc_simp}
\end{figure}

On the other hand, for $(\beta,L)=(3,10)$, 
we find that the histogram of the magnitude $u$ of the drift term
falls off fast and that the condition for the validity of the CLM
is satisfied unlike the case of the log definition.
As a result, all the observables are in complete agreement 
with the exact results for all values of $\theta$
even with $(\beta,L)=(3,10)$.
This can be seen from Fig.~\ref{fig:obs_punc_simp},
where we show our results for the punctured model
with the sine definition
for the same values of $(\beta,L)$ as the ones used
in Fig.~\ref{fig:obs_punc}.
For $(\beta,L)=(1.92,8)$
corresponding to the same $V_{\rm phys}\equiv L^2/\beta$,
however,
we actually find that the histogram has a power-law tail similarly
to the case of the log definition. Therefore, the difference
between the two definitions is merely
a small shift in the validity region of the CLM.

We also show the exact results for the punctured model
with the log and sine definitions, which tend to agree
as $\beta$ is increased with fixed $V_{\rm phys} \equiv L^2/\beta$,
which corresponds to the continuum limit.


\newpage 

\bibliographystyle{JHEP}
\bibliography{ref}

\end{document}